%% file: main.tex
\documentclass[a4paper,11pt]{article}
\pdfoutput=1 

\usepackage{jcappub}
\usepackage{mathtools}
\usepackage{color}
\usepackage[dvipsnames]{xcolor}
\usepackage{tcolorbox}
\tcbuselibrary{theorems}
\usepackage{orcidlink}
\usepackage{fontawesome5}
\usepackage{hyperref}
\usepackage{soul} 
 
\usepackage[T1]{fontenc} 

\newcommand{\threej}[6]{ \begin{pmatrix}
  #1 & #2 & #3 \\
  #4 & #5 & #6 
 \end{pmatrix}}

\newcommand{\sixj}[6]{ \begin{Bmatrix}
  #1 & #2 & #3 \\
  #4 & #5 & #6 
 \end{Bmatrix}}

\newcommand{\ninej}[9]{ \begin{Bmatrix}
  #1 & #2 & #3 \\
  #4 & #5 & #6 \\
  #7 & #8 & #9
 \end{Bmatrix}}

\usepackage{amsthm}

\def\x{{\bf x}}
\def\y{{\bf y}}
\def\k{{\bf k}}

\def\s{{\bf s}}
\def\r{{\bf r}}

\def\hk{{\bf \hat k}}
\def\hx{{\bf \hat x}}
\def\hy{{\bf \hat y}}
\def\hs{{\bf \hat s}}
\def\hr{{\bf \hat r}}
\def\hn{{\bf \hat n}}

\def\L{{\mathcal L}}
\def\O{{\mathcal O}}

\def\hP{{\widehat P}}

\def\tQ{{\widetilde Q}}
\def\tV{{\widetilde V}}

\def\tdelta{{\tilde\delta}}

\newcommand\nn{\nonumber}

\renewcommand{\thefootnote}{\fnsymbol{footnote}}

\title{Optimal and exact wide-angle power spectrum estimation}

\author[1,2,3]{Noah Sailer\footnote{Both authors contributed equally to this work.}}
\author[4]{Kendrick Smith\footnotemark[1]}
\author[5]{Yurii Kvasiuk}
\author[6]{Alex Lagu\"e}
\author[4]{Selim Hotinli}

\affiliation[1]{Kavli Institute for Particle Astrophysics and Cosmology, Stanford University and SLAC National Accelerator Laboratory, Stanford, CA 94305, USA}
\affiliation[2]{Berkeley Center for Cosmological Physics, Department of Physics, University of California, Berkeley, CA 94720, USA}
\affiliation[3]{Lawrence Berkeley National Laboratory, Berkeley, CA 94720, USA}
\affiliation[4]{Perimeter Institute for Theoretical Physics, Waterloo ON N2L 2Y5, Canada}
\affiliation[5]{Department of Physics, University of Wisconsin-Madison, Madison, WI 53706, USA}
\affiliation[6]{Department of Physics and Astronomy, University of Pennsylvania, Philadelphia, PA 19104, USA}

\emailAdd{nsailer@stanford.edu}
\emailAdd{kmsmith@perimeterinstitute.ca}

\abstract{
What is the optimal power spectrum estimator on ultra-large scales where the plane-parallel approximation breaks down? Conventional estimators, such as the Yamamoto estimator, are only optimal in the plane-parallel limit, while their associated window functions are typically approximated by truncating a slowly converging infinite series. We address two outstanding challenges in the analysis of wide-angle power spectra. First, we derive the optimal estimator for a broad class of clustering signals and show that it is equivalent to a previously proposed two-$\ell$ generalization of the Yamamoto estimator. Second, we show how to write the exact two-$\ell$ window function as a finite number of terms that can be efficiently evaluated using FFTs. Our results apply to a wide range of observables, including redshift-space distortions (RSDs) and large-scale radial-velocity reconstruction from the kinetic Sunyaev–Zel'dovich effect. Focusing on linear-theory RSDs, we validate the finite window-function representation numerically and show that the two-$\ell$ estimator can yield order-unity improvements in the signal-to-noise ratio of ultra-large-scale power spectrum measurements.
}

\begin{document}
\maketitle
\renewcommand{\thefootnote}{\arabic{footnote}}
\setcounter{footnote}{0}

\flushbottom



\input SecIntro
\input SecPreliminaries
\input SecRank1Response
\input SecUseCases
\input SecRSD
\input SecOptimality
\input SecWindowCondensed
\input SecConclusions

\acknowledgments
We thank Emanuele Castorina, Arnaud de Mattia, Henry Grasshorn Gebhardt, Takahiko Matsubara, Oliver Philcox and Roman Scoccimarro for helpful comments on an earlier version of this draft. 
NS was supported by a Porat Fellowship at the Kavli Institute for Particle Astrophysics and Cosmology and by Lawrence Berkeley National Laboratory during the early stages of this work.
KMS was supported by an NSERC Discovery Grant, by the Daniel Family Foundation, and by the Centre for the Universe at Perimeter Institute.
Research at Perimeter Institute is supported by the Government of Canada through Industry Canada and by the Province of Ontario through the Ministry of Research \& Innovation.
YK acknowledges support from the U.S. Department of Energy, Office of Science, Office of High Energy Physics under Award Number DE-SC0017647, the support by the National Science Foundation (NSF) under Grant Number 2307109 and 2509873 and the Wisconsin Alumni Research Foundation. 
AL acknowledges support from NASA grant 21-ATP21-0145.
SCH was supported by the P. J. E. Peebles Fellowship at Perimeter Institute. 


\bibliographystyle{JHEP}
\bibliography{main.bib}

\appendix

\input AppSpecFunc
\input AppDerivations
\input AppTables

\end{document}

%% file: SecIntro.tex
\section{Introduction}
\label{sec:introduction}

Two point statistics are central in extracting cosmological information from modern large-scale structure surveys such as DESI \cite{2016arXiv161100036D}, Euclid \cite{2025A&A...697A...1E}, Roman \cite{2015arXiv150303757S}, Rubin \cite{2018arXiv180901669T} and SPHEREx \cite{2025arXiv251102985B}. Large-scale linear modes contain largely ``unprocessed'' information about the early universe and provide a promising probe of primordial non-Gaussianity through the scale-dependent bias it induces \cite{2008PhRvD..77l3514D}. On the ultra-large scales where these signals are most prominent, ``wide-angle effects'' must be accounted for in modern surveys \cite{2015MNRAS.452.3704S,2015MNRAS.447.1789Y,Reimberg:2015jma,Beutler:2018vpe,Benabou:2024tmn}, which arise from the breakdown of approximate translational invariance (the ``plane-parallel'', or ``distant-observer'' approximation) on large angular scales. 

In the plane-parallel approximation a fixed line-of-sight direction $\hn$ is shared by all galaxies and the redshift-space two point correlation function (2PCF), $\langle \delta_1(\x_1)\delta_2(\x_2)\rangle$, becomes a function of two variables (e.g. $k$ and $\mu\equiv \hk\cdot\hn$) \cite{1987MNRAS.227....1K,Lilje1989,1992ApJ...385L...5H,1990MNRAS.242..428M}. The Yamamoto estimator \cite{Yamamoto:2005dz} and its FFT-based implementations \cite{Bianchi:2015oia,Scoccimarro:2015bla,2017MNRAS.464.3121W,Hand:2017irw,2018MNRAS.473.2737S} are built with the plane-parallel approximation in mind, returning a single function $\hP_\ell(k)$ for each Legendre multipole of the signal. Beyond the plane-parallel limit, however, a single $\ell$ no longer suffices to describe the data. This follows from a simple counting argument: since redshift-space distortions (RSD) preserve statistical isotropy (and break homogeneity), the most general 2PCF consistent with redshift-space symmetries depends on three numbers, rather than the two of the plane-parallel limit. The single-$\ell$ Yamamoto estimator thus measures a compressed statistic, which as we will show is suboptimal on sufficiently large scales. Furthermore, conventional approaches for incorporating wide-angle effects into the Yamamoto estimator window function rely on formally infinite series, which as we will show is unnecessary.

Wide-angle corrections to the plane-parallel approximation were computed in linear theory by Refs.~\cite{Hamilton:1995px,1996ApJ...462...25Z,1998ApJ...498L...1S,1998ASSL..231..185H,Bharadwaj_1999,1994MNRAS.266..219F,1995MNRAS.275..483H} using a variety of basis functions to represent the redshift-space density contrast $\delta_g$. 
In particular, Ref.~\cite{1998ApJ...498L...1S} showed that $\delta_g$ can be written as a finite sum of ``spherical tensors'', which is closely related to this work.
Refs.~\cite{1994MNRAS.266..219F,1995MNRAS.275..483H} introduced a spherical Fourier Bessel (SFB) decomposition, whose standard (discrete) basis serves as the baseline approach for SPHEREx analyses \cite{Gebhardt:2021mds,Gebhardt:2021tme,GrasshornGebhardt:2023mlv,Benabou:2023ldb,2024PhRvD.110f3524K,Benabou:2024tmn,Wen:2024hqj,Wen:2026fzo,Bruton:2026ykj} and was later generalized to an over-complete basis \cite{Castorina:2017inr}.
The redshift-space 2PCF has also been expressed in a basis of tri-polar spherical harmonics \cite{2004ApJ...614...51S,2008MNRAS.389..292P,PhysRevD.78.103512,2010MNRAS.409.1525R} and extended to include light-cone evolution and Alcock-Paczynski distortions \cite{APeffect,2000ApJ...535....1M,2021PhRvD.103l3534S}.
Relativistic corrections including Doppler, Sachs-Wolfe, lensing convergence, time delay, and integrated Sachs-Wolfe effects have been incorporated both in configuration and Fourier space \cite{2009PhRvD..80h3514Y,2011PhRvD..84d3516C,2011PhRvD..84f3505B,Beutler:2020evf,2012JCAP...10..025B,2014JCAP...08..022R,2018JCAP...03..019T,Castorina:2021xzs,2025JCAP...07..063G}.
Going beyond linear theory, Ref.~\cite{2018MNRAS.479..741C} computed wide-angle corrections within the Zel'dovich approximation. 
Similar calculations have been extended to galaxy bispectra \cite{2023JCAP...09..030P,Benabou:2023ldb} and peculiar-velocity statistics \cite{Castorina:2019hyr}.

Our approach to representing 2PCFs is as follows.
Although the space of 2PCFs consistent with redshift-space symmetries (i.e. rotational invariance) is large, we will show that many physically relevant signals can be decomposed into sums of \textbf{rank-1 2PCFs} \cite{1998ApJ...498L...1S}:
\begin{equation}
\label{eq:rk1_sig}
\big\langle \delta_1(\x_1) \delta_2(\x_2) \big\rangle =
\int_{\k} e^{i\k\cdot(\x_1-\x_2)} 
F_1(x_1) F_2(x_2) P(k) \,
\L_{\ell_1^S}(\hk\cdot\hx_1)
\L_{\ell_2^S}(\hk\cdot\hx_2),
\end{equation}
for radial functions\footnote{It is not strictly necessary to include the radial functions $F_i$ in Eq.~\eqref{eq:rk1_sig} as they can be absorbed into $W_i$ in Eq.~\eqref{eq:Pll}. In practice these radial functions capture the evolution of e.g. linear bias on the light-cone, which we choose to distinguish from analyst-dependent choices for the weight functions $W_i$.} $(F_1,F_2)$ and Legendre multipoles $(\ell^S_1,\ell^S_2)$, which we will sometimes refer to as the ``spins'' of the two-$\ell$ signal. 
We define \textbf{finite-rank 2PCFs} as the class of correlation functions that can be expressed as a \textit{finite} linear combination of rank-1 2PCFs.
Linear theory RSD, many relativistic contributions to galaxy number counts and the reconstruction of large-scale radial velocities with the kinetic Sunyaev-Zel'dovich effect fall into this class.
Within the class of finite-rank signals, the natural estimator is indexed by \textit{two} Legendre multipoles rather than one. This two-$\ell$ generalization of the Yamamoto estimator was first proposed by Ref.~\cite{Scoccimarro:2015bla} and subsequently applied to data in Ref.~\cite{2020MNRAS.498.2492G}. We adopt the following normalization of the two-$\ell$ estimator \cite{2026arXiv260504864X}
\begin{align}
\label{eq:Pll}
\hP_{\ell^E_1\ell^E_2}(k) \equiv \frac{2(\ell^E_1+\ell_2^E)+1}{V} 
\int \frac{d\Omega_k}{4\pi} &\int_{\x_1\x_2} 
  e^{-i\k\cdot(\x_1-\x_2)} \,
  W_1(\x_1) W_2(\x_2) \,
  \delta_1(\x_1) \delta_2(\x_2) \nn\\&\hspace{3cm}
  \times\L_{\ell^E_1}(\hk\cdot\hx_1)
    \L_{\ell^E_2}(\hk\cdot\hx_2),
\end{align}
where $V$ is the appropriately weighted survey volume, and $W_i(\x)$ encodes the survey selection function together with any weights (e.g. FKP \cite{FKP}) applied to $\delta_i(\x)$. 
The FFT-able Yamamoto estimator is recovered by setting $(\ell^E_1,\ell^E_2)=(0,\ell)$ in Eq.~(\ref{eq:Pll}) \cite{Hand:2017irw}.
As we will see, in addition to the two-$\ell$ estimator being more computationally efficient than its single-$\ell$ counterpart \cite{Scoccimarro:2015bla,2020MNRAS.498.2492G,2026arXiv260504864X}, it is also optimal for finite-rank signals.

\vspace{0.4cm}
\par
\noindent 
Our main results are:
\begin{itemize}
    \item For finite-rank signals, the optimal quadratic estimator inherits the spin structure of the signal itself. As a consequence, two-$\ell$ estimators outperform the conventional single-$\ell$ Yamamoto estimator for signals such as the linear-RSD hexadecapole, which arises from a $(\ell_1^S,\ell_2^S)=(2,2)$ rather than a $(0,4)$ structure. On sufficiently large scales the gains of a two-$\ell$ estimator become significant (\S\ref{ssec:sims} and Fig.~\ref{fig:grf}).
    \item The response of the two-$\ell$ estimator (Eq.~\ref{eq:Pll}) to a rank-1 signal (Eq.~\ref{eq:rk1_sig}) admits a closed-form expression involving a \textit{finite} number of terms (Eq.~\ref{eq:finite_window}). The computational cost of this window function is comparable to conventional wide-angle expansions of the Yamamoto window function, but avoids making approximations by truncating an infinite series (as in current approaches). Note that the two-$\ell$ estimator includes the Yamamoto estimator as the special case $\ell_1^E=0$.
\end{itemize}

The remainder of this work is organized as follows.
In \S\ref{sec:preliminaries} we review the notation and conventions adopted throughout. 
In \S\ref{sec:rnk1_response} we contrast our finite representation of the window function (Eq.~\ref{eq:finite_window}) with preexisting approaches.
We review the classification of relevant cosmological signals as finite- or infinite-rank in \S\ref{sec:use_cases}. In \S\ref{sec:rsd} we specialize to the case of linear RSD and numerically compare our approach to both the distant-observer and generalized SFB expansions. In the same section we quantify the statistical improvement of a two-$\ell$ estimator over Yamamoto.
In \S\ref{sec:optimality} we prove that Eq.~\eqref{eq:Pll} is the optimal estimator for finite rank signals.
We relegate the derivation of Eq.~\eqref{eq:finite_window} to \S\ref{sec:window_function}, show that its cost scales as $\mathcal{O}(N_{\rm pix}\log N_{\rm pix})$ and discuss practical implementation details in the same section.
Special function identities and intermediate results are provided in the appendices.

%% file: SecPreliminaries.tex
\section{Preliminaries and notation}
\label{sec:preliminaries}

If $\x$ is any three-vector, we use the shorthand notation $x=|\x|$ for its length, and $\hx=\x/x$ for its direction.
We always use a coordinate system where the observer is at the origin.
Thus, the unit vector $\hx = \x/|\x|$ is the line-of-sight direction, pointed away from the observer.
For observable quantities on the lightcone $x$ also serves as a time coordinate.

Our Fourier conventions are:
\begin{equation}
f(\x) = \int \frac{d^3\k}{(2\pi)^3} f(\k) e^{i\k\cdot\x}
 \quad{\rm and}\quad
f(\k) = \int d^3\x \, f(\x) e^{-i\k\cdot\x}.
\end{equation}
We use the following compressed notation for integrals:
\begin{equation}
\int_{\x} \equiv \int d^3\x 
 \quad{\rm and}\quad
\int_{\k} \equiv \int \frac{d^3\k}{(2\pi)^3} 
= \int \frac{k^2 \, dk}{2\pi^2} \, \frac{d\Omega_k}{4\pi},
\end{equation}
where $\int d\Omega_k$ denotes an integral over three-vectors $\k$ with $|\k|=k$, normalized to $\int d\Omega_k = 4\pi$. 

We frequently use Wigner 3j, 6j, and 9j symbols:
\begin{equation}
\threej{\ell_1}{\ell_2}{\ell_3}{m_1}{m_2}{m_3}
 \hspace{1.5cm}
\sixj{\ell_1}{\ell_2}{\ell_3}{\ell_1'}{\ell_2'}{\ell_3'}
 \hspace{1.5cm}
\ninej{l_{11}}{l_{12}}{l_{13}}{l_{21}}{l_{22}}{l_{23}}{l_{31}}{l_{32}}{l_{33}}.
\end{equation}
Their properties are reviewed in Appendix \ref{app:special_functions}.
We define the shorthand notation:
\begin{equation}
C_{\ell_1\ell_2\ell_3} \equiv
\threej{\ell_1}{\ell_2}{\ell_3}{0}{0}{0}
\quad{\rm and}\quad
G_{\ell_1\ell_2\ell_3} \equiv
\sqrt{\frac{(2\ell_1+1)(2\ell_2+1)(2\ell_3+1)}{4\pi}}
C_{\ell_1\ell_2\ell_3}
\label{eq:Clll_def}.
\end{equation}
Note that $C_{\ell_1\ell_2\ell_3}$ and $G_{\ell_1\ell_2\ell_3}$ are fully symmetric in $(\ell_1,\ell_2,\ell_3)$ and vanish if $(\ell_1+\ell_2+\ell_3)$ is odd.

To avoid notational confusion with power-spectra multipoles, we denote Legendre polynomials by $\L_\ell(\mu)$.

%% file: SecRank1Response.tex
\section{Response to a rank-1 signal}
\label{sec:rnk1_response}

In this section we briefly review the Yamamoto estimator's response to a rank-1 2PCF (Eq.~\ref{eq:rk1_sig}) in the distant-observer (\S\ref{sec:distant_observer_expansion}) and generalized spherical Fourier Bessel (\S\ref{sec:sfb}) expansions. Following the convention of Ref.~\cite{Hand:2017irw}, we take Eq.~\eqref{eq:Pll} with $(\ell_1^E,\ell_2^E)=(0,\ell)$ as our definition of the Yamamoto estimator.
Throughout we adopt the following normalization of the (generalized) rank-1 window function
\begin{equation}
\label{eq:window}
\big\langle\hP_{\ell^E_1\ell^E_2}(k)\big\rangle
=
\frac{2(\ell^E_1+\ell^E_2)+1}{V}
\int \frac{k'^2\,dk'}{2\pi^2}
W_{\ell^E_1\ell^E_2}^{\ell_1^S\ell_2^S}(k,k')\,P(k'),
\end{equation}
where the superscripts $E$ and $S$ denote estimator- and signal-multipoles respectively. 
Within both of these formalisms the analytic expression for the window function is written as an infinite series. 
However, as we claim in \S\ref{sec:finite_window} and show in \S \ref{sec:window_function_derivation} and Appendix \ref{sec:L4}, these series can be resummed to a finite expression.

\subsection{Distant-observer expansion}
\label{sec:distant_observer_expansion}

The distant-observer expansion assumes that the separation between galaxies, $\s \equiv \x_2-\x_1$, is small compared to their distance from the observer (either $x_1$ or $x_2$). 
Equivalently, the expansion parameter may be taken as the opening angle $\theta = s/x_2 \ll 1$. 
Under this assumption, a rank-1 2PCF (Eq.~\ref{eq:rk1_sig}) can be Taylor expanded in powers of $\theta^n$, and at each order in this expansion, the angular dependence of the coefficients can be expressed as a sum of Legendre polynomials \cite{Reimberg:2015jma}
\begin{equation}
    \langle\delta(\x_1)\delta(\x_2)\rangle
    =
    F_1(x_1)F_2(x_2)
    \sum_{n=0}^\infty
    \left(\frac{s}{x_2}\right)^n
    \sum_\ell\L_\ell(\hx_2\cdot\hs)\sum_{L} C^{(n)\ell_1^S\ell_2^S}_{\ell L} \int\frac{k^2dk}{2\pi^2}P(k)j_{L}(ks).
    \label{eq:2pcf_distant_observer}
\end{equation}
This construction alone implies two important properties: first, the distant-observer representation contains an infinite number of terms, and second, convergence of the expansion is only guaranteed for sufficiently small opening angles ($\theta \lesssim 60^\circ$).
Substituting Eq.~\eqref{eq:2pcf_distant_observer} into the expectation value of the Yamamoto estimator leads to the following expression for the window function \cite{2017MNRAS.466.2242B,Beutler:2018vpe,2017MNRAS.464.3121W}
\begin{equation}
W^{\ell_1^S\ell_2^S}_{0\ell}(k,k')
    =
    \sum_{n=0}^\infty
    \sum_{\ell'\ell''L}
    i^\ell
C^{(n)\ell_1^S\ell_2^S}_{\ell' L } 
C_{\ell\ell'\ell''}^2 
\int ds \, s^{n+2} \, j_\ell(ks) \, 
j_L(k's)
Q_{\ell''}^{(n)}(s) 
\label{eq:distant_observer_window_rank1}
\end{equation}
where $C_{\ell_1\ell_2\ell_3}$ is given by Eq.~\eqref{eq:Clll_def}, and $Q_L^{(n)}(s)$ is defined by:
\begin{equation}
Q_L^{(n)}(s) \equiv (2L+1)
 \int d\Omega_s \, d^3\x_1 \,
 \L_L(\hx_2 \cdot \hs) F_1(x_1)W_1(\x_1)\frac{F_2(x_2)W_2(\x_2)}{x_2^n} 
 \Big|_{\x_2=\x_1+\s}.
 \label{eq:Qn_def}
\end{equation}
The coefficients $C^{(n)\ell_1^S\ell_2^S}_{\ell L } $ appearing in Eqs.~\eqref{eq:2pcf_distant_observer} and \eqref{eq:distant_observer_window_rank1}
have been explicitly computed up to $n=4$ in the context of linear redshift-space distortions \cite{Benabou:2024tmn}, and can be algorithmically computed for larger $n$ (see e.g. \href{https://github.com/joshua-benabou/wide_angle_expansion}{\faGithub}). 
For $n=0$ the coefficients take a particularly simple form:
\begin{equation}
    C^{(0)\ell_1^S\ell_2^S}_{\ell L} = i^{-L}\delta_{\ell L} (2L+1)C^2_{\ell_1^S\ell_2^SL},
    \label{eq:n0_coefficients}
\end{equation}
where $\delta_{\ell L}$ denotes a Kronecker delta function.
By construction, the leading-order contribution ($n=0$) reproduces the plane-parallel limit \cite{1987MNRAS.227....1K}.
Consequentially, the distant-observer representation of the window function is guaranteed to converge at fixed $n$ for sufficiently large $k$.

\subsection{Generalized spherical Fourier Bessel expansion}
\label{sec:sfb}

An alternative representation of the window function is obtained with the spherical Fourier Bessel (SFB) expansion \cite{1994MNRAS.266..219F,1995MNRAS.275..483H}.
Here we consider its generalization to an over-complete basis, since it admits an analytic mapping to the Yamamoto estimator \cite{Castorina:2017inr}.
In the \textit{generalized} SFB (gSFB) expansion the weighted three-dimensional density contrast, $W_i(\x)\delta_i(\x)$, is expanded in an over-complete basis constructed from spherical harmonics and spherical Bessel functions  
\begin{equation}
    \delta^{(i)L}_{\ell m}(k)
    \equiv
    \int_\x
    j_{L}(kx)
    Y^*_{\ell m}(\hx)
    W_i(\x)\delta_i(\x).
    \label{eq:sfb_basis}
\end{equation}
Ref. \cite{Castorina:2017inr} showed that the Yamamoto estimator (Eq.~\ref{eq:Pll} with $\ell_1=0$ and $\ell_2=\ell$) can be written as a sum over pseudo-$C_\ell$ cross-spectra of the gSFB coefficients\footnote{This can be derived by expanding the plane-wave factor and Legendre polynomial appearing in the Yamamoto estimator using Eqs. \eqref{eq:rayleigh_expansion} and \eqref{eq:legpoly_ylm} respectively, performing the angular integral over $\hk$ using Eq.~\eqref{eq:yyy} and finally applying Eq.~\eqref{eq:yy3j}. Note that the term in square brackets is real-valued, leaving an ambiguity as to which gSFB coefficient is complex conjugated.}
\begin{equation}
    \hP_\ell(k)=\frac{(4\pi)^2}{V} \sum_{\ell_1\ell_2}
    i^{\ell_2-\ell_1}
    G^2_{\ell\ell_1\ell_2}
    \bigg[\frac{1}{2\ell_1+1}\sum_{m_1} \delta^{(1)\ell_1*}_{\ell_1m_1}(k)\,\delta^{(2)\ell_2}_{\ell_1 m_1}(k)\bigg],
    \label{eq:sfb_pseudo_cl}
\end{equation}
where $G_{\ell_1\ell_2\ell_3}$ is given by Eq.~\eqref{eq:Clll_def}. In Appendix \ref{sec:sfb_window} we show that the resulting gSFB representation of the rank-1 window function is
\begin{equation}
\begin{aligned}
    W^{\ell_1^S\ell_2^S}_{0\ell}(k,k')&=\frac{(4\pi)^2}{2\ell+1}
    \sum_{\ell_1\ell_2}
    i^{\ell_2-\ell_1}
    G^2_{\ell_1\ell_2\ell}
    \sum_{L_1L_2L_3}
     i^{L_1-L_2}(2L_1+1)(2L_2+1)(2L_3+1)\\&\hspace{2cm}\times C^2_{\ell_1^SL_1L_3}C^2_{\ell_2^SL_2L_3}
    \sum_{LM} C^2_{\ell_1L_3L} F^{(1)\ell_1L_1*}_{LM}(k,k')
    F^{(2)\ell_2L_2}_{LM}(k,k'),
    \\
    &\hspace{-2cm}{\rm where}\quad
    F^{(i)LL'}_{\ell m}(k,k')
    \equiv
    \int_{\x}
    j_L(kx)
    j_{L'}(k'x)
    Y^*_{\ell m}(\hx)
    W_i(\x)
    F_i(x).
    \label{eq:sfb_window}
\end{aligned}
\end{equation}
Just as for the distant-observer case, the gSFB representation carries an infinite sum. To be precise, both the sums over $\ell_1$ and $L$ are infinite in general, while all remaining sums are finite for fixed $(\ell_1,L)$. 
However, since $j_L(z)$ is suppressed by $z^L/(2L+1)!!$ for $z\ll L$, the integrals $F^{(i)LL'}_{\ell m}(k,k')$ vanish rapidly for $L\gtrsim kR$ and $L'\gtrsim k'R$, where $R$ is the radial extent of the survey. The sums over $\ell_1$ and $L_1$ are therefore effectively truncated at scales set by $kR$ and $k'R$ respectively. In the regime of interest $k\sim k'$, this gives rapid convergence at low $k$, while the cost for accurate convergence becomes increasingly demanding for larger $k$ (see Fig.~\ref{fig:full_sky_comparison}).

The quantity $F^{(i)LL'}_{\ell m}$ is analogous to the gSFB coefficients (Eq.~\ref{eq:sfb_basis}), and for a full-sky survey only the $(\ell,m)=(0,0)$ mode contributes to the window function. It follows that for a full-sky survey only a single infinite sum over $\ell_1$ is required. Since $F^{(i)LL'}_{\ell m}$ is a spherical harmonic transform of the (signal-weighted) survey selection function, the relationship between the window function for masked- and full-sky geometries is governed by the same coupling kernel used in pseudo-$C_\ell$ analyses \cite{Castorina:2017inr,Wen:2024hqj,2002ApJ...567....2H}.

\subsection{Finite expression}
\label{sec:finite_window}

We now present a central result of this work: a finite representation of the window function.
For maximum generality, we consider the response of the rank-one estimator in Eq.\ (\ref{eq:Pll}) to the rank-one signal in Eq.\ (\ref{eq:rk1_sig}). The window function was defined in \S\ref{sec:window_function_derivation} and has four spins $(\ell_1^E, \ell_2^E, \ell_1^S, \ell_2^S)$.
In \S\ref{sec:window_function_derivation} and Appendix \ref{sec:L4}, we will show that the window function is given by:
\begin{tcolorbox}[ams align]
\label{eq:finite_window}
W^{\ell_1^S\ell_2^S}_{\ell_1^E\ell_2^E}(k,k') &= 
 \hspace{-0.2cm}
 \sum_{\substack{\ell^E_3\ell^S_3 \\ L_1L_2L_3}}
 \hspace{-0.2cm}
 i^{\ell_3^E-\ell_3^S}  
 (2\ell_3^E+1) (2\ell_3^S+1) \,
C_{\ell_1^E \ell_2^E \ell_3^E} 
C_{\ell_1^S \ell_2^S \ell_3^S} 
C_{\ell_1^E \ell_1^S L_1}
C_{\ell_2^E \ell_2^S L_2}
C_{\ell_3^E \ell_3^S L_3}
\nn \\ & \hspace{1cm} \times
\ninej{\ell^E_1}{\ell^E_2}{\ell^E_3}{\ell^S_1}{\ell^S_2}{\ell^S_3}{L_1}{L_2}{L_3}
\int ds \, s^2 \, 
j_{\ell_3^E}(ks) j_{\ell_3^S}(k's) 
Q_{L_1L_2L_3}(s),
\end{tcolorbox}
\noindent
where the function $Q_{L_1L_2L_3}(s)$ is given by
\begin{align}
Q_{L_1L_2L_3}(s) &= 
(4\pi)^{3/2} \sqrt{(2L_1+1)(2L_2+1)(2L_3+1)} \nn \\
& \hspace{1cm} \times \int d\Omega_s \, d^3\x_1 \,
  W_1(\x_1) W_2(\x_2) 
  F_1(x_1) F_2(x_2) \nn \\
& \hspace{1cm} \times
  \sum_{M_1M_2M_3}
  \threej{L_1}{L_2}{L_3}{M_1}{M_2}{M_3}
  Y_{L_1M_1}(\hx_1) Y_{L_2M_2}(\hx_2) Y_{L_3M_3}(\hs)
  \Big|_{\x_2=\x_1+\s} .
\label{eq:QLLL_alternate}
\end{align}
For the special case of the Yamamoto estimator, Eq.~\eqref{eq:finite_window} simplifies to
\begin{align}
W^{\ell_1^S\ell_2^S}_{0\ell}(k,k') &=
\sum_{\substack{\ell^S_3 L_2L_3}}
 \hspace{-0.2cm}
 (-1)^{L_2}
 i^{\ell+\ell_3^S}  
\frac{2\ell_3^S+1}{2\ell_1^S+1} \,
C_{\ell_1^S \ell_2^S \ell_3^S} 
C_{\ell \ell_2^S L_2}
C_{\ell \ell_3^S L_3}
\nn \\ & \hspace{1cm} \times
\sixj{\ell_2^S}{\ell_3^S}{\ell_1^S}{L_3}{L_2}{\ell}
\int ds \, s^2 \, 
j_{\ell}(ks) j_{\ell_3^S}(k's) 
Q_{\ell_1^SL_2L_3}(s),
\label{eq:finite_window_yamamoto}
\end{align}
where we used Eqs. \eqref{eq:special_3j} and \eqref{eq:special_9j}.
Unlike the distant-observer and gSFB representations, the sums appearing in Eqs.\ \eqref{eq:finite_window} and \eqref{eq:finite_window_yamamoto} are finite. This follows from selection rules for Wigner 6j (Eq.~\ref{eq:6j_triangle_conditions}) and Wigner 9j (Eq.~\ref{eq:9j_selection_rules}) symbols.\footnote{To write this out explicitly, using the 9j selection rules (Eq. \ref{eq:9j_selection_rules}), the sums appearing in Eq.~\eqref{eq:finite_window} reduce to the manifestly finite form:
\begin{equation}
    \sum_{\substack{\ell^E_3\ell^S_3 \\ L_1L_2L_3}}
    =
    \sum_{\ell_3^E=|\ell_1^E-\ell_2^E|}^{\ell_1^E+\ell_2^E}
    \,\,\sum_{\ell_3^S=|\ell_1^S-\ell_2^S|}^{\ell_1^S+\ell_2^S}
    \,\,\sum_{L_1=|\ell_1^E-\ell_1^S|}^{\ell_1^E+\ell_1^S}
    \,\,\sum_{L_2=|\ell_2^E-\ell_2^S|}^{\ell_2^E+\ell_2^S}
    \,\,\sum_{L_3={\rm max}(|\ell^E_3-\ell^S_3|,|L_1-L_2|)}^{{\rm min}(\ell_3^E+\ell_3^S,L_1+L_2)}.
\end{equation}
}
The sums over $M_i$ in Eq.~\eqref{eq:QLLL_alternate} run from $-L_i$ to $L_i$.

Note that the quantity $Q_{L_1L_2L_3}(s)$ defined in Eq.~\eqref{eq:QLLL_alternate} is similar to the quantity $Q_\ell^{(n)}(s)$ appearing in the distant-observer expansion (Eq.~\ref{eq:Qn_def}). Using Eqs. \eqref{eq:special_3j} and \eqref{eq:legpoly_ylm}, we see that: 
\begin{equation}
Q_\ell^{(n)}(s) = \frac{(-1)^\ell}{\sqrt{2\ell+1}} Q_{0\ell\ell}(s) 
\label{eq:QnL_QLLL}
\end{equation}
where $Q_{0\ell\ell}(s)$ is computed with $W_2(\x_2) \to W_2(\x_2)/x^n_2$.
We provide an alternative Fourier-space representation of $Q_{L_1L_2L_3}$ in \S\ref{sec:fourier_Q_computational_cost} and show how to evaluate it efficiently in that same subsection. Recursion relations and analytic formulae for simplified geometries are given in subsections \ref{sec:Q_recursion} and \ref{sec:simplified_geometries} respectively.

Specializing to the Yamamoto estimator ($\ell_1^E=0$), it's illuminating to compare our final expression (\ref{eq:finite_window_yamamoto}) for the window function to the usual distant-observer expansion (\ref{eq:distant_observer_window_rank1}). Both expressions have the same LHS (the window function), and an RHS which is a sum of similar terms ($s$-integrals over $Q$-functions). The distant-observer approximation is an infinite series, which is slow to converge on large scales. In Eq.\ (\ref{eq:finite_window_yamamoto}), we've shown that the infinite series is unnecessary -- the window function can be computed in a different way that leads to a finite number of terms. This finite representation also generalizes from the Yamamoto estimator to the ``two-$\ell$'' estimator (Eq.\ \ref{eq:finite_window}).

%% file: SecUseCases.tex
\section{What signals have finite rank?}
\label{sec:use_cases}

In Eq.~\eqref{eq:finite_window}, we have derived a finite representation for the window function $W(k,k')$, under the assumption that the signal two-point function is finite-rank (as defined below Eq.~\ref{eq:rk1_sig}).
In this section we provide an overview of finite- and infinite-rank cosmological signals to give a sense for when Eq.~\eqref{eq:finite_window} is applicable in practice. 

Since in this work we are primarily interested in ultra large-scale signals such as primordial non-Gaussianity, we focus primarily on observational probes that are highly sensitive to these scales. These probes include, but are not limited to, galaxy number counts (\S\ref{sec:galaxy_number_counts}) and kSZ-reconstructed large-scale radial velocities (\S\ref{sec:radial_velocities}). We briefly note here that our formalism could likely be extended to additional tracers of the large-scale structure such as the Lyman-$\alpha$ forest, galaxy shapes and line intensity mapping. 

To aid the discussion of finite-rank 2PCFs it is useful to introduce a \textbf{finite-rank field}:
\begin{equation}
    f_N(\x)
    =
    \sum_{\ell=0}^{N}
    \int_\k
    e^{i\k\cdot\x}\,
    \L_\ell(\hk\cdot\hx)\,
    T_\ell(k,x)\,
    \mathcal{R}(\k),
\label{eq:finite_rank_field}
\end{equation}
where the subscript $N$ indicates that $f_N$ has rank $N$, $T_\ell(k,x)$ is a transfer function evaluated at the redshift corresponding to the comoving distance $x=|\x|$, and $\mathcal{R}$ is the primordial curvature perturbation.

While in the remainder of the main text we specialize to the case where the transfer function factorizes, i.e. $T_\ell(k,x) = T_\ell(k)\,F_\ell(x)$, the more general definition in Eq.~\eqref{eq:finite_rank_field} is useful for classifying a broader range of signals.
In \S\ref{sec:window_function_derivation} we show that the window function appropriate for non-factorizable transfer functions is identical to Eq.~\eqref{eq:finite_window} but with $F_i(x_i)\to F_i(x_i,k')$ in the subsequent expression for $Q_{L_1L_2L_3}(s)\to Q_{L_1L_2L_3}(s,k')$.
Alternatively, should the transfer function exhibit only weak non-factorizability, one could always approximate this behavior with a linear combination of factorizable terms. We further note that for many fields of interest (e.g. the matter density), the transfer function is in practice nearly factorizable for the redshifts and scales covered by modern LSS surveys. 

Note that the cross-correlation of two finite-rank fields yields a finite rank 2PCF. In particular,  Ref.~\cite{1998ApJ...498L...1S} originally showed that the linear theory RSD 2PCF could be written as a finite sum of rank-1 2PCFs (see also \S\ref{sec:rsd}).
We further note that scalar derivatives of finite-rank fields, such as derivatives along the line-of-sight direction ($\partial_x$) and angular Laplacians ($\Delta_\Omega\equiv x^2 (\delta_{ij} - \hat{n}_i \hat{n}_j)\partial_i\partial_j$), are themselves finite-rank fields.
More precisely, if $f_N$ has rank $N$ then $\partial_x f_N$ has rank $(N+1)$,  $\partial^2_x f_N$ and $\Delta_\Omega f_N$ have rank $(N+2)$, and so on.
However, integrals of rank-0 fields along the line-of-sight lead to negative powers of $\hk\cdot\hx$ and are no longer finite-rank.

\subsection{Galaxy number counts}
\label{sec:galaxy_number_counts}

Fundamentally, galaxy surveys observe the number of galaxies in a redshift interval $dz$ within a solid angle $d\Omega_n$, which we denote $\Delta_g(\hn,z)\,\bar{n}_g(z)\,dz\,d\Omega_n$, where $\bar{n}_g(z)$ indicates the mean density as a function of redshift.
The relationship between the light-cone observable $\Delta_g(\hn,z)$ and dynamical fields defined on spatial hypersurfaces was first worked out in linear theory by Refs.~\cite{2009PhRvD..80h3514Y,2010PhRvD..82h3508Y,2011PhRvD..84f3505B, 2011PhRvD..84d3516C}, while Refs.~\cite{2018JCAP...10..024S,2020JCAP...11..064G,Castorina:2021xzs,2023PhRvL.131k1201F} expanded on this work to also include observer terms. 
The gauge-invariant $\Delta_g$ can be split into local, observer and integrated contributions:
\begin{equation}
    \Delta_g(\hn,z)
    =
    \Delta_{\rm loc}(\x)
    +
    \Delta_{\rm obs}(\x)
    +
    \Delta_{\rm int}(\x),
    \label{eq:Delta_g}
\end{equation}
where $\x(\hat{\mathbf n},z) = x(z)\hat{\mathbf n}$, and the individual contributions are given by\footnote{The literature adopts different sign conventions for $\hn$ (and hence the radial velocity field), with some authors defining $\hn$ parallel to the photon's momentum, while others use $\hn$ as the line of sight direction. In particular, Eqs.~\eqref{eq:Delta_g}--\eqref{eq:Delta_int} are equivalent to Eq.~(A1) from Ref.~\cite{2023PhRvL.131k1201F} with $v_r=-v_\parallel$.}
\begin{align}
    \Delta_{\rm loc}(\x)
    &=
    b_g(x) D_m(\x) - \frac{\partial_x v_r(\x)}{\mathcal{H}(x)}  - \mathcal{R}_v(x) v_r(\x) 
    +(5s(x)-2)\Phi(\x) + \frac{\partial_\tau\Phi(\x)}{\mathcal{H}(x)}
    \notag\\&\hspace{2cm}
    + (\mathcal{R}_v(x)+1)\Psi(\x)
    + (b_e(x)-3)\mathcal{H}(x)V(\x)
    \label{eq:Delta_loc}
    \\
    \Delta_{\rm obs}(\x)
    &=
    \big(5s(x)-2\big)
    \left[\frac{V(\mathbf{0})}{x}-v_{r}(\mathbf{0})\right]
    +
    \mathcal{R}_v(x)\left[\mathcal{H}(0)V(\mathbf{0})+v_{r}(\mathbf{0})-\Psi(\mathbf{0})\right]
    \label{eq:Delta_obs}
    \\
    \Delta_{\rm int}(\x)
    &=
    \int_{0}^{1} dt\,
    \left[
    \big(5s(x)-2\big)
    \left(
    \frac{1-t}{2t}
    \Delta_{\Omega}
    -
    1
    \right)
    +
    x\,\mathcal{R}_v(x)\,
    \partial_\tau
    \right]
    \Big(
    \Psi(t\x)+\Phi(t\x)
    \Big).
    \label{eq:Delta_int}
\end{align}
The term $D_m(\x)$ appearing in Eq.~\eqref{eq:Delta_loc} denotes the matter density contrast in comoving gauge, and is related to the Newtonian gauge density contrast through $D_m = \delta_m + 3\mathcal{H}\nabla^{-2}(\mathbf{\nabla}\cdot\mathbf{v})$, where $\mathcal{H} \equiv \partial_\tau\ln a$ is the conformal Hubble parameter, $\tau$ indicates conformal time, and $\mathbf{v}$ is the peculiar velocity field in Newtonian gauge. 
All other dynamical fields appearing in Eqs.~\eqref{eq:Delta_loc}--\eqref{eq:Delta_int} are in Newtonian gauge.
As in the introduction to this section, the operator $\Delta_\Omega$ in Eq.~\eqref{eq:Delta_int} denotes the angular Laplacian on the sphere.
The quantity $V$ appearing in Eqs.~\eqref{eq:Delta_loc} and \eqref{eq:Delta_obs} is the velocity potential, which is defined such that ${\mathbf v} = - \boldsymbol{\nabla}V$.
The redshift-dependent coefficients $(b_g,\,s,\,\mathcal{R}_v, \,b_e)$ appearing in Eqs.~\eqref{eq:Delta_loc}--\eqref{eq:Delta_int} vary with the galaxy sample being considered. Simple expressions for the magnification bias $s$, the evolution bias $b_e$ and the coefficient $\mathcal{R}_v$ exist for idealized magnitude-limited surveys \cite{2011PhRvD..84d3516C,2023PhRvL.131k1201F}. However, for more realistic selections such as those used by the DESI survey, additional effects must be taken into account when estimating these coefficients from the data \cite{2023JCAP...11..097Z,2025JCAP...03..059F}.

The decomposition of $\Delta_g$ into local, observer and integrated contributions is useful for classifying the rank of each of these contributions.
First, consider the local contribution $\Delta_{\rm loc}(\x)$ in Eq.~\eqref{eq:Delta_loc}.
Most of the terms ($D_m$, $\Phi$, $\partial_\tau\Phi$, $\Psi$, $V$) do not involve radial derivatives and are manifestly rank-zero fields.
As discussed in \S\ref{sec:radial_velocities}, the radial velocity field $v_r$ is a rank-1 field, and its radial derivative $\partial_r v_r$ (which arises in the Kaiser RSD term) is a rank-2 field.
Combining all terms, $\Delta_{\rm loc}(\x)$ is a finite-rank field.

In the presence of local primordial non-Gaussianities (PNG) the local contribution is modified to include a scale-dependent bias contribution \cite{2008PhRvD..77l3514D}
\begin{equation}
    \Delta_{\rm loc}(\x)
    \to
    \Delta_{\rm loc}(\x)
    + \frac{3\Omega_{\rm m}H_0^2 b_\phi(x)f_{\rm NL}}{2D(x)T(k)k^2}D_m(\x),
\end{equation}
where $D(x)$ is the linear growth factor, $T(k)$ is the matter transfer function and $b_\phi(x)$ is a sample-dependent coefficient quantifying the response of the galaxy number density to a local variation in $\sigma_8$. The additional term generated by local PNG is a rank-0 field, and so $\Delta_{\rm loc}$ remains finite-rank.
The cross-correlation of any two local contributions yields a finite-rank signal, even in the presence of local PNG. 

Next, consider the observer contribution $\Delta_{\rm obs}(\x)$ in Eq.~\eqref{eq:Delta_obs}.
Each term looks like a finite-rank field ($V$, $\Psi$, or $v_r$) -- however, it is evaluated at the observer's position and is multiplied by a $x$-dependent coefficient.
This does not quite qualify as a finite-rank field according to our definition~\eqref{eq:finite_rank_field}, but it is similar enough that we can also derive a finite representation of the window function.
The details of this calculation are in Appendix \ref{app:finite_window_observer_terms}, including cross-correlations between local and observer terms.

Finally, consider the integrated contribution $\Delta_{\rm int}(\x)$ in Eq.~\eqref{eq:Delta_int}.
These terms include line-of-sight integrals over rank-0 fields and are thus infinite-rank fields.
In this case, the machinery in this paper does not produce a finite representation of the window function, but we plan to revisit this in future work.
In practice, the integrated contribution consists of terms (magnification bias and ISW) which are subdominant on sub-horizon scales, compared to the local and observer contributions.

We conclude that the local and observer contributions to galaxy number counts lead to finite representations of the window function.
The three cases of local-local, local-observer, and observer-observer correlations are given in Eqs.~\eqref{eq:finite_window},~\eqref{eq:finite_window_loc_obs} and~\eqref{eq:finite_window_obs_obs} respectively. The inclusion of integrated terms is more complicated, and we defer this to future work.

\subsection{Remote dipole reconstruction}
\label{sec:radial_velocities}

The reconstruction of large-scale velocities from the kinetic Sunyaev–Zel'dovich (kSZ) effect has more-recently emerged as a promising method to reduce sample variance on ultra-large scales and hence improve local $f_{\rm NL}$ constraints \cite{2018PhRvD..98l3501D,2018arXiv181013423S,2019PhRvD.100h3508M}. 
Within the past two years kSZ velocity reconstructions from existing CMB and LSS data have gone from mere hints of a signal to high-confidence detections \cite{2024arXiv240500809B,2025JCAP...05..057M,2025arXiv251115701M,2025PhRvL.134o1003L,2025arXiv250621657H,2025arXiv250621684L,2026arXiv260404867C}, and in the coming years joint analyses of kSZ measurements from Simons Observatory \cite{2025JCAP...08..034A} and galaxies from DESI \cite{2016arXiv161100036D}, Euclid \cite{2025A&A...697A...1E} and LSST \cite{2018arXiv180901669T} have the promise to considerably improve upon $f_{\rm NL}$ constraints derived from galaxy surveys alone \cite{2019PhRvD.100h3508M}. 

Quadratic estimators built from the kSZ effect and galaxy number counts reconstruct the remote CMB dipole field projected onto an observer's line-of-sight direction \cite{2018PhRvD..98l3501D}. 
On sufficiently small scales the dipole is dominated by peculiar velocities, and so we refer to the projected remote dipole as an effective (radial) velocity field $v_{\rm eff}(\x)$.
In Newtonian gauge and neglecting anisotropic stress, so that $\Phi=\Psi$, the remote dipole field takes the form \cite{2017JCAP...02..040T}
\begin{align}
    v_{\rm eff}(\x)
    &=
    i\int_\k
    e^{i\k\cdot\x}
    \mathcal{L}_1(\hk\cdot\hx)
    \Big[T_{\rm D}(k,x)+T_{\rm SW}(k,x)+T_{\rm ISW}(k,x)\Big]
    T_\Psi(k)
    \Psi_{\rm init}(\k)
    \label{eq:veff_ksz}
    \\
    T_{\rm D}(k,x)&= kD_v(x_\star)j_0\big(k(x_\star-x)\big)
    -2kD_v(x_\star)j_2\big(k(x_\star-x)\big)-kD_v(x)
    \label{eq:transfer_doppler}
    \\
    T_{\rm SW}(k,x)&=3\left(2D_\Psi(x_\star)-\frac{3}{2}\right)j_1\big(k(x_\star-x)\big)
    \label{eq:transfer_sw}
    \\
    T_{\rm ISW}(k,x)&=6\int_{a_\star}^{a(x)}
    da' \frac{dD_\Psi}{da'}j_1\Big(k\Big(x_\star-x(a')\Big)\Big),
    \label{eq:transfer_isw}
\end{align}
where $T_\Psi$ ($D_\Psi$) is the transfer (growth) function for the primordial potential $\Psi_{\rm init}$, $x_\star$ is the comoving distance to the surface of last-scattering, and the subscripts D, SW and ISW stand for Doppler, Sachs-Wolfe and integrated SW contributions respectively.
The Euler equation for cold dark matter, i.e. $\partial_\tau(a\,\mathbf{v}) = ia\k\Psi$ \cite{1995ApJ...455....7M}, allows one to relate the peculiar velocity field to the primordial gravitational potential through a velocity growth function, i.e. $\mathbf{v}(\x) = -D_v(x)\mathbf{\nabla}\Psi_{\rm init}(\x)$, that is given by Eq. 2.13 of Ref.~\cite{2017JCAP...02..040T}.
In particular, we note that the $-kD_v(x)$ term in Eq.~\eqref{eq:transfer_doppler} corresponds to the radial peculiar velocity field, $v_r(\x) \equiv \hx\cdot\mathbf{v}(\x)$.
As $k$ becomes sufficiently large terms multiplied by spherical Bessel functions in Eqs.~\eqref{eq:transfer_doppler}--\eqref{eq:transfer_isw} become suppressed, so that $v_{\rm eff}\simeq v_r$ on small scales.

Eq.~\eqref{eq:veff_ksz} makes it clear that the projected remote dipole field is rank-1. 
The dominant Doppler contribution $(v_r)$ is a separable rank-1 field and can readily be modeled using our formalism.
The appearance of spherical Bessel functions in Eqs.~\eqref{eq:transfer_doppler}--\eqref{eq:transfer_isw} imply that the remaining contributions to $v_{\rm eff}$ are rank-1 but with non-factorizable transfer functions. 
As argued in \S \ref{sec:window_function_derivation}, Eq.~\eqref{eq:finite_window} can be straightforwardly generalized to incorporate these terms. 

\subsection{Higher-order and observational effects}
\label{sec:additional_caveats}

In \S\ref{sec:galaxy_number_counts} and \S\ref{sec:radial_velocities} we argued that the dominant contributions to galaxy number counts and radial velocity reconstruction are finite rank fields. This discussion was framed entirely within linear theory and neglected several observational complications. In the quasi-linear regime the perturbative galaxy power spectrum can be written as a polynomial in $\mu\equiv \hk\cdot\hx$ and thus the signal remains finite rank \cite{2018MNRAS.479..741C}. However, when going to even smaller scales non-linear effects such as Fingers-of-God (FoG) become important and can no longer be treated as a perturbative expansion in $\mu$. In such cases the signal becomes infinite rank, and we no longer expect our formalism for the window function to be applicable. 
At the same time the galaxy power spectrum becomes increasingly harder to model, and so it is unlikely that excluding these scales from an analysis due to the non-applicability of Eq.~\eqref{eq:finite_window} will considerably worsen constraints on cosmological parameters.

In addition to nonlinearities, observational effects can cause an otherwise finite-rank signal to be infinite rank.
Photometric-redshift errors serve as a prime example, which effectively convolve the observed density field with the photo-$z$ error distribution along the line-of-sight direction. 
The resulting expression includes integrals over rank-0 fields, and thus even the local contribution to $\Delta_g$ becomes infinite rank. 
In the limit where photo-$z$ errors are small one can Fourier transform this expression and truncate a Taylor series at a finite order in $\mu$, in which case the smearing of the density field is written as an approximately local operator and the resulting field remains finite rank. 
If the photo-$z$ errors are large, this approach becomes intractable due to the large number of terms. In this case, the finite-rank machinery in this paper is not a good fit, and a natural alternative would be to define a few radial bins (or basis functions) and use angular power spectra $C_l^{ij}$ instead of 3D power spectra.

%% file: SecRSD.tex
\section{Case study: linear redshift-space distortions}
\label{sec:rsd}

In this section we take the formalism outlined in \S\ref{sec:finite_window} for computing the response of the (generalized) Yamamoto estimator to a rank-1 signal and apply it to the case of redshift-space distortions in linear theory \cite{1987MNRAS.227....1K}. 

\subsection{Finite representation of the window function}
\label{ssec:rsd_finite_repr}

In linear theory, the galaxy density contrast in redshift-space can be expressed as a rank-2 field \cite{1998ApJ...498L...1S}
\begin{equation}
\delta_g(\x)
 = \sum_{\ell}
 F^{\rm RSD}_\ell(x)
 \int_\k
 e^{i\k\cdot\x}
\, \L_\ell(\hk\cdot\hx)
 \delta_{m,0}(\k),
 \label{eq:kaiser_rsd}
\end{equation}
where $\delta_{m,0}(\k)$ is the present-day matter density contrast and
\begin{equation}
F^{\rm RSD}_\ell(x) \equiv \begin{cases}
 D(x)\big(b_g(x) + f(x)/3\big) & \mbox{if } \ell=0 \\
 2D(x)f(x)/3 & \mbox{if } \ell=2 \\
 0 & \mbox{otherwise},
\end{cases}
\label{eq:sig_rsd_Fl}
\end{equation}
where $D$ is the linear growth factor normalized to unity today, $b_g$ is the linear bias and $f\equiv d\ln D/d\ln a$ is the linear growth rate. The redshift-space galaxy auto-correlation function, $\langle \delta_g(\x_1)\delta_g(\x_2)\rangle$, can therefore be expanded as the sum of four rank-1 signals with $(\ell_1^S,\ell_2^S) = (0,0),\,(0,2),\,(2,0)$ and $(2,2)$ (e.g. Eq. 7 of \cite{1998ApJ...498L...1S}). It follows that the RSD window function can be expanded as a sum of four rank-1 window functions.

Let $W^{\rm RSD}_\ell$ denote the window function relating the Yamamoto estimator's expectation value to the present day matter power spectrum, i.e.
\begin{equation}
    \langle\hP_\ell(k)\rangle
    =
    \frac{2\ell+1}{V}
    \int\frac{k'^2\,dk'}{2\pi^2}
    W^{\rm RSD}_\ell(k,k')
    \,P_{m,0}(k'),
\end{equation}
where $P_{m,0}(k')$ is the present-day matter power spectrum in real-space.
The notation of \S\ref{sec:rnk1_response} suppresses the dependence of the window function on the radial functions $(F_1,F_2)$.
Here we make this dependence explicit by defining $W_\ell^{\ell_1\ell_2}[F_1,F_2]$
as a shorthand for Eq.~\eqref{eq:finite_window} with $(\ell_1^S,\ell_2^S) = (\ell_1,\ell_2)$ and $(\ell_1^E,\ell_2^E) = (0,\ell)$, with the latter choice matching the convention of Ref.~\cite{Hand:2017irw}.
We temporarily suppress the dependence on $k$ and $k'$ to make the notation less burdensome. Expanding the redshift-space 2PCF as a sum of rank-1 signals leads to\footnote{Note that $W^{\ell_1\ell_2}_\ell[F_1,F_2]\neq W^{\ell_2\ell_1}_\ell[F_2,F_1]$ in general, since the Yamamoto estimator's response to the cross-correlation of two \textit{different fields} depends on the choice of line-of-sight used in the Legendre polynomial, i.e. the choice between $(\ell_1^E,\ell_2^E)=(\ell,0)$ or $(0,\ell)$.}:
\begin{equation}
    W^{\rm RSD}_\ell
    =
    W^{00}_\ell[F^{\rm RSD}_0,F^{\rm RSD}_0]
    +
    W^{02}_\ell[F^{\rm RSD}_0,F^{\rm RSD}_2]
    +
    W^{20}_\ell[F^{\rm RSD}_2,F^{\rm RSD}_0]
    +
    W^{22}_\ell[F^{\rm RSD}_2,F^{\rm RSD}_2].
\end{equation}
Using Eq.~\eqref{eq:finite_window} we can write each of these rank-1 window functions as a finite linear combination of $Q_{L_1,L_2,L_3}(s;F_1,F_2)$'s, where again we chose to make the dependence of $Q$ on the radial functions $(F_1,F_2)$ explicit here for clarity. Doing so for the RSD monopole window function gives
\begin{equation}
\begin{aligned}
    &W^{\rm RSD}_{\ell=0}(k,k')
    =
    \int s^2\,ds\,j_0(ks)
    \Bigg\{
    j_0(k's)
    \bigg[
    Q_{000}\Big(s;F^{\rm RSD}_0,F^{\rm RSD}_0\Big)
    +
    \frac{\sqrt{5}}{125}Q_{220}\Big(s;F^{\rm RSD}_2,F^{\rm RSD}_2\Big)
    \bigg]
    \\
    &+
    \frac{\sqrt{5}}{25}
    j_2(k's)
    \bigg[
    \frac{\sqrt{14}}{35}
    Q_{222}\Big(s;F^{\rm RSD}_2,F^{\rm RSD}_2\Big)
    -
    Q_{022}\Big(s;F^{\rm RSD}_0,F^{\rm RSD}_2\Big)
    -
    Q_{202}\Big(s;F^{\rm RSD}_2,F^{\rm RSD}_0\Big)
    \bigg]
    \\
    &\hspace{9cm}+
    \frac{\sqrt{70}}{875}
    j_4(k's)
    Q_{224}\Big(s;F^{\rm RSD}_2,F^{\rm RSD}_2\Big)
    \Bigg\}.
    \label{eq:finite_monopole_window}
\end{aligned}
\end{equation}

In general Eq.~\eqref{eq:finite_monopole_window} requires 6 $Q$-evaluations. This expression simplifies further in the effective redshift approximation, where the radial functions $F^{\rm RSD}_\ell$ are assumed to be redshift-independent. 
In this approximation we may replace $Q(s;F_1,F_2)=F_1F_2\times Q(s;1,1)$ in Eq.~\eqref{eq:finite_monopole_window}, such that the $Q$'s are uniquely determined by the triplet $(L_1,L_2,L_3)$. 
Recursion relations (see \S\ref{sec:Q_recursion}) allow us to write $Q_{222}$ as a linear combination of $Q_{220}$, $Q_{202}$, $Q_{022}$ and $Q_{000}$, reducing the number of required $Q$-evaluations to five. For the special case of a galaxy auto-correlation $Q_{L_1L_2L_3}(s;1,1)$ is symmetric (up to a phase $(-1)^{L_3}$) under the exchange $L_1\leftrightarrow L_2$, further reducing the number of required $Q$-evaluations to four.

Similar expressions to Eq.~\eqref{eq:finite_monopole_window} are straightforward to compute for the quadrupole $(\ell=2)$ and hexadecapole $(\ell=4)$, however, the resulting expressions contain 22 and 26 terms respectively rather than six, and so we chose not to write them down here (see Appendix \ref{app:window_ell_2_4}). Together, the monopole and quadrupole window functions contain 16 unique $Q$'s, while the hexadecapole introduces another 8 unique $Q$'s. In the effective redshift approximation recursion relations reduce the total number (for $\ell=0,2,4$) of required $Q$-evaluations from 24 to 11, or 10 for the special case of a galaxy auto-correlation.

Recursion relations become increasingly beneficial for two-$\ell$ estimators and higher-rank signals. For example, the window function of the $(2,2)$ estimator introduces 26 $Q$'s that are not already present in the $\ell=0,2$ window functions, however, 6 of those $Q$'s can be removed using recursion relations. In the effective redshift approximation, recursion relations reduce the total number of $Q$-evaluations required for the monopole, quadrupole and $(2,2)$ estimator from 36 to 13.


\subsection{Numerical validation of the window function}
\label{sec:numerics}

\begin{figure}[!h]
    \centering
    \includegraphics[width=0.6\linewidth]{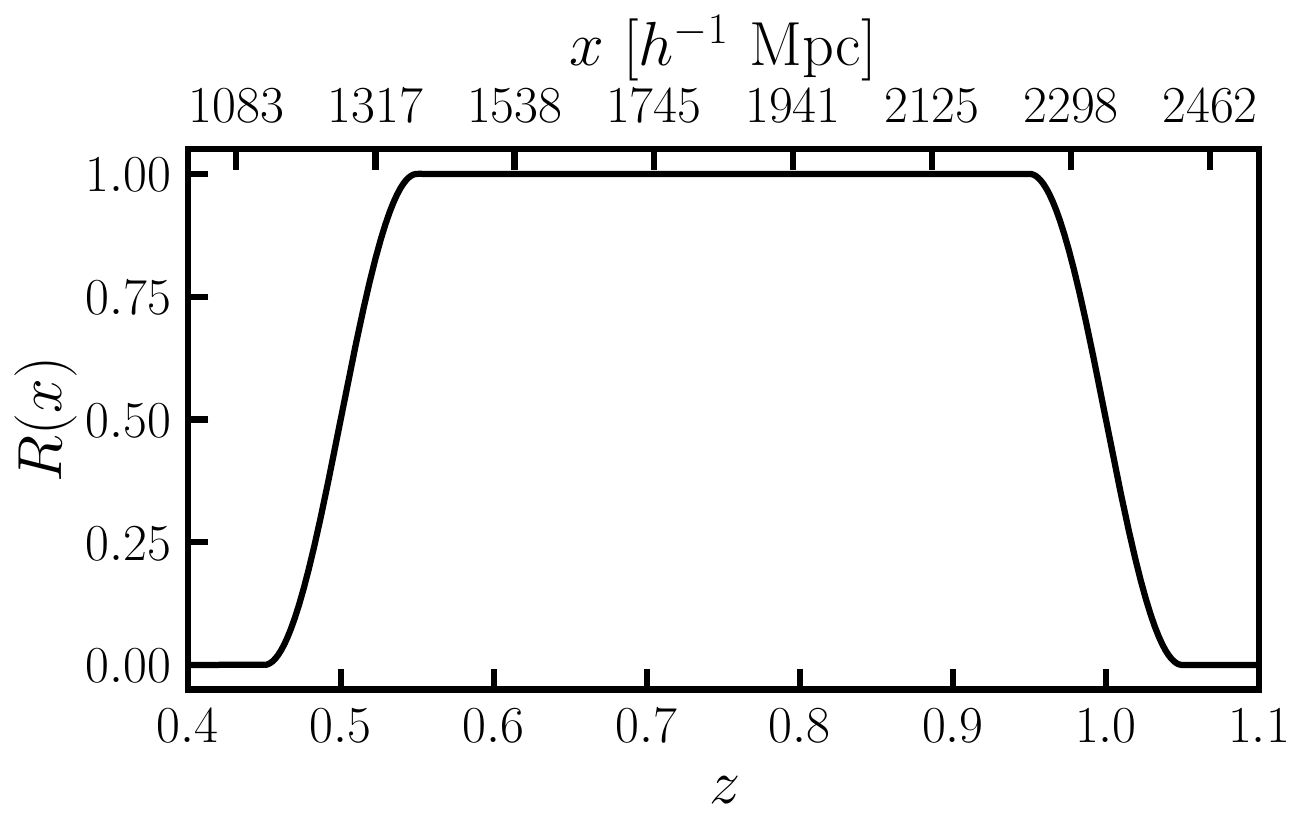}
    \caption{The radial window function used for numerical tests. See the main text in \S\ref{sec:numerics} for details.
    }
    \label{fig:R_of_x}
\end{figure}

Here we numerically compare Eq.~\eqref{eq:finite_window} with the distant-observer (\S\ref{sec:distant_observer_expansion}) and gSFB (\S\ref{sec:sfb}) expansions for the special case of linear-theory redshift space distortions. We consider a full-sky survey whose window function, $W(\x)\equiv R(x)$, approximately covers the redshift range $0.5\lesssim z \lesssim 1$. To mitigate numerical artifacts arising from sharp boundary conditions we multiply the window function edges by a cosine filter, $\big(1\pm\cos(\pi(z-z_\pm)/\Delta z)\big)/2$, with $\Delta z = 0.1$, $z_-=0.45$ and $z_+ = 0.95$. The resulting radial window is shown in Fig.~\ref{fig:R_of_x}. For simplicity we adopt an effective redshift approximation with $b_g(z) = 1$, $f(z) = 0.80$, and $D(z) = 0.68$. Throughout we assume a \textit{Planck} 2018 best-fit cosmology \cite{2020A&A...641A...6P} to convert redshifts to distances and to compute power spectra. 

\begin{figure}[!h]
    \centering
    \includegraphics[width=\linewidth]{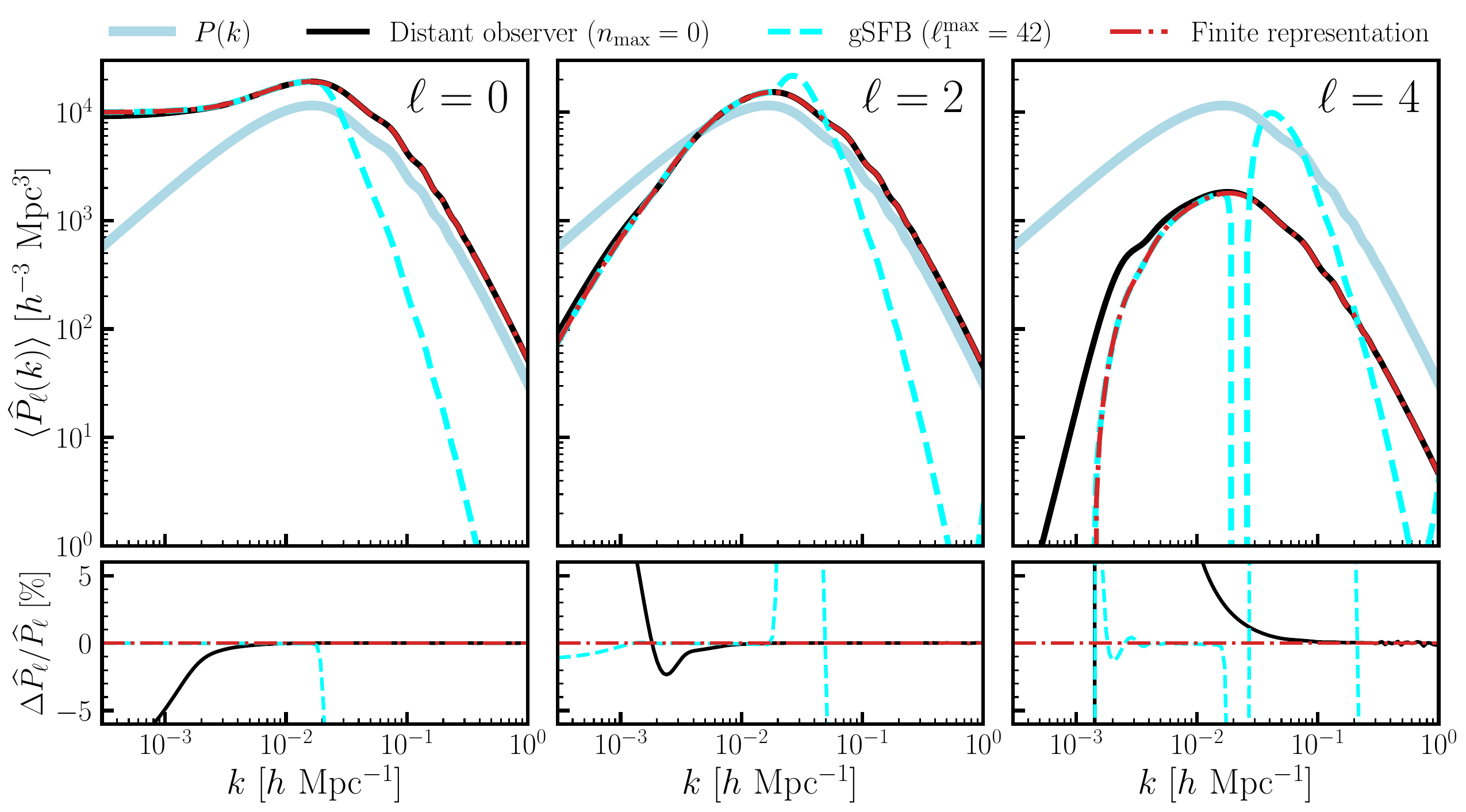}
    \caption{Expectation value of the Yamamoto estimator within the distant observer expansion (black), generalized spherical Fourier Bessel expansion (cyan) and finite representation (red) approaches to modeling the window function.
    For the distant observer case we truncate the sum at $n_\textrm{max}=0$, corresponding to the plane-parallel limit, while for the gSFB case we truncate the sum at $\ell^\textrm{max}_1=42$.
    We consider a full-sky survey and take the radial window function to be a smoothed tophat ranging approximately from $0.5<z<1$ (see Fig.~\ref{fig:R_of_x}).
    We assume that the power spectrum $P(k)\equiv D^2(z) P_{m,0}(k)$ with $D(z)=0.68$ (light blue), linear bias $b_g(z)=1$ and growth rate $f(z)=0.8$ are redshift-independent. 
    }
    \label{fig:full_sky_comparison}
\end{figure}

The present-day matter power spectrum, $P_{m,0}(k')$, is evaluated with \texttt{CLASS} \cite{2011JCAP...07..034B} on a logarithmic grid of 64,000 points spanning $10^{-8} < k'\,/\,(h\,{\rm Mpc}^{-1}) < 10^2$. We plot $P(k)\equiv D^2(z)P_{m,0}(k)$ in light blue in Fig.~\ref{fig:full_sky_comparison}. For each of the three formalisms we evaluate the window function on a logarithmically-spaced $k$-grid with 160 points spanning $3\times10^{-4} < k\,/\,(h\,{\rm Mpc}^{-1}) < 5$, and convolve the result with $P_{m,0}(k')$ over the aforementioned $k'$-range to obtain $\langle \hP_\ell(k)\rangle$ for $\ell=0,2$ and $4$. We note that the comoving horizon scale, $k_H=aH$, corresponds to $\simeq3\times10^{-4}\,h\,{\rm Mpc}^{-1}$ at $z_{\rm eff}=0.75$. Hankel transforms are evaluated with \texttt{scipy.fft.fht} on 32,768 logarithmically-spaced points spanning $10^{-18} < x / (h^{-1}\,{\rm Mpc}) < 10^{18}$. The corresponding $k$-grid used for Hankel transforms spans $6.7\times10^{-21} \lesssim k / (h\,{\rm Mpc}^{-1}) \lesssim 6.7\times10^{15}$ with the same number of logarithmically-space points. Interpolations between $k$-grids are performed using cubic splines when necessary.

We compute the finite window function using Eq.~\eqref{eq:finite_window} following the discussion at the beginning of this section. To compute $Q_{L_1L_2L_3}(s)$ we use the analytic full-sky expression given by Eq.~\eqref{eq:full_sky_q}, which we evaluate using a series of Hankel transforms. We finally evaluate Eq.~\eqref{eq:finite_monopole_window} for fixed $k$ using a Hankel transform, and interpolate the result onto the $k'$-grid used to evaluate the matter power spectrum (and similarly for $\ell=2,4$). The distant observer window function is computed up to $n_{\rm max}=0$ using  Eq.~\eqref{eq:distant_observer_window_rank1} with the coefficients given by Eq.~\eqref{eq:n0_coefficients}. We evaluate $Q^{(0)}_\ell$ using  Eq.~\eqref{eq:QnL_QLLL}. The remainder of the numerical implementation is analogous to that used for the finite window computation. We compute the gSFB window function using Eq.~\eqref{eq:sfb_window}. As was noted in \S\ref{sec:sfb}, for the case of a full-sky survey Eq.~\eqref{eq:sfb_window} contains only a single infinite sum over $\ell_1$, which we choose to truncate at $\ell_1^{\rm max}=42$. The resulting terms in Eq.~\eqref{eq:sfb_window} are then evaluated using Hankel transforms. 

In Fig.~\ref{fig:full_sky_comparison} we show the window-convolved monopole (left panels), quadrupole (middle panels) and hexadecapole (right panels) for the distant observer expansion (solid black), the gSFB expansion (dashed cyan) and the finite representation (dot dashed red) of the window function. Top panels show window-convolved power spectra alongside the ``raw'' $P(k)$ (light blue), while bottom panels show differences relative to the finite representation in percent units. Wide-angle corrections to the plane-parallel ($n_{\rm max}=0$) limit are $>5\%$ for $k\lesssim 10^{-3}$, $2\times10^{-3}$ and $10^{-2}$ $h\,{\rm Mpc}^{-1}$ for $\ell=0$, 2 and 4 respectively. 
We find excellent agreement between the gSFB and finite representations on large scales, but note residual $\sim1\%$ differences in the quadrupole for $k\lesssim 4\times10^{-4}$ $h\,{\rm Mpc}^{-1}$. On small scales the gSFB approximation rapidly breaks down for $k\gtrsim 2\times 10^{-2}$ $h\,{\rm Mpc}^{-1}$, while the plane-parallel limit becomes exact. 
The scale at which the gSFB expansion breaks down corresponds to $k\simeq \ell_1^{\rm max}/2R$, where $R=1\,h^{-1}\,{\rm Gpc}$ is the radial extent of the survey (Fig.~\ref{fig:R_of_x}).
We note that the window-convolved hexadecapole has a zero-crossing at $k\simeq 1.5\times 10^{-3}$ $h\,{\rm Mpc}^{-1}$, and that the finite representation has $\mathcal{O}(0.1\%)$ numerical instability for $k\gtrsim0.3$ $h\,{\rm Mpc}^{-1}$.

\subsection{Quantifying the improvement over the Yamamoto estimator}
\label{ssec:sims}

In this section, we apply our estimator to simplified simulations based on Gaussian random fields, generated with the \texttt{kszx} \href{https://github.com/kmsmith137/kszx}{\faGithub} package. Our simulations include linear RSDs, lightcone evolution, and a radial selection function $R(x)$, but do not include the ``discreteness'' of a real galaxy field. These simplified simulations are sufficient to capture the wide-angle two-point function of a galaxy survey, but do not include all contributions to higher-point statistics.

We will use these simplified simulations for two purposes: (1) to numerically validate our window function $W(k,k')$ from the previous section, and (2) to compare statistical errors between our optimal hexadecapole estimator $\hP_{22}$ and the Yamamoto estimator $\hP_4$. For (1), note that the window function only depends on the two-point function $\langle \delta_g(\x) \delta_g(\y) \rangle$, so Gaussian simulations suffice. For (2), we will mainly be interested in large scales, where a Gaussian field should be a reasonable approximation (and in any case, deviations from Gaussianity should affect the estimators $\hP_{22}$ and $\hP_4$ similarly).

In more detail, we simulate 256 Gaussian random fields (GRFs) on a $1024^3$ cubic box with a $19.7\,h^{-1}\,{\rm Gpc}$ side length, resulting in a Nyquist frequency of $k_{\rm Nyq}= 0.16\,h\,{\rm Mpc}^{-1}$. To construct each realization we generate a real-space density contrast drawn from the power spectrum plotted in Fig.~\ref{fig:full_sky_comparison} (light blue) and perform spin-weighted FFTs using the \texttt{kszx.fft\_c2r} routine to construct the Kaiser RSD field (Eq.~\ref{eq:kaiser_rsd}). The Yamamoto estimator is computed with spin-weighted FFTs following the discussion in \S\ref{sec:finite_and_fast}. Measured power spectra are binned in 10 logarithmically-spaced bandpowers spanning $3.2\times 10^{-4}\,h\,{\rm Mpc}^{-1} < k < k_{\rm Nyq}/2$. Analytic predictions (red curves in Fig.~\ref{fig:full_sky_comparison}) for the window-convolved power spectra are interpolated onto the $k$-grid of the box and binned in the same manner as the data to obtain bandpower-averaged analytic predictions. 

\begin{figure}[!h]
    \centering
    \includegraphics[width=\linewidth]{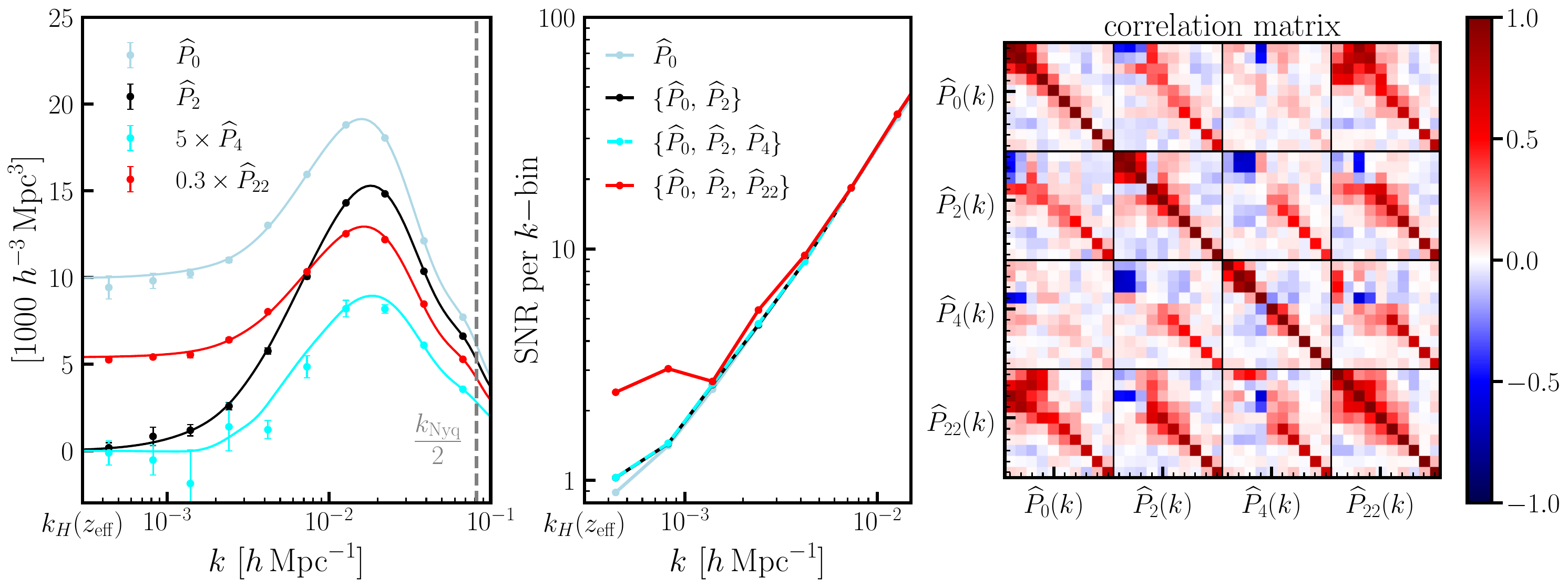}
    \caption{In the left panel we compare analytic predictions for the window-convolved Yamamoto multipoles (light blue, black and cyan) and $(2,2)$ estimator (red) against the mean of 256 simulated measurements (data points), with the errors indicating the uncertainty on the mean. In the middle panel we show the single-realization SNR per bandpower for the monopole (light blue), monopole and quadrupole (black), and when adding either the hexadecapole (cyan) or $(2,2)$ estimator (red). In both the left and middle panels the comoving horizon scale at $z_{\rm eff}=0.75$, $k_H(z_{\rm eff}) \simeq 3\times10^{-4}\,h\,{\rm Mpc}^{-1}$, corresponds to the left edge of the figures. The right panel shows the correlation matrix of the four estimators.
    }
    \label{fig:grf}
\end{figure}

The left panel of Fig.~\ref{fig:grf} compares analytic predictions for the window-convolved monopole (light blue), quadrupole (black) and hexadecapole (cyan) against the simulation mean (data points), with error bars indicating the uncertainty on the mean. In addition to the standard Yamamoto multipoles we perform a similar comparison for the $\hP_{22}(k)$ estimator (red). We find deviations between the simulated- and predicted-means of at most $2.8\sigma$.\footnote{The $2.8\sigma$ bias is for the mean of the 256 GRFs, and would be $\simeq 0.2\sigma$ for a single realization. The window function for a discrete box is not identical to the continuous calculation, thus, a small bias is expected to arise for a sufficiently large number of simulations. We expect that this bias can be removed with a more precise treatment of discreteness effects, and plan to explore this in future work.}

The middle panel of Fig.~\ref{fig:grf} shows the signal-to-noise ratio (SNR) per bandpower for a single realization of the monopole (light blue), monopole and quadrupole (black), and when adding either the hexadecapole (cyan) or $(2,2)$ estimator (red). The $(2,2)$ estimator is more optimal than the $(0,4)$ estimator on large scales (see \S\ref{ssec:estimator_optimal}), resulting in factor of two improvements in SNR for the first two bandpowers. For $k\simeq 1.4$, 2.4 and $4.2\times10^{-3}\,\,h\,{\rm Mpc}^{-1}$ these gains are still appreciable (3, 14 and 6\% improvement respectively), while for larger $k$ the gains are subpercent. We repeated these tests with a larger $99\,h^{-1}\,{\rm Gpc}$ box size and found similar results: factor of two improvements for $k\lesssim9\times10^{-4}\,h\,{\rm Mpc}^{-1}$, $10-20\%$ improvements for $9\times10^{-4} h\,{\rm Mpc}^{-1}\lesssim k\lesssim 3\times10^{-3}\,h\,{\rm Mpc}^{-1}$ and $\leq 5\%$ improvements for larger $k$.

We show the correlation matrix of the four estimators in the right panel of Fig.~\ref{fig:grf}. In the plane-parallel limit the $(2,2)$ estimator reduces to a linear combination of $\hP_0(k)$, $\hP_2(k)$ and $\hP_4(k)$. As a result, the (2,2) estimator is more highly correlated (than $\hP_4(k)$ is) with $\hP_0(k)$ and $\hP_2(k)$. Appropriate linear combinations of $\hP_0(k)$, $\hP_2(k)$ and $\hP_{22}(k)$ have been proposed to recover $\hP_4(k)$ in the plane-parallel limit, which better decorrelates the data\footnote{Refs.~\cite{Scoccimarro:2015bla,2026arXiv260504864X} have shown that $\hP_{4b}\equiv (35/18)\hP_{22}-\hP_2-(7/2)\hP_0$ has a lower SNR than $\hP_4$, however, this does not imply that the two-$\ell$ estimator is suboptimal. The relevant quantity is the SNR of the full data vector $\{\hP_0,\hP_2,\hP_{4b}\}$ (equivalent to the SNR of $\{\hP_0,\hP_2,\hP_{22}\}$), which has higher SNR than $\{\hP_0,\hP_2,\hP_{4}\}$} \cite{Scoccimarro:2015bla}.

%% file: SecOptimality.tex
\section{Two $\ell$'s are better than one}
\label{sec:optimality}

In \S\ref{ssec:sims} we numerically showed that the two-$\ell$ estimator (Eq.~\ref{eq:Pll}) is more optimal than the single-$\ell$ Yamamoto estimator, with gains becoming appreciable on ultra large scales. 
More precisely, for a galaxy survey we get larger SNR if we estimate the hexadecapole using $\hP_{22}(k)$ instead of $\hP_4(k)$, provided that $\hP_0(k)$ and $\hP_2(k)$ are also included. Note that the two sets of estimators are equivalent in the plane-parallel approximation, but differ on large scales.

In the introduction we sketched a simple counting argument for why this should be the case: the 2PCF consistent with redshift-space symmetries is a function of three numbers, while Yamamoto is only indexed by two numbers and thus measures a compressed statistic.
In this section, we explain this phenomenon more formally.

In \S\ref{ssec:estimator_optimal} we sketch a formal proof showing that the optimal estimator for a finite-rank signal with spins $(\ell_1^S, \ell_2^S)$ is the ``two-$\ell$'' estimator $\hP_{\ell_1^S \ell_2^S}(k)$. That is, the estimator is optimal if its spins $(\ell_1^E, \ell_2^E)$ are matched to the spins  $(\ell_1^S, \ell_2^S)$ of the underlying signal.
This ``spin-matching'' criterion explains why $\hP_{22}(k)$ is the optimal hexadecapole estimator for a galaxy field with RSDs. As shown in \S\ref{ssec:rsd_finite_repr}, the underlying signal consists of terms with spins (0,0), (0,2), (2,0), and (2,2). The optimal estimators are therefore $\hP_{00}(k)$,  $\hP_{20}(k)$,  $\hP_{02}(k)$,  and $\hP_{22}(k)$. 
In particular, $\hP_{22}(k)$ is the optimal hexadecapole estimator. The Yamamoto hexadecapole estimator $\hP_{04}(k)$ is not ``spin-matched'' to the signal and is therefore suboptimal.

As another example, consider the cross correlation between a galaxy field with RSDs (rank 2, \S\ref{sec:rsd}) and a kSZ radial velocity reconstruction (rank 1, \S\ref{sec:radial_velocities}). By spin-matching, the optimal estimators would be $\hP_{01}(k)$ and $\hP_{21}(k)$, and the Yamamoto estimator $\hP_{03}(k)$ would be a suboptimal estimator for the octopole.

In \S\ref{sec:finite_and_fast}, we show that in addition to being optimal, the two-$\ell$ estimator is also less computationally expensive to evaluate than Yamamoto (see also \cite{Scoccimarro:2015bla,2020MNRAS.498.2492G,2026arXiv260504864X}).
Taken together, these arguments make a compelling case for the two-$\ell$ estimator as the more natural approach to finite-rank signals.

\subsection{Two-$\ell$ estimators are optimal (if the signal is finite-rank)}
\label{ssec:estimator_optimal}

We show that if the signal is finite-rank, then the optimal quadratic estimator for its amplitude is finite-rank with the same spin structure: $(\ell_1^E, \ell_2^E) = (\ell_1^S, \ell_2^S)$. This result is a direct generalization of the optimal quadratic estimator formalism developed for the CMB by Ref.~\cite{Tegmark97}.

We begin by recalling that for zero-mean Gaussian random fields, all of the information is encoded in the covariance $C_{12}(\x_1,\x_2)\equiv\langle\delta_1(\x_1)\delta_2(\x_2)\rangle$.
For simplicity we suppose that the covariance is the sum of a rank-1 signal component $S_{12}(\x_1,\x_2)$, with the same form as Eq.~\eqref{eq:rk1_sig}, and an uncorrelated noise component. Following Ref.~\cite{Tegmark97} the optimal quadratic estimator for a parameter $P_\alpha$ entering the signal covariance is\footnote{Here we ignore noise bias since it complicates the discussion without changing the conclusion (see \S\ref{sec:noise_bias}).}
\begin{equation}
    \hP_\alpha
    =
    \sum_\beta
    F^{-1}_{\alpha\beta}
    \frac{1}{2}(C^{-1}\mathbf{d})^{\rm T}
    \frac{\partial S}{\partial P_\beta}
    (C^{-1}
    \mathbf{d}),
    \quad{\rm where}\quad
    F_{\alpha\beta}
    =
    \frac{1}{2}{\rm Tr}\left[C^{-1}\frac{\partial S}{\partial P_\alpha}C^{-1}\frac{\partial S}{\partial P_\beta}\right]
\end{equation}
and $\mathbf{d}$ denotes the data vector, which in the general case is comprised of field values for both $\delta_1(\x)$ and $\delta_2(\x)$, and $C = \langle \mathbf{d}\mathbf{d}^{\rm T}\rangle$ is the covariance. Here we assume that $P_\alpha$ is the amplitude of a bandpower centered about some wavenumber $k_\alpha$, such that the underlying power spectrum can be written as a linear combination thereof: $P(k) = \sum_\alpha P_\alpha B(k,k_\alpha)$, where $B(k,k_\alpha)$ is some function that is narrowly peaked about $k\sim k_\alpha$. 
For the continuum notation used throughout this paper, and focusing solely on quadratic estimates built from the cross-correlation $\delta_1\times\delta_2$, this becomes \cite{2006MNRAS.370..343E}
\begin{equation}
\begin{aligned}
    \hP_\alpha
    \propto
    \sum_\beta
    F^{-1}_{\alpha\beta}
    \int_{\x_1\x_2}
    \tilde{\delta}_1(\x_1)\frac{\partial S_{12}(\x_1,\x_2)}{\partial P_\beta}
    \tilde{\delta}_2(\x_2)
    \quad{\rm where}\quad \tilde{\delta}_i(\x)\equiv \int_{\x'}C^{-1}_{ii}(\x,\x')\delta_i(\x').
    \label{eq:qml3d}
\end{aligned}
\end{equation}
Note that the derivative of the signal covariance with respect to the bandpower amplitude is still a rank-1 2PCF with the same spins as the original signal:
\begin{equation}
    \frac{\partial S_{12}(\x_1,\x_2)}{\partial P_\alpha}
    =
    (-1)^{\ell_1+\ell_2}
    \int \frac{k^2\,dk}{2\pi^2}B(k,k_\alpha)
    \int \frac{d\Omega_k}{4\pi}
    e^{-i\k\cdot(\x_1-\x_2)}
    F_1(x_1)F_2(x_2)
    \L_{\ell_1}(\mu_1)
    \L_{\ell_2}(\mu_2),
\end{equation}
where we used $\L_\ell(-\mu)=(-1)^\ell\L_\ell(\mu)$ to take $\k\to-\k$ in the exponential. Upon substituting this expression into Eq.~\eqref{eq:qml3d} we see that the optimal quadratic estimator for $P_\alpha$ takes the form of Eq.~\eqref{eq:Pll} but with $W_i(\x_i)\delta_i(\x_i)$ replaced by $F_i(x_i)\tilde{\delta}_i(\x_i)$. Equivalently, the optimal estimator for a rank-1 signal is a rank-1 estimator with the same spins as the signal, but with a more complicated weighting function than the local multiplication assumed in Eq.~\eqref{eq:Pll}. Should the (inverse) covariance matrix be diagonal in pixel space, Eq.~\eqref{eq:qml3d} becomes equivalent to Eq.~\eqref{eq:Pll} with $W_i(\x_i) \propto F_i(x_i)C^{-1}_{ii}(\x_i)$ \cite{FKP}.

The correspondence between the signal and estimator spin structure is exact for the full inverse-variance weighting (Eq.~\ref{eq:qml3d}). When adopting a diagonal approximation to the covariance some information can leak into surrounding spin pairs, including pairs with odd spins even if the signal (such as linear RSD, \S\ref{sec:rsd}) only carries even spins \cite{Beutler:2018vpe,Philcox:2021tfv,Castorina:2017inr}. 
In addition to multipole leakage, odd-spin estimators are required when relativistic contributions (\S\ref{sec:galaxy_number_counts}) are included, since these genuinely generate odd-spin signals \cite{2009JCAP...11..026M}. The numerical results in \S\ref{ssec:sims} are consistent with this leakage being a subdominant effect, such that the standard spin-matched estimator still yields considerable gains over its single-$\ell$ counterpart. 

It is straightforward to generalize the argument above to the case where the signal $S_{12}(\x_1,\x_2)$ is composed of a finite sum of rank-1 2PCFs (i.e. a finite-rank signal). Just as before, the derivative of the signal with respect to the bandpower amplitude $P_\alpha$ has the exact same spin structure as the original signal. Upon substituting this result into Eq.~\eqref{eq:qml3d} we find that the optimal quadratic estimator for a finite-rank signal is a linear combination of rank-1 estimators, with the estimators being built from the same spin pairs as the signal. Subtleties related to the weighting operation, i.e. local multiplication vs the $C^{-1}$ treatment, are the same as the rank-1 case.



\subsection{Two-$\ell$ estimators are fast}
\label{sec:finite_and_fast}

Here we present an algorithm to evaluate the two-$\ell$ estimator with a few FFTs, generalizing \cite{Scoccimarro:2015bla,Hand:2017irw}.
The definition~\eqref{eq:Pll} of $\hP_{\ell_1\ell_2}(k)$ can be rewritten as \cite{Scoccimarro:2015bla}
\begin{equation}
\hP_{\ell_1\ell_2}(k) =
\frac{2(\ell_1 + \ell_2) + 1}{V}
 \int \frac{d\Omega_k}{4\pi} \,
 \tdelta_{1}^{\ell_1}(\k)^* \,
 \tdelta_{2}^{\ell_2}(\k),
 \label{eq:Pll_fourier}
\end{equation}
where the Fourier-space field
\begin{align}
\delta_i^\ell(\k) &\equiv 
 \int_{\x} e^{i\k\cdot\x} \,
 W_i(\x) \delta_i(\x) 
 \left(
 \frac{4\pi}{2\ell+1} \sum_{m=-\ell}^\ell
 Y_{\ell m}^*(\hk) Y_{\ell m}(\hx)
 \right) \nn \\
 &= \frac{4\pi}{2\ell+1} \sum_{m=-\ell}^\ell
 Y_{\ell m}^*(\hk)
 \underbrace{\int_{\x} e^{i\k\cdot\x} \, W_i(\x) \delta_i(\x) Y_{\ell m}(\hx)}_{\rm one\  FFT},
\end{align}
is computed with $(2\ell+1)$ FFTs. By using a basis of real-valued spherical harmonics, we can use $(2\ell+1)$ \texttt{r2c} FFTs, which are twice as fast and memory-efficient as \texttt{c2c} FFTs \cite{Hand:2017irw}.

It follows that $\hP_{\ell_1\ell_2}(k)$ requires at most $(2\ell_1+1)\times(2\ell_2+1)$ FFTs for the cross-correlation of two tracers, or $2\,{\rm max}(\ell_1,\ell_2)+1$ FFTs for an auto-correlation. 
For the RSD case, the Yamamoto monopole, quadrupole and hexadecapole together require $\sum_{\ell\in\{0,2,4\}}(2\ell+1)=15$ FFTs, while the monopole, quadrupole and $(2,2)$ estimators require only $\sum_{\ell\in\{0,2\}}(2\ell+1)=6$. More generally, for a rank-$\ell_{\rm max}$ signal the Yamamoto estimator requires $(2\ell_{\rm max}+1)^2$ FFTs to capture the full 2PCF information content in the plane-parallel limit, whereas the two-$\ell$ estimator requires only $(\ell_{\rm max}+1)^2$.

%% file: SecWindowCondensed.tex
\section{Finite window function: derivation, cost and practical formulae}
\label{sec:window_function}

\subsection{Derivation of Eq.~\eqref{eq:finite_window}}
\label{sec:window_function_derivation}

We begin the derivation by substituting a rank-1 2PCF (Eq.~\ref{eq:rk1_sig}) into the expectation value of the generalized Yamamoto estimator (Eq.~\ref{eq:Pll}). Comparing the resulting expression to the definition of the window function (Eq.~\ref{eq:window}) leads to the following:
\begin{equation}
\begin{aligned}
W_{\ell^E_1\ell^E_2}^{\ell_1^S\ell_2^S}(k,k') &= 
\int \frac{d\Omega_k}{4\pi} \frac{d\Omega_{k'}}{4\pi} \int_{\x_1\x_2} 
  e^{-i(\k-\k')\cdot(\x_1-\x_2)} \,
  W_1(\x_1) W_2(\x_2)F_1(x_1) F_2(x_2) \,
  \\&\hspace{3cm}\times
  \L_{\ell^E_1}(\hk\cdot\hx_1)
  \L_{\ell^E_2}(\hk\cdot\hx_2)
  \L_{\ell^S_1}(\hk'\cdot\hx_1)
  \L_{\ell^S_2}(\hk'\cdot\hx_2).
  \label{eq:window_first_step}
\end{aligned}
\end{equation}
To make progress we expand all Legendre polynomials into spherical harmonics with Eq.~\eqref{eq:legpoly_ylm}. Doing so allows one to pull all $(\hk,\hk')$-dependent terms in the second line of Eq.~\eqref{eq:window_first_step} outside the integrals over $\x_1$ and $\x_2$, implying that the window function can be computed with a finite number of FFTs. Going even further, one can recouple the resulting angular momenta of the Legendre polynomial expansion using Wigner 3j and 9j identities to express the result as a finite sum over products of four spherical harmonics in $\hk$, $\hk'$, $\hx_1$ and $\hx_2$. 
We relegate the details of this calculation to Appendix \ref{sec:L4}, which result in the identity Eq.~\eqref{eq:L4}. Applying this identity to Eq.~\eqref{eq:window_first_step} leads to 
\begin{align}
W_{\ell_1^E\ell_2^E}^{\ell_1^S\ell_2^S}(k,k')
 &= 
\sum_{\substack{\ell^E_3\ell^S_3\\L_1L_2L_3 }}
(4\pi)^2 (2L_3+1) 
\sqrt{(2\ell_3^E+1)(2\ell_3^S+1)(2L_1+1)(2L_2+1)}
\nn \\
& \hspace{0.5cm} \times 
C_{\ell_1^E \ell_2^E \ell_3^E} 
C_{\ell_1^S \ell_2^S \ell_3^S} 
C_{\ell_1^E \ell_1^S L_1}
C_{\ell_2^E \ell_2^S L_2}
\ninej{\ell^E_1}{\ell^E_2}{\ell^E_3}{\ell^S_1}{\ell^S_2}{\ell^S_3}{L_1}{L_2}{L_3}\nn\\
&\hspace{0.5cm} \times
\sum_{\substack{m^E_3 m^S_3\\ M_1 M_2 M_3}}
\threej{\ell^E_3}{\ell^S_3}{L_3}{m^E_3}{m^S_3}{M_3}
\threej{L_1}{L_2}{L_3}{M_1}{M_2}{M_3}
\nn \\
& \hspace{0.5cm} \times
\int_{\x_1\x_2} 
W_1(\x_1)  \, W_2(\x_2) \,
F_1(x_1)\, F_2(x_2)
Y_{L_1M_1}(\hx_1) \, Y_{L_2M_2}(\hx_2) \,
   \nn \\
&  \hspace{0.5cm} \times
\int \frac{d\Omega_k}{4\pi}\frac{d\Omega_{k'}}{4\pi}
e^{-i(\k-\k')\cdot (\x_1-\x_2)}
   Y_{\ell^E_3m^E_3}^*(\hk) \, Y_{\ell^S_3m^S_3}^*(\hk').
   \label{eq:finite_window_intermediate}
\end{align}
The last line of this expression is simplified by expanding the plane wave factor into spherical harmonics. In particular, using Eq.~\eqref{eq:rayleigh_integral} and $Y_{\ell m}(-\hx)=(-1)^\ell Y_{\ell m}(\hx)$, the last line reduces to
\begin{equation}
    i^{\ell_3^E-\ell_3^S}
    j_{\ell_3^E}(ks)
    j_{\ell_3^S}(k's)
    Y^*_{\ell^E_3m^E_3}(\hs)
    Y^*_{\ell^S_3m^S_3}(\hs),
\label{eq:last_line}
\end{equation}
where $\s\equiv\x_2-\x_1$. We next perform the sums over $m_3^E$ and $m_3^S$ using Eq.~\eqref{eq:yy3j}, finding:
\begin{align}
W_{\ell_1^E\ell_2^E}^{\ell_1^S\ell_2^S}(k,k')
 &= 
\sum_{\substack{\ell^E_3\ell^S_3\\L_1L_2L_3 }}
(4\pi)^2 i^{\ell_3^E-\ell_3^S}
\sqrt{(2\ell_3^E+1)(2\ell_3^S+1)(2L_1+1)(2L_2+1)}
\nn \\
& \hspace{0.5cm} \times 
C_{\ell_1^E \ell_2^E \ell_3^E} 
C_{\ell_1^S \ell_2^S \ell_3^S} 
C_{\ell_1^E \ell_1^S L_1}
C_{\ell_2^E \ell_2^S L_2}
G_{\ell_3^E \ell_3^S L_3}
\ninej{\ell^E_1}{\ell^E_2}{\ell^E_3}{\ell^S_1}{\ell^S_2}{\ell^S_3}{L_1}{L_2}{L_3}\nn\\
& \hspace{0.5cm} \times
\int_{\x_1\x_2} 
W_1(\x_1)  \, W_2(\x_2) \,
F_1(x_1)\, F_2(x_2)\,j_{\ell_3^E}(ks)
    j_{\ell_3^S}(k's)
   \nn \\
&  \hspace{0.5cm} \times  
\sum_{\substack{M_1 M_2 M_3}}
\threej{L_1}{L_2}{L_3}{M_1}{M_2}{M_3}Y_{L_1M_1}(\hx_1) \, Y_{L_2M_2}(\hx_2) \,Y_{L_3M_3}(\hs)\Big|_{\x_2=\x_1+\s}.
\end{align}
The remaining steps are to replace integration variable $\x_2$ with the separation vector $\s$, factorize the integral over $\s$ into radial and angular components, and to rewrite $G_{\ell_3^E\ell_3^SL_3}$ in terms of $C_{\ell_3^E\ell_3^SL_3}$ using Eq.~\eqref{eq:Clll_def}. Doing so leads to Eq.~\eqref{eq:finite_window}.

We encountered fields with non-factorizable transfer functions in \S\ref{sec:use_cases}. By definition, the window function appropriate for 2PCFs of these fields is identical to Eq.~\eqref{eq:window_first_step} but with $F_i(x_i)\to F_i(x_i,k')$. The additional $k'$ dependence in the transfer function carries through the above derivation without any additional complications. Thus, the window function appropriate for non-factorizable transfer functions takes the same form as Eq.~\eqref{eq:finite_window} but with $F_i(x_i)\to F_i(x_i,k')$ in the subsequent expression for $Q_{L_1L_2L_3}(s)\to Q_{L_1L_2L_3}(s,k')$.

\subsection{Fourier-space definition and computational cost}
\label{sec:fourier_Q_computational_cost}

In Appendix \ref{app:Q_equivalence} we show that the real-space representation of $Q_{L_1L_2L_3}(s)$, given by Eq.~\eqref{eq:QLLL_alternate}, is equivalent to the following Fourier-space representation:
\begin{align}
\tV_i^{LM}(\k) &\equiv
  \int_{\x} e^{-i\k\cdot\x} W_i(\x) F_i(x) Y_{LM}(\hx)
  \label{eq:tV_def2} \\
\tQ^{L_1L_2}_{L_3M_3}(\k) &\equiv
  \sum_{M_1M_2} \threej{L_1}{L_2}{L_3}{M_1}{M_2}{M_3}
    \tV_1^{L_1M_1}(-\k) \, \tV_2^{L_2M_2}(\k)
    \label{eq:tQ_def2} \\
Q^{L_1L_2}_{L_3M_3}(\s) &\equiv
  \int_{\k} e^{i\k\cdot\s} \tQ^{L_1L_2}_{L_3M_3}(\k)
  \label{eq:QLLM_def2}  \\
Q_{L_1L_2L_3}(s) &=
(4\pi)^{3/2} \sqrt{(2L_1+1)(2L_2+1)(2L_3+1)}
\int d\Omega_s
\sum_{M_3}
 Q^{L_1L_2}_{L_3M_3}(\s)
 Y_{L_3M_3}(\hs).  \label{eq:QLLL_def2}
\end{align}
The Fourier-space representation provides a ``brute force'' algorithm for computing $Q_{L_1L_2L_3}(s)$ based on 3-d FFTs, and thus a convenient handle on its computational cost. 
Here, we assume that $W_i(\x)$ is specified as a 3-d pixelized map with $N_{\rm pix} = N^3$ pixels.
This map could be obtained by ``gridding'' a random catalog.
The chain of equations above has been organized so that all computational steps are either FFTs with cost $\O(N_{\rm pix} \log N_{\rm pix})$, or element-wise operations with cost $\O(N_{\rm pix})$ (e.g.\ multiplying a real-space map by $Y_{LM}(\hx)$, or multiplying a Fourier-space map by $Y_{LM}(\hk)$).
It follows that for general weight functions $W_i(\x)$, the window function (Eq.~\ref{eq:finite_window}) can be computed with cost $\O(N_{\rm pix} \log N_{\rm pix})$.

\subsection{Recursion relations}
\label{sec:Q_recursion}

In Appendix \ref{app:recursion_relations_q} we show that $Q_{L_1L_2L_3}(s)$ satisfies the following recursion relations in $L_i$:
\begin{equation}
    \sum_{\{L^\pm_i\}}
    (-1)^{\sum L^{\pm}_i}
    \ninej{L_1}{L_2}{L_3}{1}{1}{1}{L^{\pm}_1}{L^{\pm}_2}{L^{\pm}_3}G_{L_11L^{\pm}_1}
    G_{L_21L^{\pm}_2}
    G_{L_31L^{\pm}_3} \,\,
    Q_{L^\pm_1L^\pm_2L^\pm_3}(s)
    =
    0,
\end{equation}
where $L_i^\pm\equiv L_i\pm 1$ and the sum runs over all sign choices for the triplet $(L_1^\pm,L_2^\pm,L_3^\pm)$ with $L_i\geq 0$. In the same Appendix we show that $Q_{L_1L_2L_3}$ vanishes unless $L_1+L_2+L_3$ is even. These observations dramatically reduce the computational cost for high $L_i$, by reducing the number of independent $Q_{L_1L_2L_3}$ which need to be computed in the window function. For example, taking $(L_1,L_2,L_3)=(1,1,1)$ in the expression above yields
\begin{equation}
    \frac{\sqrt{14}}{40} Q_{222} + \frac{\sqrt{5}}{40} Q_{220} + \frac{\sqrt{5}}{40} Q_{202} + \frac{\sqrt{5}}{40} Q_{022} -  \frac{1}{8} Q_{000} = 0,
\end{equation}
whereas for $(L_1,L_2,L_3) = (1,3,3),\,(3,1,3)$ and $(3,3,1)$ we find
\begin{align}
    \frac{\sqrt{22}}{56} Q_{244} + \frac{3 \sqrt{5}}{280} Q_{242} + \frac{3 \sqrt{5}}{280} Q_{224} + \frac{\sqrt{14}}{56} Q_{044} -  \frac{3}{70} Q_{222} -  \frac{3 \sqrt{70}}{280} Q_{022} &= 0
    \\
    \frac{\sqrt{22}}{56} Q_{424} + \frac{3 \sqrt{5}}{280} Q_{422} + \frac{\sqrt{14}}{56} Q_{404} + \frac{3 \sqrt{5}}{280} Q_{224} -  \frac{3}{70} Q_{222} -  \frac{3 \sqrt{70}}{280} Q_{202} &= 0
    \\
    \frac{\sqrt{22}}{56} Q_{442} + \frac{\sqrt{14}}{56} Q_{440} + \frac{3 \sqrt{5}}{280} Q_{422} + \frac{3 \sqrt{5}}{280} Q_{242} -  \frac{3}{70} Q_{222} -  \frac{3 \sqrt{70}}{280} Q_{220} &= 0,
\end{align}
respectively.

\subsection{Simplified geometries}
\label{sec:simplified_geometries}

The window function (Eqs.~\ref{eq:finite_window} and \ref{eq:QLLL_alternate}) simplifies considerably when the geometry of the survey, encoded in the weight function $W_i(\x)$, simplifies to either a purely radial function (full-sky) or the product of a purely radial function and an angular mask (masked-sky). In Appendix \ref{app:full_sky_q} we show that for purely radial weights, which we denote $W_i(\x) = R_i(x)$, Eq.~\eqref{eq:QLLL_alternate} simplifies to
\begin{equation}
\begin{aligned}
    &Q^{\rm full-sky}_{L_1L_2L_3}(s)
    =
    32\pi\,
    i^{L_1+L_3-L_2}
    \Bigg[
    \prod_{i=1}^3
    (2L_i+1)
    \Bigg]
    C_{L_1L_2L_3}
    \int k^2dk\,
    j_{L_3}(ks)
    r^{(1)}_{L_1}(k)
    r^{(2)}_{L_2}(k)
    \label{eq:full_sky_q}
    \\&{\rm where}\quad
    r^{(i)}_L(k)
    \equiv
    \int x^2 dx\,
    F_i(x)\,R_i(x)\,j_L(kx).
\end{aligned}
\end{equation}
The full-sky $Q_{L_1L_2L_3}$ can be computed with three Hankel transforms, whose cost scales as $\mathcal{O}(N\log N)$ with the FFTLog routine \cite{2015ascl.soft12017H}. 
For a 3-d pixelized map with $N_{\rm pix} = N^3$ pixels, the computational cost of Eq.~\eqref{eq:full_sky_q} therefore scales as $\mathcal{O}(N^{1/3}_{\rm pix}\log N_{\rm pix})$, which is considerably less expensive than the $\mathcal{O}(N_{\rm pix}\log N_{\rm pix})$ scaling for a generic weight function (\S\ref{sec:fourier_Q_computational_cost}). 

In Appendix \ref{app:separable_q} we generalize this result to the case when the weight function is a product of a radial function and a mask, i.e. $W_i(\x) = R_i(x)A_i(\hx)$, finding
\begin{align}
    &Q^{\rm masked-sky}_{L_1L_2L_3}(s)
    =
    \sum_{\ell\ell_1\ell_2}
    \mathcal{M}^{L_1L_2L_3}_{\ell_1\ell_2\ell}\,\,
    C^{A_1A_2}_\ell\,\,
    Q^{\rm full-sky}_{\ell_1\ell_2L_3}(s)
    \label{eq:separable_q}
    \\
    &{\rm where}\quad
    \mathcal{M}^{L_1L_2L_3}_{\ell_1\ell_2\ell}\equiv
    (-1)^{\ell_2+L_2+L_3}
    \sqrt{\frac{(2L_1+1)(2L_2+1)}{(2\ell_1+1)(2\ell_2+1)}}
    \sixj{\ell_1}{\ell_2}{L_3}{L_2}{L_1}{\ell}
    G_{\ell_1L_1\ell} G_{\ell_2L_2\ell}
    \label{eq:masked_sky_coupling}
    \\
    &{\rm and}\quad 
    C^{A_1A_2}_\ell
    \equiv
    \frac{1}{2\ell+1}\sum_m A^{(1)}_{\ell m} A^{(2)*}_{\ell m}.
    \label{eq:mask_spectrum}
\end{align}
This result is analogous to the gSFB approach (\S\ref{sec:sfb}), where Refs.~\cite{Castorina:2017inr,Wen:2024hqj} showed that the masked-sky window function can be written as the full-sky window function convolved with the familiar pseudo-$C_\ell$ coupling kernel \cite{2002ApJ...567....2H}. 
The sum over $\ell$ in Eq.~\eqref{eq:separable_q} is formally infinite, while the triangle conditions enforced by Eq.~\eqref{eq:masked_sky_coupling} ensure that the sums over $\ell_i$ run from $|L_i-\ell|$ to $L_i+\ell$.

\subsection{Noise bias subtraction}
\label{sec:noise_bias}

Up until now we ignored noise bias for simplicity.
The primary example is Poisson noise for a galaxy field, but we model noise generally as a diagonal contribution to the 2PCF:
\begin{equation}
\big\langle \delta_1(\x_1) \delta_2(\x_2) \big\rangle
 = N(\x_1) \delta^3(\x_1-\x_2).
\end{equation}
Noise bias is straightforward to calculate, starting from the definition~\eqref{eq:Pll} of $\hP_{\ell_1\ell_2}$: 
\begin{equation}
\begin{aligned}
    \text{Bias}\Big[\big\langle \hP_{\ell_1\ell_2}(k) \big\rangle\Big] &=
\frac{2(\ell_1+\ell_2)+1}{V} 
    \int_{\x}
    W_1(\x)W_2(\x)N(\x)
    \int\frac{d\Omega_k}{4\pi}
    \mathcal{L}_{\ell_1}(\hk\cdot\hx)\mathcal{L}_{\ell_2}(\hk\cdot\hx)
    \\
    &=
    \frac{4\ell_1+1}{(2\ell_1+1)V}\delta_{\ell_1\ell_2}
    \int_{\x}
    W_1(\x)W_2(\x)N(\x).
    \label{eq:noise_bias}
\end{aligned}
\end{equation}
In the presence of noise bias the optimal \textit{unbiased} estimator is obtained by subtracting Eq.~\eqref{eq:noise_bias} from Eq.~\eqref{eq:Pll}. The spin structure of the estimator is unchanged in the presence of noise bias, and the optimal estimator remains a sum of two-$\ell$ estimators.

%% file: SecConclusions.tex
\section{Discussion and conclusions}
\label{sec:discussion_and_conclusions}

The two-$\ell$ generalization of the Yamamoto estimator (Eq.~\ref{eq:Pll}) provides the natural language for measuring finite-rank 2PCFs beyond the plane-parallel approximation. We have shown that the response of the two-$\ell$ estimator to a rank-1 2PCF (Eq.~\ref{eq:rk1_sig}) can be written exactly as a finite sum over angular momentum couplings and generalized survey moments $Q_{L_1L_2L_3}(s)$. 
This result is sufficient to capture a broad range of relevant cosmological signals: linear redshift-space distortions, local relativistic contributions to galaxy number counts and remote dipole reconstruction from the kinetic Sunyaev-Zel'dovich effect can all be expressed either exactly or to a very good approximation as finite sums of rank-1 signals (\S\ref{sec:use_cases}).

We outline practical algorithms to compute the finite window function using FFTs and Hankel transforms (\S\ref{sec:window_function}), finding that its computational cost scales as $\mathcal{O}(N_{\rm pix}\log N_{\rm pix})$ for a generic survey geometry or $\mathcal{O}(N^{1/3}_{\rm pix}\log N_{\rm pix})$ when the survey factorizes into the product of a radial selection function and angular mask, and provide recursion relations for the survey moments to further optimize the window calculation for large $L_i$. 
We numerically compare our formalism against the distant-observer and gSFB expansions in the context of linear theory redshift-space distortions (\S\ref{sec:rsd}) and find excellent agreement in the expected limits (Fig.~\ref{fig:full_sky_comparison}).

Furthermore, in \S\ref{sec:optimality} we show that the optimal estimator for finite-rank 2PCFs inherits the same spin structure as the signal itself. For signals such as the redshift-space hexadecapole, which arises from a $(\ell_1^S,\ell_2^S)=(2,2)$ rather than a single $\ell=4$ structure, the two-$\ell$ estimator is both more optimal and computationally cheaper \cite{Scoccimarro:2015bla,2020MNRAS.498.2492G,2026arXiv260504864X} than the conventional $\ell=4$ Yamamoto estimator. We verify this numerically in \S\ref{ssec:sims}, finding $\mathcal{O}(1)$ improvements to the RSD power spectrum signal-to-noise ratio when substituting a $(2,2)$ estimator for the conventional $\ell=4$ Yamamoto estimator on ultra-large scales (Fig.~\ref{fig:grf}).

In our numerical tests we restricted ourselves to a full-sky survey out of simplicity. We will present an efficient numerical implementation of Eq.~\eqref{eq:finite_window} for an arbitrary survey geometry in a forthcoming publication. Our formalism is restricted to finite-rank 2PCFs, whereas in \S\ref{sec:use_cases} we encountered integrated contributions to galaxy number counts that are not naturally captured by Eq.~\eqref{eq:finite_window}. We expect that generalizing Eq.~\eqref{eq:finite_window} to integrated terms will be non-trivial, which we leave to future work.

Throughout we specialized to the two-$\ell$ estimator given by Eq.~\eqref{eq:Pll}, in which every galaxy retains its own line-of-sight direction $\hx_i$. One could imagine generalizing this result by substituting an alternative line-of-sight direction, i.e. taking $\L_{\ell_i^S}(\hk\cdot\hx_i)\to\L_{\ell_i^S}(\hk\cdot\mathbf{\hat{d}}_i)$ for some suitable $\mathbf{\hat{d}}_i$. Alternative line-of-sight choices, including the midpoint and bisector directions, have been explored for the Yamamoto estimator, where it has been shown that one can reduce the size of wide-angle effects and the impact of multipole mixing \cite{Reimberg:2015jma,Castorina:2017inr,2014PhRvD..89h3535B,Beutler:2018vpe,Philcox:2021tfv,2023JCAP...04..067P}. These alternatives are useful because the Yamamoto estimator forces a single line-of-sight onto each galaxy pair, however, this is no longer the case for the two-$\ell$ estimator. In fact, alternative line-of-sight choices would spoil the optimality of the two-$\ell$ estimator and result in an infinite series representation for its window function, so we chose not to explore this direction further.

There are several natural directions for future work. In our numerical studies we estimated uncertainties using Monte Carlo realizations, but it would be interesting to develop approximate analytic covariance estimates \citep[e.g.][]{2004MNRAS.349..603E,2026arXiv260418581G}. A number of practical effects relevant for real survey analyses also remain to be incorporated. These include integral constraint \cite{2019JCAP...08..036D} corrections and Alcock--Paczynski distortions \cite{APeffect}. Fiber-collision mitigation presents another interesting case study. In DESI-like analyses, one common approach is to apply a $\theta$-cut that removes pairs with angular separation below some threshold $\theta_c$ \cite{2025JCAP...01..131P,2025JCAP...04..074B}. While such modifications would complicate the derivation of the finite window function, they may nevertheless remain analytically tractable.

More broadly, we expect the finite-rank perspective developed here to have applications beyond the examples considered in this work. 
Extending the formalism to higher-order statistics such as bispectra, tensor fields \cite{2024PhRvD.110f3546M} or reformulating the analysis directly in configuration space may provide complementary avenues for optimal and exact treatments of wide-angle observables.

%% file: AppSpecFunc.tex
\section{Special function reference}
\label{app:special_functions}



\subsection{Wigner 3j symbols}
\label{ssec:wigner_3j}

The Wigner 3j symbol
\begin{equation}
\threej{\ell_1}{\ell_2}{\ell_3}{m_1}{m_2}{m_3}
\end{equation}
vanishes unless $|m_i| \le l_i$ for all $i$, $\sum_i m_i=0$, and $T_{\ell_1\ell_2\ell_3}=1$, where we define:
\begin{equation}
T_{\ell_1\ell_2\ell_3} \equiv \left\{ \begin{array}{cl}
 1 & \mbox{if } (l_1+l_2+l_3) \ge 2 \cdot \max(l_1,l_2,l_3) 
    \hspace{0.5cm} 
    \mbox{(``triangle condition'')} \\
 0 & \mbox{otherwise}.
\end{array} \right.
\end{equation}
The 3j symbol is symmetric (with a sign) under permutations of columns:
\begin{equation}
\threej{\ell_2}{\ell_1}{\ell_3}{m_2}{m_1}{m_3}
=
\threej{\ell_1}{\ell_3}{\ell_2}{m_1}{m_3}{m_2}
=
(-1)^{\ell_1+\ell_2+\ell_3}
\threej{\ell_1}{\ell_2}{\ell_3}{m_1}{m_2}{m_3},
\end{equation}
and under reversing signs of the $m_i$:
\begin{equation}
\threej{\ell_1}{\ell_2}{\ell_3}{-m_1}{-m_2}{-m_3}
= (-1)^{\ell_1+\ell_2+\ell_3}
\threej{\ell_1}{\ell_2}{\ell_3}{m_1}{m_2}{m_3}.
\end{equation}
Here is an expression for a ``special'' 3j symbol:
\begin{align}
\threej{\ell}{\ell}{0}{m}{-m}{0}
  &= \frac{(-1)^{\ell+m}}{\sqrt{2\ell+1}}. \label{eq:special_3j}
\end{align}
The 3j symbol obeys the orthogonality relations:
\begin{equation}
\sum_{m_1m_2}
  \threej{\ell_1}{\ell_2}{\ell_3}{m_1}{m_2}{m_3}
  \threej{\ell_1}{\ell_2}{\ell'_3}{m_1}{m_2}{m'_3}
= \frac{T_{\ell_1\ell_2\ell_3}}{2\ell_3+1}
\delta_{\ell_3\ell_3'} \delta_{m_3m_3'}
\label{eq:3j_orthogonality}
\end{equation}
\begin{equation}
    \sum_{\ell_3 m_3}
    (2\ell_3+1)
    \threej{\ell_1}{\ell_2}{\ell_3}{m_1}{m_2}{m_3}
    \threej{\ell_1}{\ell_2}{\ell_3}{m'_1}{m'_2}{m_3}
    =
    \delta_{m_1m_1'}\delta_{m_2m_2
    '}.
\label{eq:3j_othogonality2}
\end{equation}

\subsection{Wigner 6j symbols}

The following is one way of defining the Wigner 6j symbol:
\begin{align}
\sixj{\ell_1}{\ell_2}{\ell_3}{\ell_1'}{\ell_2'}{\ell_3'}
&\equiv
\sum_{\substack{m_1m_2m_3 \\ m'_1m'_2m'_3}}
(-1)^{\ell_1+\ell_1'+m_1+m_1'}
\threej{\ell_1}{\ell_2}{\ell_3}{m_1}{m_2}{m_3}
\threej{\ell_1}{\ell_2'}{\ell_3'}{-m_1}{m_2'}{m_3'} \nn \\
& \hspace{5cm} \times
\threej{\ell_1'}{\ell_2}{\ell_3'}{m'_1}{m_2}{m'_3}
\threej{\ell_1'}{\ell_2'}{\ell_3}{-m'_1}{m'_2}{m_3}.
\label{eq:6j_def}
\end{align}
The 6j symbol is invariant under any permutation of its columns, and is also invariant under the exchange of upper and lower arguments in any two columns.
It vanishes unless the following triangle conditions are satisfied:
\begin{equation}
T_{\ell_1\ell_2\ell_3} = T_{\ell_1\ell_2'\ell_3'} = T_{\ell_1'\ell_2\ell_3'} = T_{\ell_1'\ell_2'\ell_3} = 1.
\label{eq:6j_triangle_conditions}
\end{equation}
Here are expressions for some ``special'' 6j symbols\footnote{\url{https://dlmf.nist.gov/34.5\#i}}:
\begin{equation}
\label{eq:6j_special}
\begin{aligned}
\sixj{j_1}{j_2}{j_3}{j_2-1}{j_1-1}{1}
&= (-1)^{j_1+j_2+j_3} \Biggl[
   \frac{(j_1+j_2+j_3)(j_1+j_2+j_3+1)}{(2j_1-1)(2j_1)(2j_1+1)}
\\[2pt]
&\qquad\times
   \frac{(j_1+j_2-j_3)(j_1+j_2-j_3-1)}{(2j_2-1)(2j_2)(2j_2+1)}
   \Biggr]^{1/2}
\\[6pt]
\sixj{j_1}{j_2}{j_3}{j_2-1}{j_1}{1}
&= (-1)^{j_1+j_2+j_3} \Biggl[
   \frac{2(j_1+j_2+j_3+1)(j_1+j_2-j_3)}{(2j_1)(2j_1+1)(2j_1+2)}
\\[2pt]
&\qquad\times
   \frac{(j_2+j_3-j_1)(j_1-j_2+j_3+1)}{(2j_2-1)(2j_2)(2j_2+1)}
   \Biggr]^{1/2}
\\[6pt]
\sixj{j_1}{j_2}{j_3}{j_2+1}{j_1-1}{1}
&= (-1)^{j_1+j_2+j_3} \Biggl[
   \frac{(j_1-j_2+j_3-1)(j_1-j_2+j_3)}{(2j_1-1)(2j_1)(2j_1+1)}
\\[2pt]
&\qquad\times
   \frac{(j_2+j_3-j_1+1)(j_2+j_3-j_1+2)}{(2j_2+1)(2j_2+2)(2j_2+3)}
   \Biggr]^{1/2}
\\[6pt]
\sixj{j_1}{j_2}{j_3}{j_2}{j_1}{1}
&= (-1)^{j_1+j_2+j_3+1}\,
   \frac{j_1(j_1+1)+j_2(j_2+1)-j_3(j_3+1)}
        {\bigl[(2j_1)(2j_1+1)(2j_1+2)\, j_2(j_2+1)(2j_2+1)\bigr]^{1/2}}
\end{aligned}
\end{equation}
The 6j symbol can be used to ``recouple'' angular momenta, using the identity \cite{messiah61}
\begin{align}
& \sum_{m_1} (-1)^{m_1}
 \threej{\ell_1}{\ell_2}{\ell_3}{m_1}{m_2}{m_3}
 \threej{\ell_1}{\ell_2'}{\ell_3'}{-m_1}{m_2'}{m_3'} \nn \\
& \hspace{1cm}
= \sum_{\ell_1'} (2\ell_1'+1) 
  \sixj{\ell_1}{\ell_2}{\ell_3}{\ell_1'}{\ell_2'}{\ell_3'}
  \sum_{m_1'} (-1)^{\ell_1+\ell_1'+m_1'}
  \threej{\ell_1'}{\ell_2}{\ell_3'}{m_1'}{m_2}{m_3'}
  \threej{\ell_1'}{\ell_2'}{\ell_3}{-m_1'}{m_2'}{m_3},
\label{eq:6j_recoupling}
\end{align}
or the related identity:
\begin{align}
& \sum_{m'_1m'_2m'_3} (-1)^{m_1'+m_2'+m_3'}
 \threej{\ell_1}{\ell_2'}{\ell_3'}{m_1}{-m_2'}{m_3'}
 \threej{\ell_1'}{\ell_2}{\ell_3'}{m_1'}{m_2}{-m_3'}
 \threej{\ell_1'}{\ell_2'}{\ell_3}{-m_1'}{m_2'}{m_3} \nn \\
& \hspace{1.5cm}
 = (-1)^{\ell_1'+\ell_2'+\ell_3'}
   \sixj{\ell_1}{\ell_2}{\ell_3}{\ell_1'}{\ell_2'}{\ell_3'}
   \threej{\ell_1}{\ell_2}{\ell_3}{m_1}{m_2}{m_3}.
   \label{eq:6j_recoupling2}
\end{align}
Eq.~\eqref{eq:6j_recoupling2} can be derived by summing over $m_1'$ using Eq.~\eqref{eq:6j_recoupling} and over $(m_2',m'_3
)$ using Eq.~\eqref{eq:3j_orthogonality}.

\subsection{Wigner 9j symbols}

The following is one way of defining the Wigner 9j symbol:
\begin{align}
\ninej{l_{11}}{l_{12}}{l_{13}}{l_{21}}{l_{22}}{l_{23}}{l_{31}}{l_{32}}{l_{33}}
    &=
     \sum_{m_{ij}}
\threej{l_{11}}{l_{12}}{l_{13}}{m_{11}}{m_{12}}{m_{13}} 
\threej{l_{21}}{l_{22}}{l_{23}}{m_{21}}{m_{22}}{m_{23}}
\threej{l_{31}}{l_{32}}{l_{33}}{m_{31}}{m_{32}}{m_{33}}
   \nn \\ & \hspace{1cm} \times
\threej{l_{11}}{l_{21}}{l_{31}}{m_{11}}{m_{21}}{m_{31}} 
\threej{l_{12}}{l_{22}}{l_{32}}{m_{12}}{m_{22}}{m_{32}} 
\threej{l_{13}}{l_{23}}{l_{33}}{m_{13}}{m_{23}}{m_{33}}. 
    \label{eq:9j_def}
\end{align}
The 9j symbol vanishes unless triangle conditions are satisfied along all rows and columns:
\begin{equation}
T_{\ell_{11}\ell_{12}\ell_{13}}
= T_{\ell_{21}\ell_{22}\ell_{23}}
= T_{\ell_{31}\ell_{32}\ell_{33}}
= T_{\ell_{11}\ell_{21}\ell_{31}}
= T_{\ell_{12}\ell_{22}\ell_{32}}
= T_{\ell_{13}\ell_{23}\ell_{33}}
= 1.
\label{eq:9j_selection_rules}
\end{equation}
The 9j symbol is invariant under transposing the 3-by-3 array $\ell_{ij}$.
If any pair of rows or columns is exchanged, the 9j symbol changes by a sign $(-1)^{\sum\ell_{ij}}$.

Here are some ``special'' 9j symbols:
\begin{align}
\ninej{l_{11}}{l_{12}}{l_{13}}{l_{21}}{l_{22}}{l_{13}}{l_{31}}{l_{31}}{0}
 &= \frac{(-1)^{\ell_{12}+\ell_{21}+\ell_{13}+\ell_{31}}}{\sqrt{(2\ell_{13}+1)(2\ell_{31}+1)}}
 \sixj{\ell_{11}}{\ell_{12}}{\ell_{13}}{\ell_{22}}{\ell_{21}}{\ell_{31}}
 \label{eq:special_9j}
\\ 
 \ninej{L_1}{L_2}{L_3}
{1}{1}{1}
{L_1-1}{L_2-1}{L_3-1}
&=\sqrt{
\frac{
(S-1)S(S+1)
(S-2L_3)(S-2L_2)(S-2L_1)
}{
3\prod_{x\in\{L_1,L_2,L_3\}} (2x-1)(2x)(2x+1)
}}
\label{eq:special_9j_2}
\\
&\hspace{2cm}{\rm where}\hspace{5mm} S\equiv L_1+L_2+L_3.\nn
\end{align}
Eq.~\eqref{eq:special_9j_2} can be derived from Eqs.~\eqref{eq:9j6j} and \eqref{eq:6j_special}.

The 9j symbol obeys the following identities \cite{messiah61}
\begin{align}
\label{eq:9j3j}
\threej{J_{13}}{J_{24}}{J}{M_{13}}{M_{24}}{M}
\ninej
{j_1}{j_2}{J_{12}}
{j_3}{j_4}{J_{34}}
{J_{13}}{J_{24}}{J}
&=
\sum_{m_1,m_2,m_3,m_4 \atop M_{12},M_{34}}
\threej{j_1}{j_2}{J_{12}}{m_1}{m_2}{M_{12}}
\threej{j_3}{j_4}{J_{34}}{m_3}{m_4}{M_{34}} \nn\\
&\quad \times
\threej{j_1}{j_3}{J_{13}}{m_1}{m_3}{M_{13}}
\threej{j_2}{j_4}{J_{24}}{m_2}{m_4}{M_{24}}
\threej{J_{12}}{J_{34}}{J}{M_{12}}{M_{34}}{M}
\end{align}
\begin{align}
\label{eq:9j6j}
\ninej
{j_1}{j_2}{J_{12}}
{j_3}{j_4}{J_{34}}
{J_{13}}{J_{24}}{J}
=
\sum_{g}
(-1)^{2g}(2g+1)
\sixj{j_1}{j_2}{J_{12}}{J_{34}}{J}{g}
\sixj{j_3}{j_4}{J_{34}}{j_2}{g}{J_{24}}
\sixj{J_{13}}{J_{24}}{J}{g}{j_1}{j_3}
\end{align}
In particular, the 9j symbol can be used to ``recouple'' angular momenta, using the identity:
\begin{align}
& \sum_{\substack{m_{11} m_{12} \\ m_{21} m_{22}}}
 \threej{\ell_{11}}{\ell_{12}}{\ell_{13}}{m_{11}}{m_{12}}{m_{13}}
 \threej{\ell_{21}}{\ell_{22}}{\ell_{23}}{m_{21}}{m_{22}}{m_{23}}
 \threej{\ell_{11}}{\ell_{21}}{\ell_{31}}{m_{11}}{m_{21}}{m_{31}}
 \threej{\ell_{12}}{\ell_{22}}{\ell_{32}}{m_{12}}{m_{22}}{m_{32}}
   \nn \\ & \hspace{1cm} =
 \sum_{l_{33} m_{33}} (2l_{33}+1)
   \ninej{\ell_{11}}{\ell_{12}}{\ell_{13}}{\ell_{21}}{\ell_{22}}{\ell_{23}}{\ell_{31}}{\ell_{32}}{\ell_{33}}
    \threej{\ell_{13}}{\ell_{23}}{\ell_{33}}{m_{13}}{m_{23}}{m_{33}}
    \threej{\ell_{31}}{\ell_{32}}{\ell_{33}}{m_{31}}{m_{32}}{m_{33}}.
    \label{eq:9j_recoupling}
\end{align}
Eq.~\eqref{eq:9j_recoupling} can be derived by multiplying both sides of Eq.~\eqref{eq:9j3j} by $(2\ell+1)\times$ (a 3j symbol) and summing over a $(\ell,m)$ pair using Eq.~\eqref{eq:3j_othogonality2}.

\subsection{Spherical harmonics}

The integral of three spherical harmonics is given by:
\begin{equation}
\int d\Omega \, 
Y_{\ell_1m_1}(\hx) Y_{\ell_2m_2}(\hx) Y_{\ell_3m_3}(\hx)
 = G_{\ell_1\ell_2\ell_3}
     \threej{\ell_1}{\ell_2}{\ell_3}{m_1}{m_2}{m_3}
     \label{eq:yyy}
\end{equation}
where $G_{\ell_1\ell_2\ell_3}$ is defined by Eq.~\eqref{eq:Clll_def}.
Note that $G_{\ell_1\ell_2\ell_3}$ is symmetric in $(\ell_1,\ell_2,\ell_3)$, and $G_{\ell_1\ell_2\ell_3}=0$ if $(\ell_1+\ell_2+\ell_3)$ is odd.

Here are two identities that are closely related to Eq.\ (\ref{eq:yyy}):
\begin{align}
Y_{\ell_1m_1}(\hx) Y_{\ell_2m_2}(\hx) &= 
 \sum_{\ell_3m_3} G_{\ell_1\ell_2\ell_3} 
 \threej{\ell_1}{\ell_2}{\ell_3}{m_1}{m_2}{m_3}
 Y_{\ell_3m_3}^*(\hx)
 \label{eq:ylm_ylm} 
 \\
\sum_{m_1m_2} \threej{\ell_1}{\ell_2}{\ell_3}{m_1}{m_2}{m_3}
  Y_{\ell_1m_1}(\hx) Y_{\ell_2m_2}(\hx)
   &= \frac{G_{\ell_1\ell_2\ell_3}}{2\ell_3+1} \,
   Y_{\ell_3m_3}^*(\hx)
   \label{eq:yy3j}
\end{align}
These follow from Eq.\ (\ref{eq:yyy}), orthonormality of spherical harmonics, and the orthogonality relation (\ref{eq:3j_orthogonality}) for the Wigner 3j symbol.

\subsection{Legendre polynomials}

Legendre polynomials and spherical harmonics are related by:
\begin{equation}
\L_\ell(\hx\cdot\hy) = \frac{4\pi}{2\ell+1} 
 \sum_m Y_{\ell m}(\hx) Y_{\ell m}^*(\hy).
 \label{eq:legpoly_ylm}
\end{equation}
The product of Legendre polynomials is given by:
\begin{equation}
\L_{\ell_1}(\mu) \, \L_{\ell_2}(\mu)
 = \sum_{\ell_3} (2\ell_3+1)
  C_{\ell_1\ell_2\ell_3}^2 \L_{\ell_3}(\mu)
  \label{eq:Pl_product}.
\end{equation}

\subsection{Rayleigh expansion}

The Rayleigh expansion of a plane wave is:
\begin{align}
e^{i\k\cdot\r} &= \sum_{\ell=0}^\infty  i^\ell (2\ell+1) j_\ell(kr)\L_\ell(\hk\cdot\hr)
\label{eq:rayleigh_expansion_legendre}
\\
e^{i\k\cdot\r} &= 4\pi \sum_{\ell=0}^\infty \sum_{m=-\ell}^\ell i^\ell j_\ell(kr)Y_{\ell m}(\hk) Y_{\ell m}^*(\hr)
\label{eq:rayleigh_expansion}
\end{align}
The following identity follows:
\begin{equation}
\int d\Omega_k e^{i\k\cdot\r} Y_{\ell m}^*(\hk)
  = 4\pi i^\ell j_\ell(kr) Y_{\ell m}^*(\hr).
  \label{eq:rayleigh_integral}
\end{equation}

%% file: AppDerivations.tex
\section{Derivations}

\subsection{gSFB window function}
\label{sec:sfb_window}

Here we derive Eq.~\eqref{eq:sfb_window} for the window function in the gSFB formalism. We start by taking the expectation value of Eq.~\eqref{eq:sfb_pseudo_cl}, substituting the definition of a rank-1 2PCF (Eq.~\ref{eq:rk1_sig}) and comparing with Eq.~\eqref{eq:window} to read off the gSFB window function as
\begin{align}
    W^{\ell_1^S\ell_2^S}_{0\ell}(k,k')&=\frac{(4\pi)^2}{2\ell+1}\sum_{\ell_1\ell_2}i^{\ell_2-\ell_1}G^2_{\ell\ell_1\ell_2}\frac{1}{2\ell_1+1}\sum_{m_1}
    \int\frac{d\Omega_{k'}}{4\pi}\nn\\&\hspace{1cm}\times\int_{\x_1} j_{\ell_1}(kx_1)Y_{\ell_1m_1}(\hx_1)e^{i\k'\cdot\x_1}\L_{\ell_1^S}(\hk'\cdot\hx_1)W_1(\x_1)F_1(x_1)\nn\\&\hspace{1cm}\times\int_{\x_2} j_{\ell_2}(kx_2)Y^*_{\ell_1m_1}(\hx_2)e^{-i\k'\cdot\x_2}\L_{\ell_2^S}(\hk'\cdot\hx_2)W_2(\x_2)F_2(x_2).
    \label{eq:sfb_window_step1}
\end{align}
We then perform the angular integral over $\hk'$ using the following identity
\begin{align}
    &\int\frac{d\Omega_{k'}}{4\pi} e^{i\k'\cdot(\x_1-\x_2)}\L_{\ell_1^S}(\hk'\cdot\hx_1)\L_{\ell_2^S}(\hk'\cdot\hx_2)\nn\\&\hspace{1.5cm}=\sum_{L_1L_2L_3}i^{L_1-L_2}\bigg[\prod_{i=1}^3(2L_i+1)\bigg]C^2_{\ell_1^SL_1L_3}C^2_{\ell_2^SL_2L_3} j_{L_1}(k'x_1)j_{L_2}(k'x_2)\L_{L_3}(\hx_1\cdot\hx_2).
\end{align}
To derive this, we first apply the Rayleigh expansion (Eq.~\ref{eq:rayleigh_expansion_legendre}) to the two plane-wave factors. We then simplify the resulting products of Legendre polynomials sharing the same argument using Eq.~\eqref{eq:Pl_product}.
Finally, we perform the angular integral over $\hk'$ using the addition theorem (Eq.~\ref{eq:legpoly_ylm}) and orthonormality of spherical harmonics.

Substituting this identity into Eq.~\eqref{eq:sfb_window_step1} and noting that the sum over $m_1$ can be simplified using Eq.~\eqref{eq:legpoly_ylm} gives
\begin{align}
    W^{\ell_1^S\ell_2^S}_{0\ell}(k,k')&=\frac{4\pi}{2\ell+1}\sum_{\ell_1\ell_2}i^{\ell_2-\ell_1}G^2_{\ell\ell_1\ell_2}
    \sum_{L_1L_2L_3} i^{L_1-L_2}\bigg[\prod_{i=1}^3(2L_i+1)\bigg]C^2_{\ell_1^SL_1L_3}C^2_{\ell_2^SL_2L_3}
    \nn\\&\hspace{0.5cm}\times\int_{\x_1} j_{\ell_1}(kx_1)j_{L_1}(k'x_1)W_1(\x_1)F_1(x_1)\int_{\x_2} j_{\ell_2}(kx_2)j_{L_2}(k'x_2)W_2(\x_2)F_2(x_2)\nn\\&\hspace{2cm}\times\L_{\ell_1}(\hx_1\cdot\hx_2)\L_{L_3}(\hx_1\cdot\hx_2)
\end{align}
The product of Legendre polynomials can be simplified using a combination of Eqs.~\eqref{eq:Pl_product} and \eqref{eq:legpoly_ylm}
\begin{equation}
    \L_{\ell_1}(\hx_1\cdot\hx_2)\L_{L_3}(\hx_1\cdot\hx_2)
    =
    4\pi\sum_{LM}C^2_{\ell_1L_3L}Y_{LM}(\hx_1)Y^*_{LM}(\hx_2),
\end{equation}
which recovers Eq.~\eqref{eq:sfb_window}.

\subsection{An identity involving the product of four Legendre polynomials}
\label{sec:L4}

In \S\ref{sec:window_function} we'll need the following identity involving the product of four Legendre polynomials:
\begin{align}
& \L_{\ell_{11}}(\hk_1 \cdot \hx_1) \,
\L_{\ell_{12}}(\hk_1 \cdot \hx_2) \,
\L_{\ell_{21}}(\hk_2 \cdot \hx_1) \,
\L_{\ell_{22}}(\hk_2 \cdot \hx_2) \nn \\
& \hspace{1cm} = \hspace{-0.5cm}
\sum_{\substack{\ell_{13}\ell_{23}\ell_{31}\ell_{32}\ell_{33} \\  m_{13}m_{23}m_{31}m_{32}m_{33}}} \hspace{-0.3cm}
(4\pi)^2 (2\ell_{33}+1) 
\sqrt{(2\ell_{13}+1)(2\ell_{23}+1)(2\ell_{31}+1)(2\ell_{32}+1)} 
\nn \\ & \hspace{2cm} \times
C_{\ell_{11}\ell_{12}\ell_{13}}
C_{\ell_{21}\ell_{22}\ell_{23}}
C_{\ell_{11}\ell_{21}\ell_{31}}
C_{\ell_{12}\ell_{22}\ell_{32}}
\ninej{\ell_{11}}{\ell_{12}}{\ell_{13}}{\ell_{21}}{\ell_{22}}{\ell_{23}}{\ell_{31}}{\ell_{32}}{\ell_{33}}
\threej{\ell_{13}}{\ell_{23}}{\ell_{33}}{m_{13}}{m_{23}}{m_{33}}
\nn \\ & \hspace{2cm} \times
\threej{\ell_{31}}{\ell_{32}}{\ell_{33}}{m_{31}}{m_{32}}{m_{33}} 
 Y_{\ell_{13}m_{13}}^*(\hk_1) \, Y_{\ell_{23}m_{23}}^*(\hk_2) \,
  Y_{\ell_{31}m_{31}}(\hx_1) \, Y_{\ell_{32}m_{32}}(\hx_2)
  \label{eq:L4}
\end{align}
where $C_{\ell_1\ell_2\ell_3}$ was defined in Eq.\ (\ref{eq:Clll_def}).

The rest of this section is devoted to deriving Eq.\ (\ref{eq:L4}).
Let $X$ denote the LHS of (\ref{eq:L4}).
To compute $X$, we expand each Legendre polynomial in spherical harmonics:
\begin{equation}
\L_{\ell_{ij}}(\hk_i\cdot\hx_j) = \frac{4\pi}{2\ell_{ij}+1} 
 \sum_{m_{ij}} Y_{\ell_{ij} m_{ij}}(\hk_i) Y_{\ell_{ij} m_{ij}}^*(\hx_j)
\end{equation}
obtaining:
\begin{align}
X &= A \sum_{\substack{m_{11} m_{12} \\ m_{21} m_{22}}}
 \Big( Y_{\ell_{11} m_{11}}(\hk_1) 
       Y_{\ell_{11} m_{11}}^*(\hx_1) \Big) \,
 \Big( Y_{\ell_{12} m_{12}}(\hk_1) 
       Y_{\ell_{12} m_{12}}^*(\hx_2) \Big) \nn \\
 & \hspace{1.5cm} \times
 \Big( Y_{\ell_{21} m_{21}}(\hk_2) 
       Y_{\ell_{21} m_{21}}^*(\hx_1) \Big) \,
 \Big( Y_{\ell_{22} m_{22}}(\hk_2) 
       Y_{\ell_{22} m_{22}}^*(\hx_2) \Big)
\end{align}
where the coefficient $A$ is defined by:
\begin{equation}
A \equiv \frac{(4\pi)^4}{(2\ell_{11}+1) (2\ell_{12}+1) (2\ell_{21}+1) (2\ell_{22}+1)}
\label{eq:As_def}
\end{equation}
Using the identity (\ref{eq:ylm_ylm}), we multiply spherical harmonics in pairs as follows:
\begin{align}
Y_{\ell_{11}m_{11}}(\hk_1)
  Y_{\ell_{12}m_{12}}(\hk_1)
  &= \sum_{\ell_{13}m_{13}} G_{\ell_{11}\ell_{12}\ell_{13}} 
  \threej{\ell_{11}}{\ell_{12}}{\ell_{13}}{m_{11}}{m_{12}}{m_{13}}
  Y_{\ell_{13}m_{13}}^*(\hk_1) \nn \\
Y_{\ell_{21}m_{21}}(\hk_2)
  Y_{\ell_{22}m_{22}}(\hk_2)
  &= \sum_{\ell_{23}m_{23}} G_{\ell_{21}\ell_{22}\ell_{23}} 
  \threej{\ell_{21}}{\ell_{22}}{\ell_{23}}{m_{21}}{m_{22}}{m_{23}}
  Y_{\ell_{23}m_{23}}^*(\hk_2) \nn \\
Y_{\ell_{11} m_{11}}^*(\hx_1)
  Y_{\ell_{21} m_{21}}^*(\hx_1)
  &= \sum_{\ell_{31}m_{31}} G_{\ell_{11}\ell_{21}\ell_{31}}
  \threej{\ell_{11}}{\ell_{21}}{\ell_{31}}{m_{11}}{m_{21}}{m_{31}}
  Y_{\ell_{31}m_{31}}(\hx_1) \nn \\
Y_{\ell_{12} m_{12}}^*(\hx_2)
  Y_{\ell_{22} m_{22}}^*(\hx_2)
  &= \sum_{\ell_{32}m_{32}} G_{\ell_{12}\ell_{22}\ell_{32}}
  \threej{\ell_{12}}{\ell_{22}}{\ell_{32}}{m_{12}}{m_{22}}{m_{32}}
  Y_{\ell_{32}m_{32}}(\hx_2)
\end{align}
obtaining the following expression for $X$:
\begin{align}
X &= A \sum_{\substack{\ell_{13}\ell_{23}\ell_{31}\ell_{32} \\ m_{11}m_{12}m_{21}m_{22} \\ m_{13}m_{23}m_{31}m_{32}}}  
  G_{\ell_{11} \ell_{12} \ell_{13}} 
  \threej{\ell_{11}}{\ell_{12}}{\ell_{13}}{m_{11}}{m_{12}}{m_{13}}
  G_{\ell_{11} \ell_{21} \ell_{31}} 
  \threej{\ell_{11}}{\ell_{21}}{\ell_{31}}{m_{11}}{m_{21}}{m_{31}} \nn \\
& \hspace{2.5cm} \times
  G_{\ell_{21}\ell_{22}\ell_{23}}
  \threej{\ell_{21}}{\ell_{22}}{\ell_{23}}{m_{21}}{m_{22}}{m_{23}}
  G_{\ell_{12}\ell_{22}\ell_{32}}
  \threej{\ell_{12}}{\ell_{22}}{\ell_{32}}{m_{12}}{m_{22}}{m_{32}} \nn \\
& \hspace{2.5cm} \times
  Y_{\ell_{13}m_{13}}^*(\hk_1) \, Y_{\ell_{23}m_{23}}^*(\hk_2) \,
  Y_{\ell_{31}m_{31}}(\hx_1) \, Y_{\ell_{32}m_{32}}(\hx_2)
\end{align}
Next we use the 9j recoupling identity (\ref{eq:9j_recoupling}), obtaining:
\begin{align}
X &= A
\sum_{\substack{\ell_{13}\ell_{23}\ell_{31}\ell_{32}\ell_{33} \\  m_{13}m_{23}m_{31}m_{32}m_{33}}}  
(2l_{33}+1)
  G_{\ell_{11} \ell_{12} \ell_{13}} 
  G_{\ell_{11} \ell_{21} \ell_{31}} 
  G_{\ell_{21}\ell_{22}\ell_{23}}
  G_{\ell_{12}\ell_{22}\ell_{32}}
\ninej{\ell_{11}}{\ell_{12}}{\ell_{13}}{\ell_{21}}{\ell_{22}}{\ell_{23}}{\ell_{31}}{\ell_{32}}{\ell_{33}} \nn \\
& \hspace{1cm} \times
\threej{\ell_{13}}{\ell_{23}}{\ell_{33}}{m_{13}}{m_{23}}{m_{33}}
\threej{\ell_{31}}{\ell_{32}}{\ell_{33}}{m_{31}}{m_{32}}{m_{33}} 
 Y_{\ell_{13}m_{13}}^*(\hk_1) \, Y_{\ell_{23}m_{23}}^*(\hk_2) \,
  Y_{\ell_{31}m_{31}}(\hx_1) \, Y_{\ell_{32}m_{32}}(\hx_2)
\end{align}
Eq.\ (\ref{eq:L4}) now follows by plugging in the definition (\ref{eq:As_def}) of $A$, plugging in the definition (Eq. \ref{eq:Clll_def}) of $G_{\ell_1\ell_2\ell_3}$, and simplifying.

\subsection{Equivalence of real- and Fourier-space definitions of $Q$}
\label{app:Q_equivalence}

Here we show that the Fourier-space definition of $Q_{L_1L_2L_3}(s)$ in  Eqs.~\eqref{eq:tV_def2}--\eqref{eq:QLLL_def2} is equivalent to the real-space definition in Eq.~\eqref{eq:QLLL_alternate}. We start from the Fourier-space definition.
Substituting the definition of $\tV_i^{LM}$ (Eq.~\ref{eq:tV_def2}) into the definition of $\tQ^{L_1L_2}_{L_3M_3}$ (Eq.~\ref{eq:tQ_def2}), and noting that
\begin{equation}
\tV_1^{L_1M_1}(-\k) = \int_{\x_1} e^{i\k\cdot\x_1} W_1(\x_1) F_1(x_1) Y_{L_1M_1}(\hx_1)
\end{equation}
we get:
\begin{align}
\tQ^{L_1L_2}_{L_3M_3}(\k) &= \int_{\x_1} \int_{\x_2}
  e^{i\k\cdot\x_1} e^{-i\k\cdot\x_2} \,
  W_1(\x_1) W_2(\x_2) \, F_1(x_1) F_2(x_2) \nn \\
& \hspace{2cm} \times \sum_{M_1M_2} \threej{L_1}{L_2}{L_3}{M_1}{M_2}{M_3}
  Y_{L_1M_1}(\hx_1) \, Y_{L_2M_2}(\hx_2)
  \label{eq:tQ_expanded}
\end{align}
Next, substituting into Eq.~(\ref{eq:QLLM_def2}):
\begin{align}
Q^{L_1L_2}_{L_3M_3}(\s) &= \int_{\k} \int_{\x_1} \int_{\x_2}
  e^{i\k\cdot(\s+\x_1-\x_2)} \,
  W_1(\x_1) W_2(\x_2) \, F_1(x_1) F_2(x_2) \nn \\
& \hspace{2cm} \times \sum_{M_1M_2} \threej{L_1}{L_2}{L_3}{M_1}{M_2}{M_3}
  Y_{L_1M_1}(\hx_1) \, Y_{L_2M_2}(\hx_2)
\end{align}
The $\k$-integral gives a Dirac delta function $\int_{\k} e^{i\k\cdot(\s+\x_1-\x_2)} = \delta^3(\s+\x_1-\x_2)$, which enforces $\x_2 = \x_1 + \s$.
Evaluating the $\x_2$-integral:
\begin{align}
Q^{L_1L_2}_{L_3M_3}(\s) &= \int_{\x_1}
  W_1(\x_1) W_2(\x_2) \, F_1(x_1) F_2(x_2) \nn \\
& \hspace{2cm} \times \sum_{M_1M_2} \threej{L_1}{L_2}{L_3}{M_1}{M_2}{M_3}
  Y_{L_1M_1}(\hx_1) \, Y_{L_2M_2}(\hx_2)
  \bigg|_{\x_2=\x_1+\s}
  \label{eq:QLLM_realspace}
\end{align}
Finally, substituting into Eq.~(\ref{eq:QLLL_def2}) gives:
\begin{align}
Q_{L_1L_2L_3}(s)
&= (4\pi)^{3/2} \sqrt{\prod_{i=1}^3 (2L_1+1)} \int d\Omega_s \int_{\x_1}
  W_1(\x_1) W_2(\x_2) \, F_1(x_1) F_2(x_2) \nn \\
& \hspace{1cm} \times \sum_{M_1M_2M_3} \threej{L_1}{L_2}{L_3}{M_1}{M_2}{M_3}
  Y_{L_1M_1}(\hx_1) \, Y_{L_2M_2}(\hx_2) \, Y_{L_3M_3}(\hs)
  \bigg|_{\x_2=\x_1+\s},
\end{align}
which recovers Eq.~\eqref{eq:QLLL_alternate}.

\subsection{Recursion relations}
\label{app:recursion_relations_q}


In this section, we will show that the object $Q_{L_1L_2L_3}$ (Eq.~\ref{eq:QLLL_alternate}) satisfies recursion relations in $L_i$. To aid the derivation it is useful to introduce tri-polar spherical harmonics
\begin{equation}
\Phi_{L_1L_2L_3}(\hx_1,\hx_2,\hx_3) \equiv 
 \sum_{M_1M_2M_3} \threej{L_1}{L_2}{L_3}{M_1}{M_2}{M_3}
   Y_{L_1M_1}(\hx_1) Y_{L_2M_2}(\hx_2) Y_{L_3M_3}(\hx_3)
   \label{eq:Phi_def}.
\end{equation}
Note that $\Phi_{L_1L_2L_3}(\hx_1,\hx_2,\hx_3)$ appears in the integrand of Eq.~\eqref{eq:QLLL_alternate}.

\medskip
{\bf\em Step 1: Product rule for $\Phi$-functions.} First, we will show that the product of two $\Phi_{L_1L_2L_3}(\hx_1,\hx_2,\hx_3)$ can be reduced to a linear combination of $\Phi_{L_1L_2L_3}(\hx_1,\hx_2,\hx_3)$. We start from the product rule \eqref{eq:yy3j} to get 
\begin{align}
    \Phi_{\{L_i\}}(\{\hx_i\}) \Phi_{\{L_i'\}}(\{\hx_i\}) &= \sum_{M_i,M_i',M_i'',L_i''} \threej{L_1}{L_2}{L_3}{M_1}{M_2}{M_3} \threej{L_1'}{L_2'}{L_3'}{M_1'}{M_2'}{M_3'}\times \\ &\times \prod^{3}_{i=1}\threej{L_i}{L_i'}{L_i''}{M_i}{M_i'}{M_i''}G_{L_i,L_i',L_i''}Y^*_{L_iM_i}(\hx_i)
\end{align}
Then, the $\sum_{M_i}\sum_{M_i'}$ of the product of 3j-symbols gives a 9j times a 3j, according to \eqref{eq:9j3j}, so that
 \begin{equation}
  \begin{aligned}
      \Phi_{\{L_i\}}(\{\hx_i\}) \Phi_{\{L_i'\}}(\{\hx_i\}) &= \sum_{M_i'',L_i''} \threej{L_1''}{L_2''}{L_3''}{M_1''}{M_2''}{M_3''} \,
  \ninej{L_1}{L_2}{L_3}{L_1'}{L_2'}{L_3'}{L_1''}{L_2''}{L_3''} \\
      &\quad \times \prod_{i=1}^{3} G_{L_i L_i' L_i''} \, Y^*_{L_i M_i}(\hx_i)
  \end{aligned}
  \label{eq:phi-product}
  \end{equation}
Substituting the complex-conjugate identity $Y^*_{LM} = (-1)^MY_{L-M}$ into the conjugated spherical harmonics in Eq.~\ref{eq:phi-product} introduces a phase $(-1)^{\sum M_i''}$. The $3j$ symbol appearing in~\ref{eq:phi-product} enforces $M_1'' + M_2'' + M_3'' = 0$, so this phase reduces to $1$. Separately, flipping the signs of the $M_i''$ in the $3j$ symbol introduces a phase $(-1)^{\sum_i L_i''}$. Comparing to the definition of $\Phi_{L_1L_2L_3}(\hx_1,\hx_2,\hx_3)$, we finally get:
\begin{align}
    \Phi_{\{L_i\}}(\{\hx_i\}) \Phi_{\{L_i'\}}(\{\hx_i\}) &= \sum_{L_i''} \ninej{L_1}{L_2}{L_3}{L_1'}{L_2'}{L_3'}{L_1''}{L_2''}{L_3''}(-1)^{\sum^3_{i=1} L''_i}\prod^{3}_{i=1}G_{L_iL_i'L_i''}\Phi_{L''_1L''_2L''_3}(\hx_1,\hx_2,\hx_3)
\end{align}

\medskip
{\bf\em Step 2: Recursion relation for $Q$.}
We specialize the product rule to $L_1'=L_2'=L_3'=1$.
Triangular conditions in the 9j symbol force $L_i''=\{L_i-1,L_i,L_i+1\}$. Additionally, $G_{L_i1L_i''}$ enforces $L_i \neq L_i''$, since $L_i+1+L_i''$ must be even. So we end up with at most 8 terms in the sum.
Consider the following function $\tilde Q$, defined analogously to $Q$ but with an extra factor $\Phi_{111}$:
\begin{equation}
\tilde Q_{L_1L_2L_3}(\s) = \int_{\x_1} 
  W_1(\x_1) W_2(\x_2) 
  F_1(x_1) F_2(x_2) \Phi_{L_1L_2L_3}(\hx_1,\hx_2,\hs) \Phi_{111}(\hx_1,\hx_2,\hs)\bigg|_{\x_2=\s+\x_1}
\end{equation}
Then $\tQ=0$ since $\Phi_{L_1L_2L_3}(\hx_1,\hx_2,\hs)$ vanishes when $L_1+L_2+L_3$ is odd for coplanar $(\hx_1,\hx_2,\hs)$.
Let's define $L_i\pm1\equiv L_i^{\pm}$ and for a triplet $\{L_1,L_2,L_3\}$
\begin{equation}
\label{eq:C_coef}
    \ninej{L_1}{L_2}{L_3}{1}{1}{1}{L^{\pm}_1}{L^{\pm}_2}{L^{\pm}_3}\prod^3_{i=1}G_{L_i1L^{\pm}_i}(-1)^{\sum^3_{i=1}{L^{\pm}_i}} \equiv \mathfrak{C}_{\{L^{\pm}_i\}}
\end{equation}
Then, we have 
\begin{equation}
\label{eq:Qipm}
    \sum_{\{L^{\pm}_i\}}\mathfrak{C}_{\{L^{\pm}_i\}}Q_{\{L^{\pm}_i\}} = 0\,,
\end{equation}
where the sum runs over all $2^3$ sign choices $\{L_i^\pm\}=\{L_1^\pm,L_2^\pm,L_3^\pm\}$.

$Q_{L_1L_2L_3}$ can be non-zero only if
$L_1+L_2+L_3$ is even, and $\{L_1,L_2,L_3\}$ satisfies triangle inequalities.
Moreover, in \eqref{eq:Qipm} the parity of the triplets in the sum is always flipped relative to $(L_1,L_2,L_3)$ since
\begin{equation}
(L_1^\pm+L_2^\pm+L_3^\pm)=(L_1+L_2+L_3)+(\pm1\pm1\pm1),
\end{equation}
and $(\pm1\pm1\pm1)$ is always odd. Hence \eqref{eq:Qipm} is only non-trivial when $L_1+L_2+L_3$ is odd.

\medskip
{\bf\em Step 3: reducing to extremal triples.}
We recall the relation \eqref{eq:Qipm} only consider relations the bases $(L_1,L_2,L_3)$ such that all eight neighbors $(L_1^\pm,L_2^\pm,L_3^\pm)$ are non-negative.
Define a triplet $(L_1,L_2,L_3)$ to be \emph{extremal} if it saturates the triangle inequality, i.e.
\begin{equation}
\label{eq:extremal_def}
    L_1+L_2+L_3=2\max(L_1,L_2,L_3).
\end{equation}
Then:
\begin{enumerate}
    \item[(i)] For any admissible non-extremal $(L_1,L_2,L_3)$ with $L_i\ge 2$ and $L_1+L_2+L_3$ even, $Q_{L_1L_2L_3}$ can be expressed as a linear combination of $Q_{abc}$ with strictly smaller $a+b+c$.
    \item[(ii)] Extremal $Q_{abc}$ cannot be reduced to the terms with strictly smaller $a+b+c$
\end{enumerate}

\begin{proof}

\textbf{(i)} Fix an admissible triplet $(L_1,L_2,L_3)$ with $L_i\ge 2$ and even total sum $S$. Consider the instance of \eqref{eq:Qipm} with base
\begin{equation}
\label{eq:base_downward}
    (\widetilde L_1,\widetilde L_2,\widetilde L_3)=(L_1-1,\ L_2-1,\ L_3-1).
\end{equation}
The eight neighbors are
\begin{equation}
\label{eq:neighbors_downward}
    (\widetilde L_1^\pm,\widetilde L_2^\pm,\widetilde L_3^\pm)\in\{L_1-2,L_1\}\times\{L_2-2,L_2\}\times\{L_3-2,L_3\}.
\end{equation}
In particular, $(L_1,L_2,L_3)$ appears as the $(+,+,+)$ neighbor. Every other neighbor differs from $(L_1,L_2,L_3)$ by lowering at least one index by $2$, hence has strictly smaller sum:
\begin{equation}
    (a,b,c)\neq (L_1,L_2,L_3)\quad\Longrightarrow\quad a+b+c \le S-2.
\end{equation}
We may solve for $Q_{L_1L_2L_3}$:
\begin{equation}
\label{eq:downward_solution}
    Q_{L_1L_2L_3}
    =
    -\frac{1}{\mathfrak{C}_{\{\tilde{L}^+_1,\tilde{L}^+_2,\tilde{L}^+_3\}}}
    \sum_{\substack{\{\tilde{L}_i^\pm\}\in \text{allowed}\\ \{\tilde{L}_i^\pm\}\neq (+++)}}
    \mathfrak{C}_{\{\tilde{L}_i^\pm\}}\,
    Q_{\{\tilde{L}_i^\pm\}},
\end{equation}
where the sum runs over the surviving neighbors in \eqref{eq:neighbors_downward} after pruning, excluding the $(+,+,+)$ term. Every term on the right-hand side has strictly smaller total sum, $a+b+c\le S-2$. We note that for this to be true, we need $\mathfrak{C}_{\{\tilde{L}^+_1,\tilde{L}^+_2,\tilde{L}^+_3\}}\neq 0$. That's always true if the corresponding 9j symbol is non-zero. From Eq.~\eqref{eq:special_9j_2} we can see that it's always the case for a non-extremal triplet. This proves (i).

\medskip
\textbf{(ii)}
Assume that, without loss of generality, for a given $Q_{abc}$, $a+b=c$. Then the coefficient in the denominator of \eqref{eq:downward_solution} is $\mathfrak{C}_{\{\tilde{a}^+_1,\tilde{b}^+_2,\tilde{c}^+_3\}} = 0$, because the triangular inequality is violated: 
\begin{equation}
    \tilde{a}+\tilde{b} = a-1+b-1=c-2<\tilde{c}
\end{equation}
\end{proof}
\par\noindent
Summarizing, in this appendix we have derived a recursion relation (\ref{eq:Qipm}) for the $Q$-function $Q_{L_1L_2L_3}({\bf s})$.
Using this recursion relation, any $Q$-function can be written as a linear combination of extremal (in the sense defined by Eq.\ (\ref{eq:extremal_def})) $Q$-functions.

\subsection{Full-sky $Q$}
\label{app:full_sky_q}

Here we derive Eq.~\eqref{eq:full_sky_q} for $Q_{L_1L_2L_3}(s)$, when the weight function $W_i(\x)=R_i(x)$ is purely radial.
We start with the Fourier-space definition given by Eqs.~\eqref{eq:tV_def2}--\eqref{eq:QLLL_def2}. Plugging $W_i(\x)=R_i(x)$ in to Eq.~\eqref{eq:tV_def2} and using Eq.~\eqref{eq:rayleigh_integral} to perform the angular integral over $\hx$ gives
\begin{align}
\tV_i^{LM}(\k) 
&= \int x^2\,dx\, F_i(x)R_i(x)\int d\Omega_x\,e^{-i\k\cdot\x}Y_{LM}(\hx)
\\
&=
4\pi\,i^{-L} \,Y_{LM}(\hk) \underbrace{\int x^2\,dx\,F_i(x)R_i(x) j_L(kx)}_{\equiv\,\,r^{(i)}_L(k)}.
\end{align}
We substitute this result into Eq.~\eqref{eq:tQ_def2} to find
\begin{align}
    \tQ^{L_1L_2}_{L_3M_3}(\k)
    &=
    (4\pi)^2
    i^{L_1-L_2}\,
    r^{(1)}_{L_1}(k)\,
    r^{(2)}_{L_2}(k)\,
    \sum_{M_1M_2} \threej{L_1}{L_2}{L_3}{M_1}{M_2}{M_3}
    Y_{L_1M_1}(\hk)
    Y_{L_2M_2}(\hk)
    \\
    &=
    (4\pi)^2
    i^{L_1-L_2}\,
    r^{(1)}_{L_1}(k)\,
    r^{(2)}_{L_2}(k)\,
    \frac{G_{L_1L_2L_3}}{2L_3+1}\,
    Y^*_{L_3M_3}(\hk),
\end{align}
where we used $Y_{L_1 M_1}(-\hk)=i^{2L_1}Y_{L_1 M_1}(\hk)$ in the first line and Eq.~\eqref{eq:yy3j} in the second. To obtain Eq.~\eqref{eq:QLLM_def2} we Fourier transform this expression. We perform the angular integral over $\hk$ using Eq.~\eqref{eq:rayleigh_integral}, yielding
\begin{equation}
    Q^{L_1L_2}_{L_3M_3}(\s)
    =
    \frac{(4\pi)^3}{(2\pi)^3}
    i^{L_1+L_3-L_2}
    \frac{G_{L_1L_2L_3}}{2L_3+1}\,
    Y^*_{L_3M_3}(\hs)
    \int k^2\,dk\,
    r^{(1)}_{L_1}(k)\,
    r^{(2)}_{L_2}(k)\,
    j_{L_3}(ks).
\end{equation}
We now multiply by $Y_{L_3M_3}(\hs)$ and sum over $M_3$ to find
\begin{equation}
    \sum_{M_3}
    Q^{L_1L_2}_{L_3M_3}(\s)
    Y_{L_3M_3}(\hs)
    =
    \frac{(4\pi)^2}{(2\pi)^3}
    i^{L_1+L_3-L_2}
    G_{L_1L_2L_3}\,
    \int k^2\,dk\,
    r^{(1)}_{L_1}(k)\,
    r^{(2)}_{L_2}(k)\,
    j_{L_3}(ks)
\end{equation}
where we used $\sum_m Y_{\ell m}(\hx)Y^*_{\ell m}(\hx) = (2\ell+1)/4\pi$. Upon multiplying by the prefactor in Eq.~\eqref{eq:QLLL_def2}, re-expressing $G_{L_1L_2L_3}$ in terms of $C_{L_1L_2L_3}$ (using Eq.~\ref{eq:Clll_def}) and performing the (trivial) integral over $\hs$ we find
\begin{equation}
\begin{aligned}
    Q_{L_1L_2L_3}(s)
    &=
    32\pi\,
    i^{L_1+L_3-L_2}
    \left[\prod_{i=1}^3(2L_i+1)\right]
    C_{L_1L_2L_3}
    \int k^2dk\,
    j_{L_3}(ks)
    r^{(1)}_{L_1}(k)
    r^{(2)}_{L_2}(k)
\end{aligned}
\end{equation}
This recovers Eq.~\eqref{eq:full_sky_q}.

\subsection{Separable $Q$}
\label{app:separable_q}

Here we derive Eq.~\eqref{eq:separable_q} for $Q_{L_1L_2L_3}(s)$, when the weight function $W_i(\x)=R_i(x) A_i(\hx)$ factorizes into radial and angular pieces.
We start with the Fourier-space definition given by Eqs.~\eqref{eq:tV_def2}--\eqref{eq:QLLL_def2}. Plugging $W_i(\x)=R_i(x) A_i(\hx)$ in to Eq.~\eqref{eq:tV_def2} and using Eq.~\eqref{eq:rayleigh_expansion} gives
\begin{align}
    \tV_i^{LM}(\k)
    &=
    \int x^2\,dx\, F_i(x)R_i(x)\int d\Omega_x\,e^{-i\k\cdot\x}
    A_i(\hx)\,
    Y_{LM}(\hx)
    \\
    &= 4\pi 
    \sum_{\ell m}
    i^{-\ell}\,
    r^{(i)}_\ell(k)\,
    Y_{\ell m}(\hk)
    \int d\Omega_x\,
    A_i(\hx)\,
    Y^*_{\ell m}(\hx)\,
    Y_{LM}(\hx),
\end{align}
where $r^{(i)}_\ell(k)$ is defined as in \S\ref{app:full_sky_q}.
Next we expand $A_i(\hx) = \sum_{\ell m}A^{(i)}_{\ell m}\,Y_{\ell m}(\hx)$ and perform the angular integral over $\hx$ using Eq.~\eqref{eq:yyy}
\begin{equation}
    \tV_i^{LM}(\k) 
    = 4\pi
    \sum_{\ell m}
    i^{-\ell}\,
    (-1)^m\,
    r^{(i)}_\ell(k)\,
    Y_{\ell m}(\hk)
    \sum_{\ell'm'}
    \threej{L}{\ell}{\ell'}{M}{-m}{m'}
    G_{L\ell\ell'}
    A^{(i)}_{\ell'm'}.
\end{equation}
We substitute this result into Eq.~\eqref{eq:tQ_def2} to find
\begin{equation}
\begin{aligned}
    \tQ^{L_1L_2}_{L_3M_3}(\k)
    &=
    (4\pi)^2
    \sum_{\substack{\ell_1 m_1\\\ell_2 m_2}}
    i^{\ell_1-\ell_2}
    (-1)^{m_1+m_2}
    r^{(1)}_{\ell_1}(k)
    r^{(2)}_{\ell_2}(k)
    Y_{\ell_1m_1}(\hk)
    Y_{\ell_2m_2}(\hk)
    \sum_{\substack{\ell'_1m'_1\\\ell'_2m'_2}}
    G_{L_1\ell_1\ell'_1}
    G_{L_2\ell_2\ell'_2}
    \\
    &\hspace{1cm}\times
    A^{(1)}_{\ell'_1m'_1}
    A^{(2)}_{\ell'_2m'_2}
    \sum_{M_1M_2}
    \threej{L_1}{L_2}{L_3}{M_1}{M_2}{M_3}
    \threej{L_1}{\ell_1}{\ell'_1}{M_1}{-m_1}{m'_1}
    \threej{L_2}{\ell_2}{\ell'_2}{M_2}{-m_2}{m'_2}.
\end{aligned}
\end{equation}
Next we Fourier transform this expression to obtain Eq.~\eqref{eq:QLLM_def2}, using Eq.~\eqref{eq:rayleigh_expansion} to expand the exponential and separate the radial-- and angular--components of the integral
\begin{equation}
    Q^{L_1L_2}_{L_3M_3}(\s)
    =
    \sum_{\ell m}
    i^\ell 
    Y^*_{\ell m}(\hs)
    \int \frac{k^2\,dk}{2\pi^2}\,j_\ell(ks)
    \int d\Omega_k\,
    Y_{\ell m}(\hk)\,
    \tQ^{L_1L_2}_{L_3M_3}(\k),
\end{equation}
which upon substituting our result for $\tQ$ and performing the angular integral over $\hk$ using Eq.~\eqref{eq:yyy} gives
\begin{equation}
\begin{aligned}
    Q^{L_1L_2}_{L_3M_3}(\s)
    &=
    8
    \sum_{\substack{\ell m\\\ell_1m_1\\\ell_2m_2}}
    i^{\ell+\ell_1-\ell_2}
    (-1)^{m_1+m_2}
    G_{\ell\ell_1\ell_2}
    Y^*_{\ell m}(\hs)
    \int k^2\,dk\,j_\ell(ks)\,r^{(1)}_{\ell_1}(k)\,r^{(2)}_{\ell_2}(k)
    \\&\hspace{1cm}\times
    \sum_{\substack{\ell'_1m'_1\\\ell'_2m'_2}}
    G_{L_1\ell_1\ell'_1}
    G_{L_2\ell_2\ell'_2}
    A^{(1)}_{\ell'_1m'_1}
    A^{(2)}_{\ell'_2m'_2}
    \\&\hspace{1cm}\times
    \sum_{M_1M_2}
    \threej{\ell}{\ell_1}{\ell_2}{m}{m_1}{m_2}
    \threej{L_1}{L_2}{L_3}{M_1}{M_2}{M_3}
    \threej{L_1}{\ell_1}{\ell'_1}{M_1}{-m_1}{m'_1}
    \threej{L_2}{\ell_2}{\ell'_2}{M_2}{-m_2}{m'_2}.
\end{aligned}
\end{equation}
Next we multiply by $Y_{L_3M_3}(\s)$, perform the angular integral over $\hs$ to enforce $(\ell,m)=(L_3,M_3)$, and sum over $M_3$ to obtain
\begin{equation}
\begin{aligned}
    X
    &\equiv
    \sum_{M_3}
    \int d\Omega_s 
    Y_{L_3 M_3}(\hs)
    Q^{L_1L_2}_{L_3M_3}(\s)
    \\
    &=
    8
    \sum_{\ell_1\ell_2}
    i^{L_3+\ell_1-\ell_2}
    G_{L_3\ell_1\ell_2}
    \int k^2\,dk\,j_{L_3}(ks)\,r^{(1)}_{\ell_1}(k)\,r^{(2)}_{\ell_2}(k)
    \sum_{\substack{\ell'_1m'_1\\\ell'_2m'_2}}
    G_{L_1\ell_1\ell'_1}
    G_{L_2\ell_2\ell'_2}
    A^{(1)}_{\ell'_1m'_1}
    A^{(2)}_{\ell'_2m'_2}
    \\
    &\hspace{1cm}\times
    \sum_{\substack{m_1m_2\\M_1M_2}}
    \sum_{M_3}(-1)^{M_3}
    \threej{\ell_1}{\ell_2}{L_3}{m_1}{m_2}{M_3}
    \threej{L_1}{L_2}{L_3}{M_1}{M_2}{M_3}
    \threej{L_1}{\ell_1}{\ell'_1}{M_1}{-m_1}{m'_1}
    \threej{L_2}{\ell_2}{\ell'_2}{M_2}{-m_2}{m'_2},
\end{aligned}
\end{equation}
where we used $m_1+m_2=-M_3$ to rewrite the phase in terms of $M_3$, and that $\ell_1+\ell_2+L_3$ is even (due to $G_{\ell_1\ell_2L_3}$) to rearrange the columns of the first 3j symbol.

Next we evaluate the sums over $m_i$ and $M_i$ in the last line. We first perform the sum over $M_3$ using the 6j recoupling identity (Eq.~\ref{eq:6j_recoupling}). The remaining sums collapse due to the orthogonality of the 3j symbols (Eq.~\ref{eq:3j_orthogonality}), giving
\begin{equation}
\begin{aligned}
    &\sum_{\substack{m_1m_2\\M_1M_2}}
    \sum_{M_3}(-1)^{M_3}
    \threej{\ell_1}{\ell_2}{L_3}{m_1}{m_2}{M_3}
    \threej{L_1}{L_2}{L_3}{M_1}{M_2}{M_3}
    \threej{L_1}{\ell_1}{\ell'_1}{M_1}{-m_1}{m'_1}
    \threej{L_2}{\ell_2}{\ell'_2}{M_2}{-m_2}{m'_2}
    \\&\hspace{2cm}=
    \sum_{\ell'm'}
    (-1)^{L_3+\ell'+m'}
    (2\ell'+1)
    \sixj{L_3}{L_1}{L_2}{\ell'}{\ell_2}{\ell_1}
    \sum_{m_1M_1}
    \threej{\ell'}{L_1}{\ell_1}{m'}{M_1}{-m_1}
    \threej{L_1}{\ell_1}{\ell'_1}{M_1}{-m_1}{m'_1}
    \\&\hspace{6cm}\times
    \sum_{m_2M_2}
    \threej{\ell'}{\ell_2}{L_2}{-m'}{-m_2}{M_2}
    \threej{L_2}{\ell_2}{\ell'_2}{M_2}{-m_2}{m'_2}
    \\&\hspace{2cm}=
    \sum_{\ell'm'}
    (-1)^{L_3+\ell'+m'}
    (2\ell'+1)
    \sixj{L_3}{L_1}{L_2}{\ell'}{\ell_2}{\ell_1}
    \frac{T_{L_1\ell_1\ell_1'}}{2\ell'_1+1}\delta_{\ell',\ell'_1}\delta_{m',m'_1}
    \\&\hspace{6cm}\times
    (-1)^{L_2+\ell_2+\ell'_2}
    \frac{T_{L_2\ell_2\ell'_2}}{2\ell'_2+1}\delta_{\ell',\ell'_2}\delta_{-m',m'_2}
    \\&\hspace{2cm}=
    (-1)^{\ell_2+L_2+L_3+m'_1}
    \frac{\delta_{\ell'_1,\ell'_2}\delta_{m'_1,-m'_2}}{2\ell'_1+1}
    T_{L_1\ell_1\ell'_1}
    T_{L_2\ell_2\ell'_2}
    \sixj{L_3}{\ell_1}{\ell_2}{\ell'_1}{L_2}{L_1},
\end{aligned}
\end{equation}
where in the last line we used the invariance of the 6j symbol under permutations of columns and upper and lower arguments. Substituting this expression to $X$ gives
\begin{equation}
\begin{aligned}
    X
    &=
    8
    \sum_{\ell'\ell_1\ell_2}
    i^{L_3+\ell_1-\ell_2}
    \bigg[
    \sum_{m'}
    \frac{(-1)^{m'}}{2\ell'+1}
    A^{(1)}_{\ell'm'} A^{(2)}_{\ell'-m'}
    \bigg]
    G_{L_3\ell_1\ell_2}
    \int k^2\,dk\,j_{L_3}(ks)
    r^{(1)}_{\ell_1}(k)
    r^{(2)}_{\ell_2}(k)
    \\
    &\hspace{1cm}\times
    (-1)^{\ell_2+L_2+L_3}
    \sixj{L_3}{\ell_1}{\ell_2}{\ell'}{L_2}{L_1}
    G_{L_1\ell_1\ell'}
    G_{L_2\ell_2\ell'},
    \label{eq:x_simplified}
\end{aligned}
\end{equation}
which upon substitution into the definition of $Q_{L_1L_2L_3}$ gives
\begin{equation}
\begin{aligned}
    Q_{L_1L_2L_3}(s)
    &=
    (4\pi)^{3/2}
    \sqrt{(2L_1+1)(2L_2+1)(2L_3+1)}
    \,X
    \\
    &\hspace{-1cm}=
    \sum_{\ell'\ell_1\ell_2}
    C^{A_1A_2}_{\ell'}
    \Bigg[
    (-1)^{\ell_2+L_2+L_3}
    \sixj{L_3}{\ell_1}{\ell_2}{\ell'}{L_2}{L_1}
    G_{L_1\ell_1\ell'}
    G_{L_2\ell_2\ell'}
    \sqrt{\frac{(2L_1+1)(2L_2+1)}{(2\ell_1+1)(2\ell_2+1)}
    }
    \Bigg]
    \\
    &\hspace{-0.5cm}\times
    32\pi i^{\ell_1+L_3-\ell_2}
    (2\ell_1+1)(2\ell_2+1)(2L_3+1)
    C_{L_3\ell_1\ell_2}
    \int k^2\,dk\,j_{L_3}(ks)
    r^{(1)}_{\ell_1}(k)
    r^{(2)}_{\ell_2}(k),
    \label{eq:q_separable_derived}
\end{aligned}
\end{equation}
where we substituted the definition of the mask power spectrum (Eq.~\ref{eq:mask_spectrum}) into the bracketed term in Eq.~\eqref{eq:x_simplified} and rearranged factors of $2\ell_i+1$ to make the bottom line equivalent to the full-sky expression for $Q_{\ell_1\ell_2L_3}$. This result is equivalent to Eq.~\eqref{eq:separable_q}.

To recover the full-sky expression, we set $C^{A_1A_2}_\ell = 4\pi \delta_{\ell, 0}$ and use the identities
\begin{equation}
\begin{aligned}
    \sixj{L_3}{\ell_1}{\ell_2}{0}{L_2}{L_1}
    &=
    \frac{(-1)^{L_1+L_2+L_3}\delta_{\ell_1,L_1}\delta_{\ell_2,L_2}}{\sqrt{(2L_1+1)(2L_2+1)}}
    T_{L_1L_2L_3}
    \\
    G_{L_i\ell_i0}
    &=
    (-1)^{L_i}\delta_{\ell_i,L_i}\sqrt{\frac{2L_i+1}{4\pi}}.
\end{aligned}
\end{equation}
Substituting this into Eq.~\eqref{eq:q_separable_derived} reduces the bracketed term to $\delta_{\ell_1L_1}\delta_{\ell_2L_2}/4\pi$, recovering the full-sky expression.

\subsection{Window functions for observer terms}
\label{app:finite_window_observer_terms}

In \S\ref{sec:galaxy_number_counts} we decomposed the galaxy field into local, observer, and integrated terms. Our finite representation (Eq.~\ref{eq:finite_window}) for the window function applies to local-local correlations. In this appendix, we derive analogous finite representations for local-observer and observer-observer correlations (Eqs.~\eqref{eq:finite_window_loc_obs},~\eqref{eq:finite_window_obs_obs} below).

To set up the calculation formally, our goal is to calculate the window function, defined as usual by Eq.~\eqref{eq:window}.
In the main paper, we assumed that the 2PCF is given by the ``rank-one'' form in Eq.~\eqref{eq:rk1_sig}.
Here, we assume that the 2PCF $\langle \delta(\x_1) \delta(\x_2) \rangle$ is given by either:
\begin{align}
    {\rm local}\,\,\times\,\,{\rm observer}
    &=
    \int_\k e^{i\k\cdot\x_1}
    F_1(x_1)F_2(x_2)P(k)\L_{\ell_1^S}(\hk\cdot\hx_1)\L_{\ell_2^S}(\hk\cdot\hx_2)
    \label{eq:rk1_sig_loc_obs}
    \\
    {\rm observer}\,\,\times\,\,{\rm observer}
    &=
    \int_\k
    F_1(x_1)F_2(x_2)P(k)\L_{\ell_1^S}(\hk\cdot\hx_1)\L_{\ell_2^S}(\hk\cdot\hx_2).
    \label{eq:rk1_sig_obs_obs}
\end{align} 
Below we sketch derivations for the window functions appropriate for these terms, closely following \S\ref{sec:window_function}.

\textbf{Local $\times$ observer}: The window function appropriate for a local $\times$ observer rank-1 signal (Eq.~\ref{eq:rk1_sig_loc_obs}) is identical to Eq.~\eqref{eq:finite_window_intermediate} but with 
\begin{equation}
    e^{-i(\k-\k')\cdot(\x_1-\x_2)}
    \to
    e^{-i\k\cdot(\x_1-\x_2)} e^{i\k'\cdot\x_1}
\end{equation}
in the last line. Performing the angular integrals in the last line using Eq.~\eqref{eq:rayleigh_integral} yields 
\begin{equation}
    i^{\ell_3^E+\ell_3^S}
    j_{\ell_3^E}(ks)
    j_{\ell_3^S}(k'x_1)
    Y^*_{\ell^E_3m^E_3}(\hs)
    Y^*_{\ell^S_3m^S_3}(\hx_1),
\end{equation}
rather than Eq.~\eqref{eq:last_line}.
The sum over $(m_3^E,m_3^S)$ is no longer trivial, since the spherical harmonics corresponding to these $m$'s don't point in the same direction anymore. However, note that both $Y_{L_1M_1}$ and $Y^*_{\ell^S_3m^S_3}$ point in the $\x_1$ direction. Using Eq.~\eqref{eq:ylm_ylm} we can then write
\begin{align}
W_{\ell_1^E\ell_2^E}^{\ell_1^S\ell_2^S}(k,k')
 &= 
\sum_{\substack{L\ell^E_3\ell^S_3\\L_1L_2L_3 }}
(4\pi)^2 i^{\ell_3^E+\ell_3^S} (2L_3+1) 
\sqrt{(2\ell_3^E+1)(2\ell_3^S+1)(2L_1+1)(2L_2+1)}
\nn \\
& \hspace{0.5cm} \times 
C_{\ell_1^E \ell_2^E \ell_3^E} 
C_{\ell_1^S \ell_2^S \ell_3^S} 
C_{\ell_1^E \ell_1^S L_1}
C_{\ell_2^E \ell_2^S L_2}
G_{\ell_3^S L_1 L}
\ninej{\ell^E_1}{\ell^E_2}{\ell^E_3}{\ell^S_1}{\ell^S_2}{\ell^S_3}{L_1}{L_2}{L_3}\nn\\
& \hspace{0.5cm} \times
\int s^2 \,ds\, j_{\ell_3^E}(ks)
\int d\Omega_s
\int_{\x_1} 
j_{\ell_3^S}(k'x_1)\,
W_1(\x_1)  \, W_2(\x_2) \,
F_1(x_1)\, F_2(x_2)
   \nn \\
&  \hspace{0.5cm} \times
\sum_{\substack{M M_2 m_3^E}}
(-1)^{m_3^E}
    Y_{LM}(\hx_1) \, Y_{L_2M_2}(\hx_2) \,
    Y_{\ell^E_3m^E_3}(\hs)
    \Big|_{\x_2=\x_1+\s}
    \nn\\
&\hspace{0.5cm}\times
\sum_{\substack{M_1 M_3 m_3^S}}
(-1)^{M_1}
\threej{\ell^E_3}{\ell^S_3}{L_3}{-m^E_3}{-m^S_3}{M_3}
\threej{L_1}{L_2}{L_3}{M_1}{M_2}{M_3}
\threej{\ell_3^S}{L_1}{L}{m_3^S}{M_1}{-M}
\end{align}
Using Eq.\ (\ref{eq:6j_recoupling2}), the last line reduces to
\begin{equation}
(-1)^{m_3^E+L+L_1+L_2}
\sixj{L}{L_2}{\ell_3^E}{L_3}{\ell_3^S}{L_1}
\threej{L}{L_2}{\ell_3^E}{M}{M_2}{m_3^E},
\end{equation}
from which we obtain
\begin{align}
W_{\ell_1^E\ell_2^E}^{\ell_1^S\ell_2^S}(k,k')
 &= 
\sum_{\substack{L\ell^E_3\ell^S_3\\L_1L_2L_3 }}
(4\pi)^2 i^{\ell_3^E-\ell_3^S} (2L_3+1) 
\sqrt{(2\ell_3^E+1)(2\ell_3^S+1)(2L_1+1)(2L_2+1)}
\nn \\
& \hspace{0.5cm} \times 
(-1)^{L_2}
C_{\ell_1^E \ell_2^E \ell_3^E} 
C_{\ell_1^S \ell_2^S \ell_3^S} 
C_{\ell_1^E \ell_1^S L_1}
C_{\ell_2^E \ell_2^S L_2}
G_{\ell_3^S L_1 L}
\sixj{L}{L_2}{\ell_3^E}{L_3}{\ell_3^S}{L_1}
\ninej{\ell^E_1}{\ell^E_2}{\ell^E_3}{\ell^S_1}{\ell^S_2}{\ell^S_3}{L_1}{L_2}{L_3}\nn\\
& \hspace{0.5cm} \times
\int s^2 \,ds\, j_{\ell_3^E}(ks)
\int d\Omega_s
\int_{\x_1} 
j_{\ell_3^S}(k'x_1)\,
W_1(\x_1)  \, W_2(\x_2) \nn\\
&\hspace{4cm}\times
F_1(x_1)\, F_2(x_2)
    \Phi_{LL_2\ell_3^E}(\hx_1,\hx_2,\hs)
    \Big|_{\x_2=\x_1+\s},
\label{eq:finite_window_loc_obs}
\end{align}
where the tri-polar spherical harmonic $\Phi_{L_1L_2L_3}(\hx_1,\hx_2,\hx_3)$ is given by Eq.~\eqref{eq:Phi_def}.

\textbf{Observer $\times$ observer}: The window function appropriate for an observer $\times$ observer rank-1 signal (Eq.~\ref{eq:rk1_sig_obs_obs}) is identical to Eq.~\eqref{eq:finite_window_intermediate} but with 
\begin{equation}
    e^{-i(\k-\k')\cdot(\x_1-\x_2)}
    \to
    e^{-i\k\cdot(\x_1-\x_2)} 
\end{equation}
in the last line. Performing the angular integrals in the last line using Eq.~\eqref{eq:rayleigh_integral} yields 
\begin{equation}
    i^{\ell_3^E}
    j_{\ell_3^E}(ks)
    Y^*_{\ell^E_3m^E_3}(\hs)
    \frac{\delta_{\ell_3^S,0}\delta_{m_3^S,0}}{
    \sqrt{4\pi}
    },
\end{equation}
rather than Eq.~\eqref{eq:last_line}.
Plugging in this result gives
\begin{align}
W_{\ell_1^E\ell_2^E}^{\ell_1^S\ell_2^S}(k,k')
 &= 
\sum_{\substack{\ell^E_3L_1L_2L_3 }}
(4\pi)^{3/2} (2L_3+1) 
\sqrt{(2\ell_3^E+1)(2L_1+1)(2L_2+1)}
\nn \\
& \hspace{0.5cm} \times 
C_{\ell_1^E \ell_2^E \ell_3^E} 
C_{\ell_1^S \ell_2^S 0} 
C_{\ell_1^E \ell_1^S L_1}
C_{\ell_2^E \ell_2^S L_2}
\ninej{\ell^E_1}{\ell^E_2}{\ell^E_3}{\ell^S_1}{\ell^S_2}{0}{L_1}{L_2}{L_3}\nn\\
&\hspace{0.5cm} \times
\sum_{\substack{m^E_3M_1 M_2 M_3}}
\threej{\ell^E_3}{0}{L_3}{m^E_3}{0}{M_3}
\threej{L_1}{L_2}{L_3}{M_1}{M_2}{M_3}
\nn \\
& \hspace{0.5cm} \times
\int_{\x_1\x_2} 
W_1(\x_1)  \, W_2(\x_2) \,
F_1(x_1)\, F_2(x_2)
Y_{L_1M_1}(\hx_1) \, Y_{L_2M_2}(\hx_2) \,
   \nn \\
&  \hspace{0.5cm} \times
i^{\ell_3^E}
    j_{\ell_3^E}(ks)
    Y^*_{\ell^E_3m^E_3}(\hs)
\end{align}
which can be dramatically simplified using Eqs. \eqref{eq:special_3j} and \eqref{eq:special_9j}:
\begin{align}
W_{\ell_1^E\ell_2^E}^{\ell_1^S\ell_2^S}(k,k')
 &= 
 \delta_{\ell_1^S\ell_2^S}
 \frac{(-1)^{\ell_1^E}}{2\ell_1^S+1}
\sum_{\substack{L_1L_2L_3 }}
(-1)^{L_1}
i^{3L_3}
C_{\ell_1^E \ell_2^E L_3} 
C_{\ell_1^E \ell_1^S L_1}
C_{\ell_2^E \ell_1^S L_2}
\sixj{\ell_1^E}{\ell_2^E}{L_3}{L_2}{L_1}{\ell_1^S}
\nn \\
& \hspace{4.5cm} \times
\int s^2\,ds\,j_{L_3}(ks)
Q_{L_1L_2L_3}(s).
\label{eq:finite_window_obs_obs}
\end{align}
This result can equivalently be written as 
\begin{align}
W_{\ell^E_1\ell^E_2}^{\ell_1^S\ell_2^S}(k,k') &= 
\frac{\delta_{\ell_S^1\ell_S^2}}{2\ell_1^S+1}
\sum_{L_1L_2L_3}
(4\pi)^{3/2} \sqrt{\prod(2L_i+1)} \,
C_{\ell_1^E\ell_1^SL_1} C_{\ell_2^E\ell_2^SL_2} C_{\ell_1^E\ell_2^EL_3} 
\sixj{L_1}{L_2}{L_3}{\ell_2^E}{\ell_1^E}{\ell_1^S}
 \nn \\
& \hspace{1cm} \times \int \frac{d\Omega_k}{4\pi} \sum_{M_1M_2M_3}
  \threej{L_1}{L_2}{L_3}{M_1}{M_2}{M_3} \tV_1^{L_1M_1}(\k) \tV_2^{L_2M_2}(-\k) Y_{L_3M_3}(\hk),
  \label{eq:obs_obs2}
\end{align}
where $\tV_i^{LM}(\k)$ is defined by Eq.~\eqref{eq:tV_def2}.

%% file: AppTables.tex
\section{Window functions for the RSD quadrupole and hexadecapole}
\label{app:window_ell_2_4}

In \S\ref{sec:rsd} we specialized to the case of linear theory redshift-space distortions and explicitly wrote down the window function for the Yamamoto estimator monopole (Eq.~\ref{eq:finite_monopole_window}) using our general formula Eq.~\eqref{eq:finite_window}. Adopting the shorthand notation $Q_{L_1L_2L_3}^{\ell_1\ell_2}(s)$ for $Q_{L_1L_2L_3}(s;F^{\rm RSD}_{\ell_1}, F^{\rm RSD}_{\ell_2})$, the analogous expressions for the quadrupole and hexadecapole are:
\allowdisplaybreaks
\begin{align}
  W^{\rm RSD}_{\ell=2}(k,k') = \int ds\, s^2\, j_{2}(ks)\,\Bigg[ & -\frac{\sqrt{5}}{25}\,j_{0}(k's)\,Q^{00}_{022}(s) + \frac{1}{5}\,j_{2}(k's)\,Q^{02}_{000}(s) + \frac{2\sqrt{5}}{175}\,j_{2}(k's)\,Q^{02}_{022}(s) \nn \\
    & \hspace{-3cm} + \frac{2}{105}\,j_{2}(k's)\,Q^{02}_{044}(s) + \frac{\sqrt{5}}{125}\,j_{2}(k's)\,Q^{20}_{220}(s) - \frac{\sqrt{70}}{875}\,j_{2}(k's)\,Q^{20}_{222}(s) \nn \\
    & \hspace{-3cm} + \frac{\sqrt{70}}{875}\,j_{2}(k's)\,Q^{20}_{224}(s) + \frac{2\sqrt{5}}{875}\,j_{2}(k's)\,Q^{22}_{220}(s) - \frac{2\sqrt{715}}{25025}\,j_{4}(k's)\,Q^{22}_{246}(s)\nn \\
    & \hspace{-3cm} + \left(- \frac{\sqrt{5}}{625}\,j_{0}(k's) + \frac{2\sqrt{5}}{875}\,j_{2}(k's) - \frac{18\sqrt{5}}{4375}\,j_{4}(k's)\right)Q^{22}_{202}(s) \nn \\
    & \hspace{-3cm} + \left(\frac{\sqrt{70}}{4375}\,j_{0}(k's) + \frac{3\sqrt{70}}{42875}\,j_{2}(k's) + \frac{36\sqrt{70}}{214375}\,j_{4}(k's)\right)Q^{22}_{222}(s) \nn \\
    & \hspace{-3cm} + \left(\frac{4\sqrt{70}}{42875}\,j_{2}(k's) - \frac{2\sqrt{70}}{8575}\,j_{4}(k's)\right)Q^{22}_{224}(s) \nn \\
    & \hspace{-3cm} + \left(- \frac{\sqrt{70}}{4375}\,j_{0}(k's) + \frac{4\sqrt{70}}{42875}\,j_{2}(k's) - \frac{\sqrt{70}}{214375}\,j_{4}(k's)\right)Q^{22}_{242}(s) \nn \\
    & \hspace{-3cm} + \left(- \frac{2\sqrt{385}}{25725}\,j_{2}(k's) + \frac{2\sqrt{385}}{94325}\,j_{4}(k's)\right)Q^{22}_{244}(s)\Bigg]
\end{align}

\allowdisplaybreaks
\begin{align}
  W^{\rm RSD}_{\ell=4}(k,k') = \int ds\, s^2\, j_{4}(ks)\,\Bigg[ & \frac{1}{27}\,j_{0}(k's)\,Q^{00}_{044}(s) - \frac{2\sqrt{5}}{175}\,j_{2}(k's)\,Q^{02}_{022}(s) - \frac{20}{2079}\,j_{2}(k's)\,Q^{02}_{044}(s) \nn \\
    & \hspace{-3cm} - \frac{5\sqrt{13}}{1859}\,j_{2}(k's)\,Q^{02}_{066}(s) - \frac{\sqrt{70}}{1575}\,j_{2}(k's)\,Q^{20}_{242}(s) + \frac{2\sqrt{385}}{10395}\,j_{2}(k's)\,Q^{20}_{244}(s) \nn \\
    & \hspace{-3cm} - \frac{\sqrt{715}}{6435}\,j_{2}(k's)\,Q^{20}_{246}(s) + \frac{2\sqrt{5}}{875}\,j_{4}(k's)\,Q^{22}_{220}(s) + \frac{2\sqrt{7735}}{158015}\,j_{4}(k's)\,Q^{22}_{268}(s) \nn \\
    & \hspace{-3cm} + \left(\frac{4\sqrt{70}}{42875}\,j_{2}(k's) - \frac{2\sqrt{70}}{8575}\,j_{4}(k's)\right)Q^{22}_{222}(s) \nn \\
    & \hspace{-3cm} + \left(\frac{\sqrt{70}}{7875}\,j_{0}(k's) - \frac{2\sqrt{70}}{15435}\,j_{2}(k's) + \frac{3\sqrt{70}}{42875}\,j_{4}(k's)\right)Q^{22}_{224}(s) \nn \\
    & \hspace{-3cm} + \left(- \frac{2\sqrt{70}}{15435}\,j_{2}(k's) + \frac{2\sqrt{70}}{56595}\,j_{4}(k's)\right)Q^{22}_{242}(s) \nn \\
    & \hspace{-3cm} + \left(- \frac{2\sqrt{385}}{51975}\,j_{0}(k's) - \frac{26\sqrt{385}}{1120581}\,j_{2}(k's) - \frac{36\sqrt{385}}{1037575}\,j_{4}(k's)\right)Q^{22}_{244}(s) \nn \\
    & \hspace{-3cm} + \left(- \frac{8\sqrt{715}}{495495}\,j_{2}(k's) + \frac{4\sqrt{715}}{165165}\,j_{4}(k's)\right)Q^{22}_{246}(s) \nn \\
    & \hspace{-3cm} + \left(\frac{\sqrt{715}}{32175}\,j_{0}(k's) - \frac{8\sqrt{715}}{495495}\,j_{2}(k's) + \frac{6\sqrt{715}}{3578575}\,j_{4}(k's)\right)Q^{22}_{264}(s) \nn \\
    & \hspace{-3cm} + \left(\frac{\sqrt{10010}}{143143}\,j_{2}(k's) - \frac{2\sqrt{10010}}{715715}\,j_{4}(k's)\right)Q^{22}_{266}(s) \Bigg]
\end{align}

%% file: main.bib
@article{Bianchi:2015oia,
    author = "Bianchi, Davide and Gil-Mar\'\i{}n, H\'ector and Ruggeri, Rossana and Percival, Will J.",
    title = "{Measuring line-of-sight dependent Fourier-space clustering using FFTs}",
    eprint = "1505.05341",
    archivePrefix = "arXiv",
    primaryClass = "astro-ph.CO",
    doi = "10.1093/mnrasl/slv090",
    journal = "Mon. Not. Roy. Astron. Soc.",
    volume = "453",
    number = "1",
    pages = "L11--L15",
    year = "2015"
}

@ARTICLE{2019JCAP...08..036D,
       author = {{de Mattia}, Arnaud and {Ruhlmann-Kleider}, Vanina},
        title = "{Integral constraints in spectroscopic surveys}",
      journal = {\jcap},
         year = 2019,
        month = aug,
       volume = {2019},
       number = {8},
          eid = {036},
        pages = {036},
          doi = {10.1088/1475-7516/2019/08/036},
archivePrefix = {arXiv},
       eprint = {1904.08851},
 primaryClass = {astro-ph.CO},
       adsurl = {https://ui.adsabs.harvard.edu/abs/2019JCAP...08..036D}
}

@ARTICLE{2026arXiv260418581G,
       author = {{Goldstein}, Samuel and {Smith}, Kendrick M. and {Giri}, Utkarsh and {M{\"u}nchmeyer}, Moritz},
        title = "{If at First You Don't Succeed, Trispectrum: I. Estimating the Matter Power Spectrum Covariance with Higher-Order Statistics}",
      journal = {arXiv e-prints},
         year = 2026,
        month = apr,
          eid = {arXiv:2604.18581},
        pages = {arXiv:2604.18581},
archivePrefix = {arXiv},
       eprint = {2604.18581},
 primaryClass = {astro-ph.CO},
       adsurl = {https://ui.adsabs.harvard.edu/abs/2026arXiv260418581G}
}

@ARTICLE{2018MNRAS.473.2737S,
       author = {{Sugiyama}, Naonori S. and {Shiraishi}, Maresuke and {Okumura}, Teppei},
        title = "{Limits on statistical anisotropy from BOSS DR12 galaxies using bipolar spherical harmonics}",
      journal = {\mnras},
         year = 2018,
        month = jan,
       volume = {473},
       number = {2},
        pages = {2737-2752},
          doi = {10.1093/mnras/stx2333},
archivePrefix = {arXiv},
       eprint = {1704.02868},
 primaryClass = {astro-ph.CO},
       adsurl = {https://ui.adsabs.harvard.edu/abs/2018MNRAS.473.2737S}
}

@article{Wen:2026fzo,
    author = "Wen, Robin Y. and Gebhardt, Henry S. Grasshorn and Heinrich, Chen and Dor{\'e}, Olivier",
    title = "{Large-scale Modeling of the Observed Power Spectrum Multipoles}",
    eprint = "2601.19438",
    archivePrefix = "arXiv",
    primaryClass = "astro-ph.CO",
    month = "1",
    year = "2026"
}

@ARTICLE{1995MNRAS.275..483H,
       author = {{Heavens}, A.~F. and {Taylor}, A.~N.},
        title = "{A spherical harmonic analysis of redshift space}",
      journal = {\mnras},
         year = 1995,
        month = jul,
       volume = {275},
       number = {2},
        pages = {483-497},
          doi = {10.1093/mnras/275.2.483},
archivePrefix = {arXiv},
       eprint = {astro-ph/9409027},
 primaryClass = {astro-ph},
       adsurl = {https://ui.adsabs.harvard.edu/abs/1995MNRAS.275..483H}
}

@ARTICLE{2000ApJ...535....1M,
       author = {{Matsubara}, Takahiko},
        title = "{The Correlation Function in Redshift Space: General Formula with Wide-Angle Effects and Cosmological Distortions}",
      journal = {\apj},
         year = 2000,
        month = may,
       volume = {535},
       number = {1},
        pages = {1-23},
          doi = {10.1086/308827},
archivePrefix = {arXiv},
       eprint = {astro-ph/9908056},
 primaryClass = {astro-ph},
       adsurl = {https://ui.adsabs.harvard.edu/abs/2000ApJ...535....1M}
}

@ARTICLE{2025arXiv251102985B,
       author = {{Bock}, James J. and {Aboobaker}, Asad M. and {Adamo}, Joseph and {Akeson}, Rachel and {Alred}, John M. and {Alibay}, Farah and {Ashby}, Matthew L.~N. and {Bach}, Yoonsoo P. and {Bleem}, Lindsey E. and {Bolton}, Douglas and {Braun}, David F. and {Bruton}, Sean and {Bryan}, Sean A. and {Chang}, Tzu-Ching and {Chen}, Shuang-Shuang and {Cheng}, Yun-Ting and {Cheshire}, IV, James R. and {Chiang}, Yi-Kuan and {Choppin de Janvry}, Jean and {Condon}, Samuel and {Cook}, Walter R. and {Cooray}, Asantha and {Crill}, Brendan P. and {Cukierman}, Ari J. and {Dore}, Olivier and {Dowell}, C. Darren and {Dubois-Felsmann}, Gregory P. and {Eifler}, Tim and {Everett}, Spencer and {Fabinsky}, Beth E. and {Faisst}, Andreas L. and {Fanson}, James L. and {Farrington}, Allen H. and {Fatahi}, Tamim and {Fazar}, Candice M. and {Feder}, Richard M. and {Frater}, Eric H. and {Grasshorn Gebhardt}, Henry S. and {Giri}, Utkarsh and {Goldina}, Tatiana and {Gorjian}, Varoujan and {Habib}, Salman and {Hart}, William G. and {Heinrich}, Chen and {Hora}, Joseph L. and {Huai}, Zhaoyu and {Hui}, Howard and {Jo}, Young-Soo and {Jeong}, Woong-Seob and {Kang}, Jae Hwan and {Kang}, Miju and {Kecman}, Branislav and {Kim}, Chul-Hwan and {Kim}, Jaeyeong and {Kim}, Minjin and {Kim}, Young-Jun and {Kim}, Yongjung and {Kirkpatrick}, J. Davy and {Kobayashi}, Yosuke and {Korngut}, Phil M. and {Krause}, Elisabeth and {Lee}, Bomee and {Lee}, Ho-Gyu and {Lee}, Jae-Joon and {Lee}, Jeong-Eun and {Lisse}, Carey M. and {Mariani}, Giacomo and {Masters}, Daniel C. and {Mauskopf}, Philip D. and {Melnick}, Gary J. and {Minasyan}, Mary H. and {Mirocha}, Jordan and {Miyasaka}, Hiromasa and {Moore}, Anne and {Moore}, Bradley D. and {Murgia}, Giulia and {Naylor}, Bret J. and {Nelson}, Christina and {Nguyen}, Chi H. and {Nguyen}, Hien T. and {Noh}, Jinyoung K. and {Padin}, Stephen and {Paladini}, Roberta and {Park}, Sung-Joon and {Penanen}, Konstantin I. and {Putnam}, Dustin S. and {Pyo}, Jeonghyun and {Ramachandra}, Nesar and {Ramanathan}, Keshav and {Rustamkulov}, Zafar and {Reiley}, Daniel J. and {Rice}, Eric B. and {Rocca}, Jennifer M. and {Seok}, Ji Yeon and {Smith}, Roger and {Stober}, Jeremy and {Susca}, Sara and {Teplitz}, Harry I. and {Thelen}, Michael P. and {Tolls}, Volker and {Torrini}, Gabriela and {Trangsrud}, Amy R. and {Unwin}, Stephen and {Velicheti}, Phani and {Wang}, Pao-Yu and {Wen}, Robin Y. and {-Werner}, Michael-W. and {Williams}, Abby E. and {Williamson}, Ross and {Wincentsen}, James and {Windhorst}, Rogier A. and {Yang}, Soung-Chul and {Yang}, Yujin and {Zemcov}, Michael},
        title = "{The SPHEREx Satellite Mission}",
      journal = {arXiv e-prints},
         year = 2025,
        month = nov,
          eid = {arXiv:2511.02985},
        pages = {arXiv:2511.02985},
          doi = {10.48550/arXiv.2511.02985},
archivePrefix = {arXiv},
       eprint = {2511.02985},
 primaryClass = {astro-ph.IM},
       adsurl = {https://ui.adsabs.harvard.edu/abs/2025arXiv251102985B}
}

@ARTICLE{2026arXiv260404867C,
       author = {{Chaussidon}, Edmond and {Hotinli}, Selim C. and {Ferraro}, Simone and {Smith}, Kendrick and {Chen}, Xinyi and {Aguilar}, J. and {Ahlen}, S. and {Bianchi}, D. and {Brooks}, D. and {Claybaugh}, T. and {Cuceu}, A. and {de la Macorra}, A. and {Dey}, B. and {Doel}, P. and {Font-Ribera}, A. and {Forero-Romero}, J.~E. and {Gazta{\~n}aga}, E. and {Gontcho}, S. Gontcho A and {Gutierrez}, G. and {Guy}, J. and {Herrera-Alcantar}, H.~K. and {Honscheid}, K. and {Howlett}, C. and {Huterer}, D. and {Ishak}, M. and {Joyce}, R. and {Kirkby}, D. and {Kremin}, A. and {Lahav}, O. and {Landriau}, M. and {Le Guillou}, L. and {Manera}, M. and {Meisner}, A. and {Miquel}, R. and {Nadathur}, S. and {Newman}, J.~A. and {Palanque-Delabrouille}, N. and {Percival}, W.~J. and {Prada}, F. and {P{\'e}rez-R{\`a}fols}, I. and {Rossi}, G. and {Samushia}, L. and {Sanchez}, E. and {Schlegel}, D. and {Schubnell}, M. and {Seo}, H. and {Silber}, J. and {Sprayberry}, D. and {Tarl{\'e}}, G. and {Weaver}, B.~A. and {Y{\`e}che}, C. and {Zhou}, R.},
        title = "{Measurement of the galaxy-velocity power spectrum of DESI tracers with the kinematic Sunyaev-Zeldovich effect using DESI DR2 and ACT DR6}",
      journal = {arXiv e-prints},
         year = 2026,
        month = apr,
          eid = {arXiv:2604.04867},
        pages = {arXiv:2604.04867},
archivePrefix = {arXiv},
       eprint = {2604.04867},
 primaryClass = {astro-ph.CO},
       adsurl = {https://ui.adsabs.harvard.edu/abs/2026arXiv260404867C}
}

@ARTICLE{FKP,
       author = {{Feldman}, Hume A. and {Kaiser}, Nick and {Peacock}, John A.},
        title = "{Power-Spectrum Analysis of Three-dimensional Redshift Surveys}",
      journal = {\apj},
         year = 1994,
        month = may,
       volume = {426},
        pages = {23},
          doi = {10.1086/174036},
archivePrefix = {arXiv},
       eprint = {astro-ph/9304022},
 primaryClass = {astro-ph},
       adsurl = {https://ui.adsabs.harvard.edu/abs/1994ApJ...426...23F}
}

@ARTICLE{2025A&A...697A...1E,
       author = {{Euclid Collaboration} and {Mellier}, Y. and {Abdurro'uf} and {Acevedo Barroso}, J.~A. and {Ach{\'u}carro}, A. and {Adamek}, J. and {Adam}, R. and {Addison}, G.~E. and {Aghanim}, N. and {Aguena}, M. and {Ajani}, V. and {Akrami}, Y. and {Al-Bahlawan}, A. and {Alavi}, A. and {Albuquerque}, I.~S. and {Alestas}, G. and {Alguero}, G. and {Allaoui}, A. and {Allen}, S.~W. and {Allevato}, V. and {Alonso-Tetilla}, A.~V. and {Altieri}, B. and {Alvarez-Candal}, A. and {Alvi}, S. and {Amara}, A. and {Amendola}, L. and {Amiaux}, J. and {Andika}, I.~T. and {Andreon}, S. and {Andrews}, A. and {Angora}, G. and {Angulo}, R.~E. and {Annibali}, F. and {Anselmi}, A. and {Anselmi}, S. and {Arcari}, S. and {Archidiacono}, M. and {Aric{\`o}}, G. and {Arnaud}, M. and {Arnouts}, S. and {Asgari}, M. and {Asorey}, J. and {Atayde}, L. and {Atek}, H. and {Atrio-Barandela}, F. and {Aubert}, M. and {Aubourg}, E. and {Auphan}, T. and {Auricchio}, N. and {Aussel}, B. and {Aussel}, H. and {Avelino}, P.~P. and {Avgoustidis}, A. and {Avila}, S. and {Awan}, S. and {Azzollini}, R. and {Baccigalupi}, C. and {Bachelet}, E. and {Bacon}, D. and {Baes}, M. and {Bagley}, M.~B. and {Bahr-Kalus}, B. and {Balaguera-Antolinez}, A. and {Balbinot}, E. and {Balcells}, M. and {Baldi}, M. and {Baldry}, I. and {Balestra}, A. and {Ballardini}, M. and {Ballester}, O. and {Balogh}, M. and {Ba{\~n}ados}, E. and {Barbier}, R. and {Bardelli}, S. and {Baron}, M. and {Barreiro}, T. and {Barrena}, R. and {Barriere}, J.-C. and {Barros}, B.~J. and {Barthelemy}, A. and {Bartolo}, N. and {Basset}, A. and {Battaglia}, P. and {Battisti}, A.~J. and {Baugh}, C.~M. and {Baumont}, L. and {Bazzanini}, L. and {Beaulieu}, J.-P. and {Beckmann}, V. and {Belikov}, A.~N. and {Bel}, J. and {Bellagamba}, F. and {Bella}, M. and {Bellini}, E. and {Benabed}, K. and {Bender}, R. and {Benevento}, G. and {Bennett}, C.~L. and {Benson}, K. and {Bergamini}, P. and {Bermejo-Climent}, J.~R. and {Bernardeau}, F. and {Bertacca}, D. and {Berthe}, M. and {Berthier}, J. and {Bethermin}, M. and {Beutler}, F. and {Bevillon}, C. and {Bhargava}, S. and {Bhatawdekar}, R. and {Bianchi}, D. and {Bisigello}, L. and {Biviano}, A. and {Blake}, R.~P. and {Blanchard}, A. and {Blazek}, J. and {Blot}, L. and {Bosco}, A. and {Bodendorf}, C. and {Boenke}, T. and {B{\"o}hringer}, H. and {Boldrini}, P. and {Bolzonella}, M. and {Bonchi}, A. and {Bonici}, M. and {Bonino}, D. and {Bonino}, L. and {Bonvin}, C. and {Bon}, W. and {Booth}, J.~T. and {Borgani}, S. and {Borlaff}, A.~S. and {Borsato}, E. and {Bose}, B. and {Botticella}, M.~T. and {Boucaud}, A. and {Bouche}, F. and {Boucher}, J.~S. and {Boutigny}, D. and {Bouvard}, T. and {Bouwens}, R. and {Bouy}, H. and {Bowler}, R.~A.~A. and {Bozza}, V. and {Bozzo}, E. and {Branchini}, E. and {Brando}, G. and {Brau-Nogue}, S. and {Brekke}, P. and {Bremer}, M.~N. and {Brescia}, M. and {Breton}, M.-A. and {Brinchmann}, J. and {Brinckmann}, T. and {Brockley-Blatt}, C. and {Brodwin}, M. and {Brouard}, L. and {Brown}, M.~L. and {Bruton}, S. and {Bucko}, J. and {Buddelmeijer}, H. and {Buenadicha}, G. and {Buitrago}, F. and {Burger}, P. and {Burigana}, C. and {Busillo}, V. and {Busonero}, D. and {Cabanac}, R. and {Cabayol-Garcia}, L. and {Cagliari}, M.~S. and {Caillat}, A. and {Caillat}, L. and {Calabrese}, M. and {Calabro}, A. and {Calderone}, G. and {Calura}, F. and {Camacho Quevedo}, B. and {Camera}, S. and {Campos}, L. and {Ca{\~n}as-Herrera}, G. and {Candini}, G.~P. and {Cantiello}, M. and {Capobianco}, V. and {Cappellaro}, E. and {Cappelluti}, N. and {Cappi}, A. and {Caputi}, K.~I. and {Cara}, C. and {Carbone}, C. and {Cardone}, V.~F. and {Carella}, E. and {Carlberg}, R.~G. and {Carle}, M. and {Carminati}, L. and {Caro}, F. and {Carrasco}, J.~M. and {Carretero}, J. and {Carrilho}, P. and {Carron Duque}, J. and {Carry}, B.},
        title = "{Euclid: I. Overview of the Euclid mission}",
      journal = {\aap},
         year = 2025,
        month = may,
       volume = {697},
          eid = {A1},
        pages = {A1},
          doi = {10.1051/0004-6361/202450810},
archivePrefix = {arXiv},
       eprint = {2405.13491},
 primaryClass = {astro-ph.CO},
       adsurl = {https://ui.adsabs.harvard.edu/abs/2025A&A...697A...1E}
}

@ARTICLE{2018arXiv180901669T,
       author = {{The LSST Dark Energy Science Collaboration} and {Mandelbaum}, Rachel and {Eifler}, Tim and {Hlo{\v{z}}ek}, Ren{\'e}e and {Collett}, Thomas and {Gawiser}, Eric and {Scolnic}, Daniel and {Alonso}, David and {Awan}, Humna and {Biswas}, Rahul and {Blazek}, Jonathan and {Burchat}, Patricia and {Chisari}, Nora Elisa and {Dell'Antonio}, Ian and {Digel}, Seth and {Frieman}, Josh and {Goldstein}, Daniel A. and {Hook}, Isobel and {Ivezi{\'c}}, {\v{Z}}eljko and {Kahn}, Steven M. and {Kamath}, Sowmya and {Kirkby}, David and {Kitching}, Thomas and {Krause}, Elisabeth and {Leget}, Pierre-Fran{\c{c}}ois and {Marshall}, Philip J. and {Meyers}, Joshua and {Miyatake}, Hironao and {Newman}, Jeffrey A. and {Nichol}, Robert and {Rykoff}, Eli and {Sanchez}, F. Javier and {Slosar}, An{\v{z}}e and {Sullivan}, Mark and {Troxel}, M.~A.},
        title = "{The LSST Dark Energy Science Collaboration (DESC) Science Requirements Document}",
      journal = {arXiv e-prints},
         year = 2018,
        month = sep,
          eid = {arXiv:1809.01669},
        pages = {arXiv:1809.01669},
          doi = {10.48550/arXiv.1809.01669},
archivePrefix = {arXiv},
       eprint = {1809.01669},
 primaryClass = {astro-ph.CO},
       adsurl = {https://ui.adsabs.harvard.edu/abs/2018arXiv180901669T}
}

@ARTICLE{2015arXiv150303757S,
       author = {{Spergel}, D. and {Gehrels}, N. and {Baltay}, C. and {Bennett}, D. and {Breckinridge}, J. and {Donahue}, M. and {Dressler}, A. and {Gaudi}, B.~S. and {Greene}, T. and {Guyon}, O. and {Hirata}, C. and {Kalirai}, J. and {Kasdin}, N.~J. and {Macintosh}, B. and {Moos}, W. and {Perlmutter}, S. and {Postman}, M. and {Rauscher}, B. and {Rhodes}, J. and {Wang}, Y. and {Weinberg}, D. and {Benford}, D. and {Hudson}, M. and {Jeong}, W.-S. and {Mellier}, Y. and {Traub}, W. and {Yamada}, T. and {Capak}, P. and {Colbert}, J. and {Masters}, D. and {Penny}, M. and {Savransky}, D. and {Stern}, D. and {Zimmerman}, N. and {Barry}, R. and {Bartusek}, L. and {Carpenter}, K. and {Cheng}, E. and {Content}, D. and {Dekens}, F. and {Demers}, R. and {Grady}, K. and {Jackson}, C. and {Kuan}, G. and {Kruk}, J. and {Melton}, M. and {Nemati}, B. and {Parvin}, B. and {Poberezhskiy}, I. and {Peddie}, C. and {Ruffa}, J. and {Wallace}, J.~K. and {Whipple}, A. and {Wollack}, E. and {Zhao}, F.},
        title = "{Wide-Field InfrarRed Survey Telescope-Astrophysics Focused Telescope Assets WFIRST-AFTA 2015 Report}",
      journal = {arXiv e-prints},
         year = 2015,
        month = mar,
          eid = {arXiv:1503.03757},
        pages = {arXiv:1503.03757},
          doi = {10.48550/arXiv.1503.03757},
archivePrefix = {arXiv},
       eprint = {1503.03757},
 primaryClass = {astro-ph.IM},
       adsurl = {https://ui.adsabs.harvard.edu/abs/2015arXiv150303757S}
}

@ARTICLE{2016arXiv161100036D,
       author = {{DESI Collaboration} and {Aghamousa}, Amir and {Aguilar}, Jessica and {Ahlen}, Steve and {Alam}, Shadab and {Allen}, Lori E. and {Allende Prieto}, Carlos and {Annis}, James and {Bailey}, Stephen and {Balland}, Christophe and {Ballester}, Otger and {Baltay}, Charles and {Beaufore}, Lucas and {Bebek}, Chris and {Beers}, Timothy C. and {Bell}, Eric F. and {Bernal}, Jos{\'e} Luis and {Besuner}, Robert and {Beutler}, Florian and {Blake}, Chris and {Bleuler}, Hannes and {Blomqvist}, Michael and {Blum}, Robert and {Bolton}, Adam S. and {Briceno}, Cesar and {Brooks}, David and {Brownstein}, Joel R. and {Buckley-Geer}, Elizabeth and {Burden}, Angela and {Burtin}, Etienne and {Busca}, Nicolas G. and {Cahn}, Robert N. and {Cai}, Yan-Chuan and {Cardiel-Sas}, Laia and {Carlberg}, Raymond G. and {Carton}, Pierre-Henri and {Casas}, Ricard and {Castander}, Francisco J. and {Cervantes-Cota}, Jorge L. and {Claybaugh}, Todd M. and {Close}, Madeline and {Coker}, Carl T. and {Cole}, Shaun and {Comparat}, Johan and {Cooper}, Andrew P. and {Cousinou}, M.-C. and {Crocce}, Martin and {Cuby}, Jean-Gabriel and {Cunningham}, Daniel P. and {Davis}, Tamara M. and {Dawson}, Kyle S. and {de la Macorra}, Axel and {De Vicente}, Juan and {Delubac}, Timoth{\'e}e and {Derwent}, Mark and {Dey}, Arjun and {Dhungana}, Govinda and {Ding}, Zhejie and {Doel}, Peter and {Duan}, Yutong T. and {Ealet}, Anne and {Edelstein}, Jerry and {Eftekharzadeh}, Sarah and {Eisenstein}, Daniel J. and {Elliott}, Ann and {Escoffier}, St{\'e}phanie and {Evatt}, Matthew and {Fagrelius}, Parker and {Fan}, Xiaohui and {Fanning}, Kevin and {Farahi}, Arya and {Farihi}, Jay and {Favole}, Ginevra and {Feng}, Yu and {Fernandez}, Enrique and {Findlay}, Joseph R. and {Finkbeiner}, Douglas P. and {Fitzpatrick}, Michael J. and {Flaugher}, Brenna and {Flender}, Samuel and {Font-Ribera}, Andreu and {Forero-Romero}, Jaime E. and {Fosalba}, Pablo and {Frenk}, Carlos S. and {Fumagalli}, Michele and {Gaensicke}, Boris T. and {Gallo}, Giuseppe and {Garcia-Bellido}, Juan and {Gaztanaga}, Enrique and {Pietro Gentile Fusillo}, Nicola and {Gerard}, Terry and {Gershkovich}, Irena and {Giannantonio}, Tommaso and {Gillet}, Denis and {Gonzalez-de-Rivera}, Guillermo and {Gonzalez-Perez}, Violeta and {Gott}, Shelby and {Graur}, Or and {Gutierrez}, Gaston and {Guy}, Julien and {Habib}, Salman and {Heetderks}, Henry and {Heetderks}, Ian and {Heitmann}, Katrin and {Hellwing}, Wojciech A. and {Herrera}, David A. and {Ho}, Shirley and {Holland}, Stephen and {Honscheid}, Klaus and {Huff}, Eric and {Hutchinson}, Timothy A. and {Huterer}, Dragan and {Hwang}, Ho Seong and {Illa Laguna}, Joseph Maria and {Ishikawa}, Yuzo and {Jacobs}, Dianna and {Jeffrey}, Niall and {Jelinsky}, Patrick and {Jennings}, Elise and {Jiang}, Linhua and {Jimenez}, Jorge and {Johnson}, Jennifer and {Joyce}, Richard and {Jullo}, Eric and {Juneau}, St{\'e}phanie and {Kama}, Sami and {Karcher}, Armin and {Karkar}, Sonia and {Kehoe}, Robert and {Kennamer}, Noble and {Kent}, Stephen and {Kilbinger}, Martin and {Kim}, Alex G. and {Kirkby}, David and {Kisner}, Theodore and {Kitanidis}, Ellie and {Kneib}, Jean-Paul and {Koposov}, Sergey and {Kovacs}, Eve and {Koyama}, Kazuya and {Kremin}, Anthony and {Kron}, Richard and {Kronig}, Luzius and {Kueter-Young}, Andrea and {Lacey}, Cedric G. and {Lafever}, Robin and {Lahav}, Ofer and {Lambert}, Andrew and {Lampton}, Michael and {Landriau}, Martin and {Lang}, Dustin and {Lauer}, Tod R. and {Le Goff}, Jean-Marc and {Le Guillou}, Laurent and {Le Van Suu}, Auguste and {Lee}, Jae Hyeon and {Lee}, Su-Jeong and {Leitner}, Daniela and {Lesser}, Michael and {Levi}, Michael E. and {L'Huillier}, Benjamin and {Li}, Baojiu and {Liang}, Ming and {Lin}, Huan and {Linder}, Eric and {Loebman}, Sarah R. and {Luki{\'c}}, Zarija and {Ma}, Jun and {MacCrann}, Niall and {Magneville}, Christophe and {Makarem}, Laleh and {Manera}, Marc and {Manser}, Christopher J. and {Marshall}, Robert and {Martini}, Paul and {Massey}, Richard and {Matheson}, Thomas and {McCauley}, Jeremy and {McDonald}, Patrick and {McGreer}, Ian D. and {Meisner}, Aaron and {Metcalfe}, Nigel and {Miller}, Timothy N. and {Miquel}, Ramon and {Moustakas}, John and {Myers}, Adam and {Naik}, Milind and {Newman}, Jeffrey A. and {Nichol}, Robert C. and {Nicola}, Andrina and {Nicolati da Costa}, Luiz and {Nie}, Jundan and {Niz}, Gustavo and {Norberg}, Peder and {Nord}, Brian and {Norman}, Dara and {Nugent}, Peter and {O'Brien}, Thomas and {Oh}, Minji and {Olsen}, Knut A.~G.},
        title = "{The DESI Experiment Part I: Science,Targeting, and Survey Design}",
      journal = {arXiv e-prints},
         year = 2016,
        month = oct,
          eid = {arXiv:1611.00036},
        pages = {arXiv:1611.00036},
          doi = {10.48550/arXiv.1611.00036},
archivePrefix = {arXiv},
       eprint = {1611.00036},
 primaryClass = {astro-ph.IM},
       adsurl = {https://ui.adsabs.harvard.edu/abs/2016arXiv161100036D}
}

@ARTICLE{2025arXiv250621657H,
       author = {{Hotinli}, Selim C. and {Smith}, Kendrick M. and {Ferraro}, Simone},
        title = "{Velocity Reconstruction from KSZ: Measuring $f_{NL}$ with ACT and DESILS}",
      journal = {arXiv e-prints},
         year = 2025,
        month = jun,
          eid = {arXiv:2506.21657},
        pages = {arXiv:2506.21657},
          doi = {10.48550/arXiv.2506.21657},
archivePrefix = {arXiv},
       eprint = {2506.21657},
 primaryClass = {astro-ph.CO},
       adsurl = {https://ui.adsabs.harvard.edu/abs/2025arXiv250621657H}
}

@ARTICLE{2025PhRvL.134o1003L,
       author = {{Lagu{\"e}}, Alex and {Madhavacheril}, Mathew S. and {Smith}, Kendrick M. and {Ferraro}, Simone and {Schaan}, Emmanuel},
        title = "{Constraints on Local Primordial Non-Gaussianity with 3D Velocity Reconstruction from the Kinetic Sunyaev-Zeldovich Effect}",
      journal = {\prl},
         year = 2025,
        month = apr,
       volume = {134},
       number = {15},
          eid = {151003},
        pages = {151003},
          doi = {10.1103/PhysRevLett.134.151003},
archivePrefix = {arXiv},
       eprint = {2411.08240},
 primaryClass = {astro-ph.CO},
       adsurl = {https://ui.adsabs.harvard.edu/abs/2025PhRvL.134o1003L}
}

@ARTICLE{2025JCAP...05..057M,
       author = {{McCarthy}, Fiona and {Battaglia}, Nicholas and {Bean}, Rachel and {Richard Bond}, J. and {Cai}, Hongbo and {Calabrese}, Erminia and {Coulton}, William R. and {Devlin}, Mark J. and {Dunkley}, Jo and {Ferraro}, Simone and {Gluscevic}, Vera and {Guan}, Yilun and {Colin Hill}, J. and {Johnson}, Matthew C. and {Kusiak}, Aleksandra and {Lagu{\"e}}, Alex and {MacCrann}, Niall and {Madhavacheril}, Mathew S. and {Moodley}, Kavilan and {Naess}, Sigurd and {Qu}, Frank J. and {Ried Guachalla}, Bernardita and {Sehgal}, Neelima and {Sherwin}, Blake D. and {Sif{\'o}n}, Crist{\'o}bal and {Smith}, Kendrick M. and {Staggs}, Suzanne T. and {van Engelen}, Alexander and {Vavagiakis}, Eve M. and {Wollack}, Edward J.},
        title = "{The Atacama Cosmology Telescope: Large-scale velocity reconstruction with the kinematic Sunyaev-Zel'dovich effect and DESI LRGs}",
      journal = {\jcap},
         year = 2025,
        month = may,
       volume = {2025},
       number = {5},
          eid = {057},
        pages = {057},
          doi = {10.1088/1475-7516/2025/05/057},
archivePrefix = {arXiv},
       eprint = {2410.06229},
 primaryClass = {astro-ph.CO},
       adsurl = {https://ui.adsabs.harvard.edu/abs/2025JCAP...05..057M}
}

@ARTICLE{2019PhRvD.100h3508M,
       author = {{M{\"u}nchmeyer}, Moritz and {Madhavacheril}, Mathew S. and {Ferraro}, Simone and {Johnson}, Matthew C. and {Smith}, Kendrick M.},
        title = "{Constraining local non-Gaussianities with kinetic Sunyaev-Zel'dovich tomography}",
      journal = {\prd},
         year = 2019,
        month = oct,
       volume = {100},
       number = {8},
          eid = {083508},
        pages = {083508},
          doi = {10.1103/PhysRevD.100.083508},
archivePrefix = {arXiv},
       eprint = {1810.13424},
 primaryClass = {astro-ph.CO},
       adsurl = {https://ui.adsabs.harvard.edu/abs/2019PhRvD.100h3508M}
}

@ARTICLE{2018arXiv181013423S,
       author = {{Smith}, Kendrick M. and {Madhavacheril}, Mathew S. and {M{\"u}nchmeyer}, Moritz and {Ferraro}, Simone and {Giri}, Utkarsh and {Johnson}, Matthew C.},
        title = "{KSZ tomography and the bispectrum}",
      journal = {arXiv e-prints},
         year = 2018,
        month = oct,
          eid = {arXiv:1810.13423},
        pages = {arXiv:1810.13423},
          doi = {10.48550/arXiv.1810.13423},
archivePrefix = {arXiv},
       eprint = {1810.13423},
 primaryClass = {astro-ph.CO},
       adsurl = {https://ui.adsabs.harvard.edu/abs/2018arXiv181013423S}
}

@ARTICLE{1998ApJ...498L...1S,
       author = {{Szalay}, Alexander S. and {Matsubara}, Takahiko and {Landy}, Stephen D.},
        title = "{Redshift-Space Distortions of the Correlation Function in Wide-Angle Galaxy Surveys}",
      journal = {\apjl},
         year = 1998,
        month = may,
       volume = {498},
       number = {1},
        pages = {L1-L4},
          doi = {10.1086/311293},
archivePrefix = {arXiv},
       eprint = {astro-ph/9712007},
 primaryClass = {astro-ph},
       adsurl = {https://ui.adsabs.harvard.edu/abs/1998ApJ...498L...1S}
}

@ARTICLE{2025JCAP...01..131P,
       author = {{Pinon}, M. and {de Mattia}, A. and {McDonald}, P. and {Burtin}, E. and {Ruhlmann-Kleider}, V. and {White}, M. and {Bianchi}, D. and {Ross}, A.~J. and {Aguilar}, J. and {Ahlen}, S. and {Brooks}, D. and {Cahn}, R.~N. and {Chaussidon}, E. and {Claybaugh}, T. and {Cole}, S. and {de la Macorra}, A. and {Dey}, B. and {Doel}, P. and {Fanning}, K. and {Forero-Romero}, J.~E. and {Gazta{\~n}aga}, E. and {Gontcho A Gontcho}, S. and {Howlett}, C. and {Kirkby}, D. and {Kisner}, T. and {Kremin}, A. and {Lambert}, A. and {Landriau}, M. and {Lasker}, J. and {Le Guillou}, L. and {Levi}, M.~E. and {Manera}, M. and {Martini}, P. and {Meisner}, A. and {Miquel}, R. and {Moustakas}, J. and {Myers}, A.~D. and {Niz}, G. and {Palanque-Delabrouille}, N. and {Percival}, W.~J. and {Poppett}, C. and {Rossi}, G. and {Sanchez}, E. and {Schlegel}, D. and {Schubnell}, M. and {Seo}, H. and {Sprayberry}, D. and {Tarl{\'e}}, G. and {Vargas-Maga{\~n}a}, M. and {Weaver}, B.~A. and {Zarrouk}, P. and {Zhou}, R. and {Zou}, H.},
        title = "{Mitigation of DESI fiber assignment incompleteness effect on two-point clustering with small angular scale truncated estimators}",
      journal = {\jcap},
         year = 2025,
        month = jan,
       volume = {2025},
       number = {1},
          eid = {131},
        pages = {131},
          doi = {10.1088/1475-7516/2025/01/131},
archivePrefix = {arXiv},
       eprint = {2406.04804},
 primaryClass = {astro-ph.CO},
       adsurl = {https://ui.adsabs.harvard.edu/abs/2025JCAP...01..131P}
}

@ARTICLE{2004MNRAS.349..603E,
       author = {{Efstathiou}, G.},
        title = "{Myths and truths concerning estimation of power spectra: the case for a hybrid estimator}",
      journal = {\mnras},
         year = 2004,
        month = apr,
       volume = {349},
       number = {2},
        pages = {603-626},
          doi = {10.1111/j.1365-2966.2004.07530.x},
archivePrefix = {arXiv},
       eprint = {astro-ph/0307515},
 primaryClass = {astro-ph},
       adsurl = {https://ui.adsabs.harvard.edu/abs/2004MNRAS.349..603E}
}

@ARTICLE{2025JCAP...04..074B,
       author = {{Bianchi}, D. and {Hanif}, M.~M.~S. and {Carnero Rosell}, A. and {Lasker}, J. and {Ross}, A.~J. and {Pinon}, M. and {de Mattia}, A. and {White}, M. and {Ahlen}, S. and {Bailey}, S. and {Brooks}, D. and {Burtin}, E. and {Chaussidon}, E. and {Claybaugh}, T. and {Cole}, S. and {de la Macorra}, A. and {Ferraro}, S. and {Font-Ribera}, A. and {Forero-Romero}, J.~E. and {Gazta{\~n}aga}, E. and {Gontcho}, S. Gontcho A. and {Gutierrez}, G. and {Guy}, J. and {Hahn}, C. and {Honscheid}, K. and {Howlett}, C. and {Juneau}, S. and {Kirkby}, D. and {Kisner}, T. and {Kremin}, A. and {Landriau}, M. and {Le Guillou}, L. and {Levi}, M.~E. and {McDonald}, P. and {Meisner}, A. and {Miquel}, R. and {Moustakas}, J. and {Palanque-Delabrouille}, N. and {Percival}, W.~J. and {Prada}, F. and {P{\'e}rez-R{\`a}fols}, I. and {Raichoor}, A. and {Rossi}, G. and {Sanchez}, E. and {Schlegel}, D. and {Schubnell}, M. and {Sharples}, R. and {Silber}, J. and {Sprayberry}, D. and {Tarl{\'e}}, G. and {Vargas-Maga{\~n}a}, M. and {Weaver}, B.~A. and {Zarrouk}, P. and {Zhou}, R. and {Zou}, H.},
        title = "{Characterization of DESI fiber assignment incompleteness effect on 2-point clustering and mitigation methods for DR1 analysis}",
      journal = {\jcap},
         year = 2025,
        month = apr,
       volume = {2025},
       number = {4},
          eid = {074},
        pages = {074},
          doi = {10.1088/1475-7516/2025/04/074},
archivePrefix = {arXiv},
       eprint = {2411.12025},
 primaryClass = {astro-ph.CO},
       adsurl = {https://ui.adsabs.harvard.edu/abs/2025JCAP...04..074B}
}

@ARTICLE{2020MNRAS.498.2492G,
       author = {{Gil-Mar{\'\i}n}, H{\'e}ctor and {Bautista}, Juli{\'a}n E. and {Paviot}, Romain and {Vargas-Maga{\~n}a}, Mariana and {de la Torre}, Sylvain and {Fromenteau}, Sebastien and {Alam}, Shadab and {{\'A}vila}, Santiago and {Burtin}, Etienne and {Chuang}, Chia-Hsun and {Dawson}, Kyle S. and {Hou}, Jiamin and {de Mattia}, Arnaud and {Mohammad}, Faizan G. and {M{\"u}ller}, Eva-Maria and {Nadathur}, Seshadri and {Neveux}, Richard and {Percival}, Will J. and {Raichoor}, Anand and {Rezaie}, Mehdi and {Ross}, Ashley J. and {Rossi}, Graziano and {Ruhlmann-Kleider}, Vanina and {Smith}, Alex and {Tamone}, Am{\'e}lie and {Tinker}, Jeremy L. and {Tojeiro}, Rita and {Wang}, Yuting and {Zhao}, Gong-Bo and {Zhao}, Cheng and {Brinkmann}, Jonathan and {Brownstein}, Joel R. and {Choi}, Peter D. and {Escoffier}, Stephanie and {de la Macorra}, Axel and {Moon}, Jeongin and {Newman}, Jeffrey A. and {Schneider}, Donald P. and {Seo}, Hee-Jong and {Vivek}, Mariappan},
        title = "{The Completed SDSS-IV extended Baryon Oscillation Spectroscopic Survey: measurement of the BAO and growth rate of structure of the luminous red galaxy sample from the anisotropic power spectrum between redshifts 0.6 and 1.0}",
      journal = {\mnras},
         year = 2020,
        month = oct,
       volume = {498},
       number = {2},
        pages = {2492-2531},
          doi = {10.1093/mnras/staa2455},
archivePrefix = {arXiv},
       eprint = {2007.08994},
 primaryClass = {astro-ph.CO},
       adsurl = {https://ui.adsabs.harvard.edu/abs/2020MNRAS.498.2492G}
}

@article{Scoccimarro:2015bla,
    author = "Scoccimarro, Roman",
    title = "{Fast Estimators for Redshift-Space Clustering}",
    eprint = "1506.02729",
    archivePrefix = "arXiv",
    primaryClass = "astro-ph.CO",
    doi = "10.1103/PhysRevD.92.083532",
    journal = "Phys. Rev. D",
    volume = "92",
    number = "8",
    pages = "083532",
    year = "2015"
}

@software{2015ascl.soft12017H,
       author = {{Hamilton}, Andrew J.~S.},
        title = "{FFTLog: Fast Fourier or Hankel transform}",
 howpublished = {Astrophysics Source Code Library, record ascl:1512.017},
         year = 2015,
        month = dec,
          eid = {ascl:1512.017},
archivePrefix = {ascl},
       eprint = {1512.017},
       adsurl = {https://ui.adsabs.harvard.edu/abs/2015ascl.soft12017H}
}

@article{Hand:2017irw,
    author = "Hand, Nick and Li, Yin and Slepian, Zachary and Seljak, Uros",
    title = "{An optimal FFT-based anisotropic power spectrum estimator}",
    eprint = "1704.02357",
    archivePrefix = "arXiv",
    primaryClass = "astro-ph.CO",
    doi = "10.1088/1475-7516/2017/07/002",
    journal = "JCAP",
    volume = "07",
    pages = "002",
    year = "2017"
}

@article{Yamamoto:2005dz,
    author = "Yamamoto, Kazuhiro and Nakamichi, Masashi and Kamino, Akinari and Bassett, Bruce A. and Nishioka, Hiroaki",
    title = "{A Measurement of the quadrupole power spectrum in the clustering of the 2dF QSO Survey}",
    eprint = "astro-ph/0505115",
    archivePrefix = "arXiv",
    doi = "10.1093/pasj/58.1.93",
    journal = "Publ. Astron. Soc. Jap.",
    volume = "58",
    pages = "93--102",
    year = "2006"
}

@ARTICLE{2026arXiv260504864X,
       author = {{Xie}, Yunchen and {Zhao}, Ruiyang and {Gu}, Gan and {Wang}, Xiaoma and {Mu}, Xiaoyong and {Wang}, Yuting and {Zhao}, Gong-Bo and {Beutler}, Florian and {Peacock}, John A.},
        title = "{Efficient estimators for power spectrum and bispectrum multipole measurements}",
      journal = {arXiv e-prints},
         year = 2026,
        month = may,
          eid = {arXiv:2605.04864},
        pages = {arXiv:2605.04864},
          doi = {10.48550/arXiv.2605.04864},
archivePrefix = {arXiv},
       eprint = {2605.04864},
 primaryClass = {astro-ph.CO},
       adsurl = {https://ui.adsabs.harvard.edu/abs/2026arXiv260504864X}
}

@article{Reimberg:2015jma,
    author = "Reimberg, Paulo H. F. and Bernardeau, Francis and Pitrou, Cyril",
    title = "{Redshift-space distortions with wide angular separations}",
    eprint = "1506.06596",
    archivePrefix = "arXiv",
    primaryClass = "astro-ph.CO",
    doi = "10.1088/1475-7516/2016/01/048",
    journal = "JCAP",
    volume = "01",
    pages = "048",
    year = "2016"
}

@article{Castorina:2019hyr,
    author = "Castorina, Emanuele and White, Martin",
    title = "{Wide angle effects for peculiar velocities}",
    eprint = "1911.08353",
    archivePrefix = "arXiv",
    primaryClass = "astro-ph.CO",
    reportNumber = "CERN-TH-2019-191",
    doi = "10.1093/mnras/staa2129",
    journal = "Mon. Not. Roy. Astron. Soc.",
    volume = "499",
    number = "1",
    pages = "893--905",
    year = "2020"
}

@article{Philcox:2021tfv,
    author = "Philcox, Oliver H. E. and Slepian, Zachary",
    title = "{Beyond the Yamamoto approximation: Anisotropic power spectra and correlation functions with pairwise lines of sight}",
    eprint = "2102.08384",
    archivePrefix = "arXiv",
    primaryClass = "astro-ph.CO",
    doi = "10.1103/PhysRevD.103.123509",
    journal = "Phys. Rev. D",
    volume = "103",
    number = "12",
    pages = "123509",
    year = "2021"
}

@article{Castorina:2017inr,
    author = "Castorina, Emanuele and White, Martin",
    title = "{Beyond the plane-parallel approximation for redshift surveys}",
    eprint = "1709.09730",
    archivePrefix = "arXiv",
    primaryClass = "astro-ph.CO",
    doi = "10.1093/mnras/sty410",
    journal = "Mon. Not. Roy. Astron. Soc.",
    volume = "476",
    number = "4",
    pages = "4403--4417",
    year = "2018"
}

@article{Beutler:2018vpe,
    author = "Beutler, Florian and Castorina, Emanuele and Zhang, Pierre",
    title = "{Interpreting measurements of the anisotropic galaxy power spectrum}",
    eprint = "1810.05051",
    archivePrefix = "arXiv",
    primaryClass = "astro-ph.CO",
    doi = "10.1088/1475-7516/2019/03/040",
    journal = "JCAP",
    volume = "03",
    pages = "040",
    year = "2019"
}

@article{Castorina:2021xzs,
    author = "Castorina, Emanuele and di Dio, Enea",
    title = "{The observed galaxy power spectrum in General Relativity}",
    eprint = "2106.08857",
    archivePrefix = "arXiv",
    primaryClass = "astro-ph.CO",
    reportNumber = "CERN-TH-2021-087",
    doi = "10.1088/1475-7516/2022/01/061",
    journal = "JCAP",
    volume = "01",
    number = "01",
    pages = "061",
    year = "2022"
}

@ARTICLE{2020JCAP...11..064G,
       author = {{Grimm}, Nastassia and {Scaccabarozzi}, Fulvio and {Yoo}, Jaiyul and {Biern}, Sang Gyu and {Gong}, Jinn-Ouk},
        title = "{Galaxy power spectrum in general relativity}",
      journal = {\jcap},
         year = 2020,
        month = nov,
       volume = {2020},
       number = {11},
          eid = {064},
        pages = {064},
          doi = {10.1088/1475-7516/2020/11/064},
archivePrefix = {arXiv},
       eprint = {2005.06484},
 primaryClass = {astro-ph.CO},
       adsurl = {https://ui.adsabs.harvard.edu/abs/2020JCAP...11..064G}
}

@ARTICLE{1995ApJ...455....7M,
       author = {{Ma}, Chung-Pei and {Bertschinger}, Edmund},
        title = "{Cosmological Perturbation Theory in the Synchronous and Conformal Newtonian Gauges}",
      journal = {\apj},
         year = 1995,
        month = dec,
       volume = {455},
        pages = {7},
          doi = {10.1086/176550},
archivePrefix = {arXiv},
       eprint = {astro-ph/9506072},
 primaryClass = {astro-ph},
       adsurl = {https://ui.adsabs.harvard.edu/abs/1995ApJ...455....7M}
}

@ARTICLE{2018JCAP...10..024S,
       author = {{Scaccabarozzi}, Fulvio and {Yoo}, Jaiyul and {Biern}, Sang Gyu},
        title = "{Galaxy two-point correlation function in general relativity}",
      journal = {\jcap},
         year = 2018,
        month = oct,
       volume = {2018},
       number = {10},
          eid = {024},
        pages = {024},
          doi = {10.1088/1475-7516/2018/10/024},
archivePrefix = {arXiv},
       eprint = {1807.09796},
 primaryClass = {astro-ph.CO},
       adsurl = {https://ui.adsabs.harvard.edu/abs/2018JCAP...10..024S}
}

@ARTICLE{Benabou:2024tmn,
       author = {{Benabou}, Joshua N. and {Sands}, Isabel and {Gebhardt}, Henry S. Grasshorn and {Heinrich}, Chen and {Dor{\'e}}, Olivier},
        title = "{Wide-angle effects in the power spectrum multipoles in next-generation redshift surveys}",
      journal = {\prd},
         year = 2024,
        month = oct,
       volume = {110},
       number = {8},
          eid = {083526},
        pages = {083526},
          doi = {10.1103/PhysRevD.110.083526},
archivePrefix = {arXiv},
       eprint = {2404.04811},
 primaryClass = {astro-ph.CO},
       adsurl = {https://ui.adsabs.harvard.edu/abs/2024PhRvD.110h3526B}
}

@ARTICLE{2002ApJ...567....2H,
       author = {{Hivon}, Eric and {G{\'o}rski}, Krzysztof M. and {Netterfield}, C. Barth and {Crill}, Brendan P. and {Prunet}, Simon and {Hansen}, Frode},
        title = "{MASTER of the Cosmic Microwave Background Anisotropy Power Spectrum: A Fast Method for Statistical Analysis of Large and Complex Cosmic Microwave Background Data Sets}",
      journal = {\apj},
         year = 2002,
        month = mar,
       volume = {567},
       number = {1},
        pages = {2-17},
          doi = {10.1086/338126},
archivePrefix = {arXiv},
       eprint = {astro-ph/0105302},
 primaryClass = {astro-ph},
       adsurl = {https://ui.adsabs.harvard.edu/abs/2002ApJ...567....2H}
}

@article{Wen:2024hqj,
    author = "Wen, Robin Y. and Gebhardt, Henry S. Grasshorn and Heinrich, Chen and Dor\'e, Olivier",
    title = "{Exact Modeling of Power Spectrum Multipole through Spherical Fourier-Bessel Basis}",
    eprint = "2404.04812",
    archivePrefix = "arXiv",
    primaryClass = "astro-ph.CO",
    month = "4",
    year = "2024"
}

@article{Beutler:2020evf,
    author = "Beutler, Florian and Di Dio, Enea",
    title = "{Modeling relativistic contributions to the halo power spectrum dipole}",
    eprint = "2004.08014",
    archivePrefix = "arXiv",
    primaryClass = "astro-ph.CO",
    doi = "10.1088/1475-7516/2020/07/048",
    journal = "JCAP",
    volume = "07",
    number = "07",
    pages = "048",
    year = "2020"
}

@ARTICLE{2012JCAP...10..025B,
       author = {{Bertacca}, Daniele and {Maartens}, Roy and {Raccanelli}, Alvise and {Clarkson}, Chris},
        title = "{Beyond the plane-parallel and Newtonian approach: wide-angle redshift distortions and convergence in general relativity}",
      journal = {\jcap},
         year = 2012,
        month = oct,
       volume = {2012},
       number = {10},
          eid = {025},
        pages = {025},
          doi = {10.1088/1475-7516/2012/10/025},
archivePrefix = {arXiv},
       eprint = {1205.5221},
 primaryClass = {astro-ph.CO},
       adsurl = {https://ui.adsabs.harvard.edu/abs/2012JCAP...10..025B}
}

@ARTICLE{2018JCAP...03..019T,
       author = {{Tansella}, Vittorio and {Bonvin}, Camille and {Durrer}, Ruth and {Ghosh}, Basundhara and {Sellentin}, Elena},
        title = "{The full-sky relativistic correlation function and power spectrum of galaxy number counts. Part I: theoretical aspects}",
      journal = {\jcap},
         year = 2018,
        month = mar,
       volume = {2018},
       number = {3},
          eid = {019},
        pages = {019},
          doi = {10.1088/1475-7516/2018/03/019},
archivePrefix = {arXiv},
       eprint = {1708.00492},
 primaryClass = {astro-ph.CO},
       adsurl = {https://ui.adsabs.harvard.edu/abs/2018JCAP...03..019T}
}

@ARTICLE{2018MNRAS.479..741C,
       author = {{Castorina}, Emanuele and {White}, Martin},
        title = "{The Zeldovich approximation and wide-angle redshift-space distortions}",
      journal = {\mnras},
         year = 2018,
        month = sep,
       volume = {479},
       number = {1},
        pages = {741-752},
          doi = {10.1093/mnras/sty1437},
archivePrefix = {arXiv},
       eprint = {1803.08185},
 primaryClass = {astro-ph.CO},
       adsurl = {https://ui.adsabs.harvard.edu/abs/2018MNRAS.479..741C}
}

@ARTICLE{2024PhRvD.110f3546M,
       author = {{Matsubara}, Takahiko},
        title = "{Integrated perturbation theory for cosmological tensor fields. IV. Full-sky formulation}",
      journal = {\prd},
         year = 2024,
        month = sep,
       volume = {110},
       number = {6},
          eid = {063546},
        pages = {063546},
          doi = {10.1103/PhysRevD.110.063546},
archivePrefix = {arXiv},
       eprint = {2405.09038},
 primaryClass = {astro-ph.CO},
       adsurl = {https://ui.adsabs.harvard.edu/abs/2024PhRvD.110f3546M}
}

@ARTICLE{2009JCAP...11..026M,
       author = {{McDonald}, Patrick},
        title = "{Gravitational redshift and other redshift-space distortions of the imaginary part of the power spectrum}",
      journal = {\jcap},
         year = 2009,
        month = nov,
       volume = {2009},
       number = {11},
          eid = {026},
        pages = {026},
          doi = {10.1088/1475-7516/2009/11/026},
archivePrefix = {arXiv},
       eprint = {0907.5220},
 primaryClass = {astro-ph.CO},
       adsurl = {https://ui.adsabs.harvard.edu/abs/2009JCAP...11..026M}
}

@ARTICLE{2025JCAP...07..063G,
       author = {{Guedezounme}, S{\^e}cloka L. and {Jolicoeur}, Sheean and {Maartens}, Roy},
        title = "{Primordial non-Gaussianity {\textemdash} the effects of relativistic and wide-angle corrections to the power spectrum}",
      journal = {\jcap},
         year = 2025,
        month = jul,
       volume = {2025},
       number = {7},
          eid = {063},
        pages = {063},
          doi = {10.1088/1475-7516/2025/07/063},
archivePrefix = {arXiv},
       eprint = {2412.06553},
 primaryClass = {astro-ph.CO},
       adsurl = {https://ui.adsabs.harvard.edu/abs/2025JCAP...07..063G}
}

@ARTICLE{2021PhRvD.103l3534S,
       author = {{Shiraishi}, Maresuke and {Akitsu}, Kazuyuki and {Okumura}, Teppei},
        title = "{Alcock-Paczynski effects on wide-angle galaxy statistics}",
      journal = {\prd},
         year = 2021,
        month = jun,
       volume = {103},
       number = {12},
          eid = {123534},
        pages = {123534},
          doi = {10.1103/PhysRevD.103.123534},
archivePrefix = {arXiv},
       eprint = {2103.08126},
 primaryClass = {astro-ph.CO},
       adsurl = {https://ui.adsabs.harvard.edu/abs/2021PhRvD.103l3534S}
}

@ARTICLE{2023JCAP...09..030P,
       author = {{Pardede}, Kevin and {Di Dio}, Enea and {Castorina}, Emanuele},
        title = "{Wide-angle effects in the galaxy bispectrum}",
      journal = {\jcap},
         year = 2023,
        month = sep,
       volume = {2023},
       number = {9},
          eid = {030},
        pages = {030},
          doi = {10.1088/1475-7516/2023/09/030},
archivePrefix = {arXiv},
       eprint = {2302.12789},
 primaryClass = {astro-ph.CO},
       adsurl = {https://ui.adsabs.harvard.edu/abs/2023JCAP...09..030P}
}

@ARTICLE{2006MNRAS.370..343E,
       author = {{Efstathiou}, G.},
        title = "{Hybrid estimation of cosmic microwave background polarization power spectra}",
      journal = {\mnras},
         year = 2006,
        month = jul,
       volume = {370},
       number = {1},
        pages = {343-362},
          doi = {10.1111/j.1365-2966.2006.10486.x},
archivePrefix = {arXiv},
       eprint = {astro-ph/0601107},
 primaryClass = {astro-ph},
       adsurl = {https://ui.adsabs.harvard.edu/abs/2006MNRAS.370..343E}
}

@article{Tegmark97,
  title = {How to measure CMB power spectra without losing information},
  author = {Tegmark, Max},
  journal = {Phys. Rev. D},
  volume = {55},
  issue = {10},
  pages = {5895--5907},
  numpages = {0},
  year = {1997},
  month = {May},
  publisher = {American Physical Society},
  doi = {10.1103/PhysRevD.55.5895},
  url = {https://link.aps.org/doi/10.1103/PhysRevD.55.5895}
}

@ARTICLE{2017JCAP...02..040T,
       author = {{Terrana}, Alexandra and {Harris}, Mary-Jean and {Johnson}, Matthew C.},
        title = "{Analyzing the cosmic variance limit of remote dipole measurements of the cosmic microwave background using the large-scale kinetic Sunyaev Zel'dovich effect}",
      journal = {\jcap},
         year = 2017,
        month = feb,
       volume = {2017},
       number = {2},
          eid = {040},
        pages = {040},
          doi = {10.1088/1475-7516/2017/02/040},
archivePrefix = {arXiv},
       eprint = {1610.06919},
 primaryClass = {astro-ph.CO},
       adsurl = {https://ui.adsabs.harvard.edu/abs/2017JCAP...02..040T}
}

@ARTICLE{2018PhRvD..98l3501D,
       author = {{Deutsch}, Anne-Sylvie and {Dimastrogiovanni}, Emanuela and {Johnson}, Matthew C. and {M{\"u}nchmeyer}, Moritz and {Terrana}, Alexandra},
        title = "{Reconstruction of the remote dipole and quadrupole fields from the kinetic Sunyaev Zel'dovich and polarized Sunyaev Zel'dovich effects}",
      journal = {\prd},
         year = 2018,
        month = dec,
       volume = {98},
       number = {12},
          eid = {123501},
        pages = {123501},
          doi = {10.1103/PhysRevD.98.123501},
archivePrefix = {arXiv},
       eprint = {1707.08129},
 primaryClass = {astro-ph.CO},
       adsurl = {https://ui.adsabs.harvard.edu/abs/2018PhRvD..98l3501D}
}

@ARTICLE{2023JCAP...11..097Z,
       author = {{Zhou}, Rongpu and {Ferraro}, Simone and {White}, Martin and {DeRose}, Joseph and {Sailer}, Noah and {Aguilar}, Jessica and {Ahlen}, Steven and {Bailey}, Stephen and {Brooks}, David and {Claybaugh}, Todd and {Dawson}, Kyle and {de la Macorra}, Axel and {Dey}, Biprateep and {Doel}, Peter and {Font-Ribera}, Andreu and {Forero-Romero}, Jaime E. and {Gontcho A Gontcho}, Satya and {Guy}, Julien and {Kremin}, Anthony and {Lambert}, Andrew and {Le Guillou}, Laurent and {Levi}, Michael and {Magneville}, Christophe and {Manera}, Marc and {Meisner}, Aaron and {Miquel}, Ramon and {Moustakas}, John and {Myers}, Adam D. and {Newman}, Jeffrey A. and {Nie}, Jundan and {Percival}, Will and {Rezaie}, Mehdi and {Rossi}, Graziano and {Sanchez}, Eusebio and {Schlegel}, David and {Schubnell}, Michael and {Seo}, Hee-Jong and {Tarl{\'e}}, Gregory and {Zhou}, Zhimin},
        title = "{DESI luminous red galaxy samples for cross-correlations}",
      journal = {\jcap},
         year = 2023,
        month = nov,
       volume = {2023},
       number = {11},
          eid = {097},
        pages = {097},
          doi = {10.1088/1475-7516/2023/11/097},
archivePrefix = {arXiv},
       eprint = {2309.06443},
 primaryClass = {astro-ph.CO},
       adsurl = {https://ui.adsabs.harvard.edu/abs/2023JCAP...11..097Z}
}

@ARTICLE{2025JCAP...03..059F,
       author = {{Friedman-Shaw}, Batia and {Krolewski}, Alex and {Foglieni}, Matteo and {Afshordi}, Niayesh},
        title = "{Doppler bias: impact of peculiar velocities on color selection and the large scale structure of galaxy surveys}",
      journal = {\jcap},
         year = 2025,
        month = mar,
       volume = {2025},
       number = {3},
          eid = {059},
        pages = {059},
          doi = {10.1088/1475-7516/2025/03/059},
archivePrefix = {arXiv},
       eprint = {2410.04705},
 primaryClass = {astro-ph.CO},
       adsurl = {https://ui.adsabs.harvard.edu/abs/2025JCAP...03..059F}
}

@ARTICLE{2023JCAP...04..067P,
       author = {{Paul}, Pritha and {Clarkson}, Chris and {Maartens}, Roy},
        title = "{Wide-angle effects in multi-tracer power spectra with Doppler corrections}",
      journal = {\jcap},
         year = 2023,
        month = apr,
       volume = {2023},
       number = {4},
          eid = {067},
        pages = {067},
          doi = {10.1088/1475-7516/2023/04/067},
archivePrefix = {arXiv},
       eprint = {2208.04819},
 primaryClass = {astro-ph.CO},
       adsurl = {https://ui.adsabs.harvard.edu/abs/2023JCAP...04..067P}
}

@ARTICLE{1992ApJ...385L...5H,
       author = {{Hamilton}, A.~J.~S.},
        title = "{Measuring Omega and the Real Correlation Function from the Redshift Correlation Function}",
      journal = {\apjl},
         year = 1992,
        month = jan,
       volume = {385},
        pages = {L5},
          doi = {10.1086/186264},
       adsurl = {https://ui.adsabs.harvard.edu/abs/1992ApJ...385L...5H}
}

@INPROCEEDINGS{1998ASSL..231..185H,
       author = {{Hamilton}, A.~J.~S.},
        title = "{Linear Redshift Distortions: a Review}",
    booktitle = {The Evolving Universe},
         year = 1998,
       editor = {{Hamilton}, Donald},
       series = {Astrophysics and Space Science Library},
       volume = {231},
        month = jan,
        pages = {185},
          doi = {10.1007/978-94-011-4960-0_17},
archivePrefix = {arXiv},
       eprint = {astro-ph/9708102},
 primaryClass = {astro-ph},
       adsurl = {https://ui.adsabs.harvard.edu/abs/1998ASSL..231..185H}
}

@article{Hamilton:1995px,
    author = "Hamilton, A. J. S. and Culhane, M.",
    title = "{Spherical redshift distortions}",
    eprint = "astro-ph/9507021",
    archivePrefix = "arXiv",
    doi = "10.1093/mnras/278.1.73",
    journal = "Mon. Not. Roy. Astron. Soc.",
    volume = "278",
    pages = "73",
    year = "1996"
}

@ARTICLE{1996ApJ...462...25Z,
       author = {{Zaroubi}, Saleem and {Hoffman}, Yehuda},
        title = "{Clustering in Redshift Space: Linear Theory}",
      journal = {\apj},
         year = 1996,
        month = may,
       volume = {462},
        pages = {25},
          doi = {10.1086/177124},
       adsurl = {https://ui.adsabs.harvard.edu/abs/1996ApJ...462...25Z}
}

@ARTICLE{1987MNRAS.227....1K,
       author = {{Kaiser}, Nick},
        title = "{Clustering in real space and in redshift space}",
      journal = {\mnras},
         year = 1987,
        month = jul,
       volume = {227},
        pages = {1-21},
          doi = {10.1093/mnras/227.1.1},
       adsurl = {https://ui.adsabs.harvard.edu/abs/1987MNRAS.227....1K}
}

@article{Gebhardt:2021mds,
    author = "Gebhardt, Henry S. Grasshorn and Dor{\'e}, Olivier",
    title = "{Fabulous code for spherical Fourier-Bessel decomposition}",
    eprint = "2102.10079",
    archivePrefix = "arXiv",
    primaryClass = "astro-ph.CO",
    doi = "10.1103/PhysRevD.104.123548",
    journal = "Phys. Rev. D",
    volume = "104",
    number = "12",
    pages = "123548",
    year = "2021"
}

@article{GrasshornGebhardt:2023mlv,
    author = "Grasshorn Gebhardt, Henry S. and Dor{\'e}, Olivier",
    title = "{Validation of spherical Fourier-Bessel power spectrum analysis with log-normal simulations and eBOSS DR16 LRG EZmocks}",
    eprint = "2310.17677",
    archivePrefix = "arXiv",
    primaryClass = "astro-ph.IM",
    doi = "10.1103/PhysRevD.109.083502",
    journal = "Phys. Rev. D",
    volume = "109",
    number = "8",
    pages = "083502",
    year = "2024"
}

@article{Benabou:2023ldb,
    author = "Benabou, Joshua N. and Testa, Adriano and Heinrich, Chen and Grasshorn Gebhardt, Henry S. and Dor{\'e}, Olivier",
    title = "{Galaxy bispectrum in the spherical Fourier-Bessel basis}",
    eprint = "2312.15992",
    archivePrefix = "arXiv",
    primaryClass = "astro-ph.CO",
    doi = "10.1103/PhysRevD.109.103507",
    journal = "Phys. Rev. D",
    volume = "109",
    number = "10",
    pages = "103507",
    year = "2024"
}

@article{Gebhardt:2021tme,
    author = "Gebhardt, Henry S. Grasshorn and Dor{\'e}, Olivier",
    title = "{Harmonic analysis of isotropic fields on the sphere with arbitrary masks}",
    eprint = "2109.13352",
    archivePrefix = "arXiv",
    primaryClass = "astro-ph.CO",
    doi = "10.1088/1475-7516/2022/01/038",
    journal = "JCAP",
    volume = "01",
    number = "01",
    pages = "038",
    year = "2022"
}

@ARTICLE{2004ApJ...614...51S,
       author = {{Szapudi}, Istv{\'a}n},
        title = "{Wide-Angle Redshift Distortions Revisited}",
      journal = {\apj},
         year = 2004,
        month = oct,
       volume = {614},
       number = {1},
        pages = {51-55},
          doi = {10.1086/423168},
archivePrefix = {arXiv},
       eprint = {astro-ph/0404477},
 primaryClass = {astro-ph},
       adsurl = {https://ui.adsabs.harvard.edu/abs/2004ApJ...614...51S}
}

@ARTICLE{2008MNRAS.389..292P,
       author = {{P{\'a}pai}, P{\'e}ter and {Szapudi}, Istv{\'a}n},
        title = "{Non-perturbative effects of geometry in wide-angle redshift distortions}",
      journal = {\mnras},
         year = 2008,
        month = sep,
       volume = {389},
       number = {1},
        pages = {292-296},
          doi = {10.1111/j.1365-2966.2008.13572.x},
archivePrefix = {arXiv},
       eprint = {0802.2940},
 primaryClass = {astro-ph},
       adsurl = {https://ui.adsabs.harvard.edu/abs/2008MNRAS.389..292P}
}

@article{Lilje1989,
    author = {Lilje, Per B. and Efstathiou, G.},
    title = {Gravitationally induced velocity fields in the Universe – I. Correlation functions},
    journal = {Monthly Notices of the Royal Astronomical Society},
    volume = {236},
    number = {4},
    pages = {851-864},
    year = {1989},
    month = {02},
    issn = {0035-8711},
    doi = {10.1093/mnras/236.4.851},
    url = {https://doi.org/10.1093/mnras/236.4.851},
    eprint = {https://academic.oup.com/mnras/article-pdf/236/4/851/3924694/mnras236-0851.pdf},
}

@ARTICLE{1990MNRAS.242..428M,
       author = {{McGill}, Colin},
        title = "{The redshift projection. I - Caustics and correlation functions}",
      journal = {\mnras},
         year = 1990,
        month = feb,
       volume = {242},
        pages = {428-438},
          doi = {10.1093/mnras/242.3.428},
       adsurl = {https://ui.adsabs.harvard.edu/abs/1990MNRAS.242..428M}
}

@ARTICLE{APeffect,
       author = {{Alcock}, C. and {Paczynski}, B.},
        title = "{An evolution free test for non-zero cosmological constant}",
      journal = {\nat},
         year = 1979,
        month = oct,
       volume = {281},
        pages = {358},
          doi = {10.1038/281358a0},
       adsurl = {https://ui.adsabs.harvard.edu/abs/1979Natur.281..358A}
}

@ARTICLE{1994MNRAS.266..219F,
       author = {{Fisher}, Karl B. and {Scharf}, Caleb A. and {Lahav}, Ofer},
        title = "{A spherical harmonic approach to redshift distortion and a measurement of Omega(0) from the 1.2-Jy IRAS Redshift Survey}",
      journal = {\mnras},
         year = 1994,
        month = jan,
       volume = {266},
        pages = {219},
          doi = {10.1093/mnras/266.1.219},
archivePrefix = {arXiv},
       eprint = {astro-ph/9309027},
 primaryClass = {astro-ph},
       adsurl = {https://ui.adsabs.harvard.edu/abs/1994MNRAS.266..219F}
}

@article{Bharadwaj_1999,
doi = {10.1086/307118},
url = {https://doi.org/10.1086/307118},
year = {1999},
month = {may},
publisher = {},
volume = {516},
number = {2},
pages = {507},
author = {Bharadwaj, Somnath},
title = {Radial Redshift Space Distortions},
journal = {The Astrophysical Journal},
}

@ARTICLE{2015MNRAS.452.3704S,
       author = {{Samushia}, L. and {Branchini}, E. and {Percival}, W.~J.},
        title = "{Geometric biases in power-spectrum measurements}",
      journal = {\mnras},
         year = 2015,
        month = oct,
       volume = {452},
       number = {4},
        pages = {3704-3709},
          doi = {10.1093/mnras/stv1568},
archivePrefix = {arXiv},
       eprint = {1504.02135},
 primaryClass = {astro-ph.CO},
       adsurl = {https://ui.adsabs.harvard.edu/abs/2015MNRAS.452.3704S}
}

@ARTICLE{2010MNRAS.409.1525R,
       author = {{Raccanelli}, Alvise and {Samushia}, Lado and {Percival}, Will J.},
        title = "{Simulating redshift-space distortions for galaxy pairs with wide angular separation}",
      journal = {\mnras},
         year = 2010,
        month = dec,
       volume = {409},
       number = {4},
        pages = {1525-1533},
          doi = {10.1111/j.1365-2966.2010.17388.x},
archivePrefix = {arXiv},
       eprint = {1006.1652},
 primaryClass = {astro-ph.CO},
       adsurl = {https://ui.adsabs.harvard.edu/abs/2010MNRAS.409.1525R}
}

@article{PhysRevD.78.103512,
  title = {Nonlinear redshift-space power spectra},
  author = {Shaw, J. Richard and Lewis, Antony},
  journal = {Phys. Rev. D},
  volume = {78},
  issue = {10},
  pages = {103512},
  numpages = {16},
  year = {2008},
  month = {Nov},
  publisher = {American Physical Society},
  doi = {10.1103/PhysRevD.78.103512},
  url = {https://link.aps.org/doi/10.1103/PhysRevD.78.103512}
}

@ARTICLE{2014PhRvD..89h3535B,
       author = {{Bonvin}, Camille and {Hui}, Lam and {Gazta{\~n}aga}, Enrique},
        title = "{Asymmetric galaxy correlation functions}",
      journal = {\prd},
         year = 2014,
        month = apr,
       volume = {89},
       number = {8},
          eid = {083535},
        pages = {083535},
          doi = {10.1103/PhysRevD.89.083535},
archivePrefix = {arXiv},
       eprint = {1309.1321},
 primaryClass = {astro-ph.CO},
       adsurl = {https://ui.adsabs.harvard.edu/abs/2014PhRvD..89h3535B}
}

@ARTICLE{2011PhRvD..84f3505B,
       author = {{Bonvin}, Camille and {Durrer}, Ruth},
        title = "{What galaxy surveys really measure}",
      journal = {\prd},
         year = 2011,
        month = sep,
       volume = {84},
       number = {6},
          eid = {063505},
        pages = {063505},
          doi = {10.1103/PhysRevD.84.063505},
archivePrefix = {arXiv},
       eprint = {1105.5280},
 primaryClass = {astro-ph.CO},
       adsurl = {https://ui.adsabs.harvard.edu/abs/2011PhRvD..84f3505B}
}

@ARTICLE{2014JCAP...08..022R,
       author = {{Raccanelli}, Alvise and {Bertacca}, Daniele and {Dor{\'e}}, Olivier and {Maartens}, Roy},
        title = "{Large-scale 3D galaxy correlation function and non-Gaussianity}",
      journal = {\jcap},
         year = 2014,
        month = aug,
       volume = {2014},
       number = {8},
        pages = {022-022},
          doi = {10.1088/1475-7516/2014/08/022},
archivePrefix = {arXiv},
       eprint = {1306.6646},
 primaryClass = {astro-ph.CO},
       adsurl = {https://ui.adsabs.harvard.edu/abs/2014JCAP...08..022R}
}

@ARTICLE{2015MNRAS.447.1789Y,
       author = {{Yoo}, Jaiyul and {Seljak}, Uro{\v{s}}},
        title = "{Wide-angle effects in future galaxy surveys}",
      journal = {\mnras},
         year = 2015,
        month = feb,
       volume = {447},
       number = {2},
        pages = {1789-1805},
          doi = {10.1093/mnras/stu2491},
archivePrefix = {arXiv},
       eprint = {1308.1093},
 primaryClass = {astro-ph.CO},
       adsurl = {https://ui.adsabs.harvard.edu/abs/2015MNRAS.447.1789Y}
}

@ARTICLE{2017MNRAS.466.2242B,
       author = {{Beutler}, Florian and {Seo}, Hee-Jong and {Saito}, Shun and {Chuang}, Chia-Hsun and {Cuesta}, Antonio J. and {Eisenstein}, Daniel J. and {Gil-Mar{\'\i}n}, H{\'e}ctor and {Grieb}, Jan Niklas and {Hand}, Nick and {Kitaura}, Francisco-Shu and {Modi}, Chirag and {Nichol}, Robert C. and {Olmstead}, Matthew D. and {Percival}, Will J. and {Prada}, Francisco and {S{\'a}nchez}, Ariel G. and {Rodriguez-Torres}, Sergio and {Ross}, Ashley J. and {Ross}, Nicholas P. and {Schneider}, Donald P. and {Tinker}, Jeremy and {Tojeiro}, Rita and {Vargas-Maga{\~n}a}, Mariana},
        title = "{The clustering of galaxies in the completed SDSS-III Baryon Oscillation Spectroscopic Survey: anisotropic galaxy clustering in Fourier space}",
      journal = {\mnras},
         year = 2017,
        month = apr,
       volume = {466},
       number = {2},
        pages = {2242-2260},
          doi = {10.1093/mnras/stw3298},
archivePrefix = {arXiv},
       eprint = {1607.03150},
 primaryClass = {astro-ph.CO},
       adsurl = {https://ui.adsabs.harvard.edu/abs/2017MNRAS.466.2242B}
}

@ARTICLE{2017MNRAS.464.3121W,
       author = {{Wilson}, M.~J. and {Peacock}, J.~A. and {Taylor}, A.~N. and {de la Torre}, S.},
        title = "{Rapid modelling of the redshift-space power spectrum multipoles for a masked density field}",
      journal = {\mnras},
         year = 2017,
        month = jan,
       volume = {464},
       number = {3},
        pages = {3121-3130},
          doi = {10.1093/mnras/stw2576},
archivePrefix = {arXiv},
       eprint = {1511.07799},
 primaryClass = {astro-ph.CO},
       adsurl = {https://ui.adsabs.harvard.edu/abs/2017MNRAS.464.3121W}
}

@ARTICLE{2023PhRvL.131k1201F,
       author = {{Foglieni}, Matteo and {Pantiri}, Mattia and {Di Dio}, Enea and {Castorina}, Emanuele},
        title = "{Large Scale Limit of the Observed Galaxy Power Spectrum}",
      journal = {\prl},
         year = 2023,
        month = sep,
       volume = {131},
       number = {11},
          eid = {111201},
        pages = {111201},
          doi = {10.1103/PhysRevLett.131.111201},
archivePrefix = {arXiv},
       eprint = {2303.03142},
 primaryClass = {astro-ph.CO},
       adsurl = {https://ui.adsabs.harvard.edu/abs/2023PhRvL.131k1201F}
}

@ARTICLE{2009PhRvD..80h3514Y,
       author = {{Yoo}, Jaiyul and {Fitzpatrick}, A. Liam and {Zaldarriaga}, Matias},
        title = "{New perspective on galaxy clustering as a cosmological probe: General relativistic effects}",
      journal = {\prd},
         year = 2009,
        month = oct,
       volume = {80},
       number = {8},
          eid = {083514},
        pages = {083514},
          doi = {10.1103/PhysRevD.80.083514},
archivePrefix = {arXiv},
       eprint = {0907.0707},
 primaryClass = {astro-ph.CO},
       adsurl = {https://ui.adsabs.harvard.edu/abs/2009PhRvD..80h3514Y}
}

@ARTICLE{2010PhRvD..82h3508Y,
       author = {{Yoo}, Jaiyul},
        title = "{General relativistic description of the observed galaxy power spectrum: Do we understand what we measure?}",
      journal = {\prd},
         year = 2010,
        month = oct,
       volume = {82},
       number = {8},
          eid = {083508},
        pages = {083508},
          doi = {10.1103/PhysRevD.82.083508},
archivePrefix = {arXiv},
       eprint = {1009.3021},
 primaryClass = {astro-ph.CO},
       adsurl = {https://ui.adsabs.harvard.edu/abs/2010PhRvD..82h3508Y}
}

@ARTICLE{2011PhRvD..84d3516C,
       author = {{Challinor}, Anthony and {Lewis}, Antony},
        title = "{Linear power spectrum of observed source number counts}",
      journal = {\prd},
         year = 2011,
        month = aug,
       volume = {84},
       number = {4},
          eid = {043516},
        pages = {043516},
          doi = {10.1103/PhysRevD.84.043516},
archivePrefix = {arXiv},
       eprint = {1105.5292},
 primaryClass = {astro-ph.CO},
       adsurl = {https://ui.adsabs.harvard.edu/abs/2011PhRvD..84d3516C}
}

@ARTICLE{2025arXiv251115701M,
       author = {{McCarthy}, Fiona and {Hadzhiyska}, Boryana and {Bond}, J. Richard and {Coulton}, William R. and {Dunkley}, Jo and {Embil Villagra}, Carmen and {Johnson}, Matthew C. and {Moodley}, Kavilan and {Namikawa}, Toshiya and {Ried Guachalla}, Bernardita and {Sherwin}, Blake D. and {Sif{\'o}n}, Crist{\'o}bal and {van Engelen}, Alexander and {Vavagiakis}, Eve M. and {Wollack}, Edward J.},
        title = "{The Atacama Cosmology Telescope: Cross-correlation of kSZ and continuity equation velocity reconstruction with photometric DESI LRGs}",
      journal = {arXiv e-prints},
         year = 2025,
        month = nov,
          eid = {arXiv:2511.15701},
        pages = {arXiv:2511.15701},
          doi = {10.48550/arXiv.2511.15701},
archivePrefix = {arXiv},
       eprint = {2511.15701},
 primaryClass = {astro-ph.CO},
       adsurl = {https://ui.adsabs.harvard.edu/abs/2025arXiv251115701M}
}

@ARTICLE{2025arXiv250621684L,
       author = {{Lai}, Anderson C.~M. and {Kvasiuk}, Yurii and {M{\"u}nchmeyer}, Moritz},
        title = "{KSZ Velocity Reconstruction with ACT and DESI-LS using a Tomographic QML Power Spectrum Estimator}",
      journal = {arXiv e-prints},
         year = 2025,
        month = jun,
          eid = {arXiv:2506.21684},
        pages = {arXiv:2506.21684},
          doi = {10.48550/arXiv.2506.21684},
archivePrefix = {arXiv},
       eprint = {2506.21684},
 primaryClass = {astro-ph.CO},
       adsurl = {https://ui.adsabs.harvard.edu/abs/2025arXiv250621684L}
}

@ARTICLE{2025JCAP...08..034A,
       author = {{Abitbol}, M. and {Abril-Cabezas}, I. and {Adachi}, S. and {Ade}, P. and {Adler}, A.~E. and {Agrawal}, P. and {Aguirre}, J. and {Ahmed}, Z. and {Aiola}, S. and {Alford}, T. and {Ali}, A. and {Alonso}, D. and {Alvarez}, M.~A. and {An}, R. and {Arnold}, K. and {Ashton}, P. and {Atkins}, Z. and {Austermann}, J. and {Azzoni}, S. and {Baccigalupi}, C. and {Baleato Lizancos}, A. and {Barron}, D. and {Barry}, P. and {Bartlett}, J. and {Battaglia}, N. and {Battye}, R. and {Baxter}, E. and {Bazarko}, A. and {Beall}, J.~A. and {Bean}, R. and {Beck}, D. and {Beckman}, S. and {Begin}, J. and {Beheshti}, A. and {Beringue}, B. and {Bhandarkar}, T. and {Bhimani}, S. and {Bianchini}, F. and {Biermann}, E. and {Biquard}, S. and {Bixler}, B. and {Boada}, S. and {Boettger}, D. and {Bolliet}, B. and {Bond}, J.~R. and {Borrill}, J. and {Borrow}, J. and {Braithwaite}, C. and {Brien}, T.~L.~R. and {Brown}, M.~L. and {Bruno}, S.~M. and {Bryan}, S. and {Bustos}, R. and {Cai}, H. and {Calabrese}, E. and {Calafut}, V. and {Carl}, F.~M. and {Carones}, A. and {Carron}, J. and {Challinor}, A. and {Chanial}, P. and {Chen}, N. and {Cheung}, K. and {Chiang}, B. and {Chinone}, Y. and {Chluba}, J. and {Cho}, H.~S. and {Choi}, S.~K. and {Chu}, M. and {Clancy}, J. and {Clark}, S.~E. and {Clarke}, P. and {Cleary}, J. and {Clements}, D.~L. and {Connors}, J. and {Contaldi}, C. and {Coppi}, G. and {Corbett}, L. and {Cothard}, N.~F. and {Coulton}, W. and {Crowley}, K.~D. and {Crowley}, K.~T. and {Cukierman}, A. and {D'Ewart}, J.~M. and {Dachlythra}, K. and {Datta}, R. and {Day-Weiss}, S. and {de Haan}, T. and {Devlin}, M. and {Di Mascolo}, L. and {Dicker}, S. and {Dober}, B. and {Doux}, C. and {Dow}, P. and {Doyle}, S. and {Duell}, C.~J. and {Duff}, S.~M. and {Duivenvoorden}, A.~J. and {Dunkley}, J. and {Dutcher}, D. and {D{\"u}nner}, R. and {Edenton}, M. and {El Bouhargani}, H. and {Errard}, J. and {Fabbian}, G. and {Fanfani}, V. and {Farren}, G.~S. and {Fergusson}, J. and {Ferraro}, S. and {Flauger}, R. and {Foster}, A. and {Freese}, K. and {Frisch}, J.~C. and {Frolov}, A. and {Fuller}, G. and {Galitzki}, N. and {Gallardo}, P.~A. and {Galvez Ghersi}, J.~T. and {Ganga}, K. and {Gao}, J. and {Garrido}, X. and {Gawiser}, E. and {Gerbino}, M. and {Gerras}, R. and {Giardiello}, S. and {Gill}, A. and {Gilles}, V. and {Giri}, U. and {Gleave}, E. and {Gluscevic}, V. and {Goeckner-Wald}, N. and {Golec}, J.~E. and {Gordon}, S. and {Gralla}, M. and {Gratton}, S. and {Green}, D. and {Groh}, J.~C. and {Groppi}, C. and {Guan}, Y. and {Gupta}, N. and {Gudmundsson}, J.~E. and {Hagstotz}, S. and {Hargrave}, P. and {Haridas}, S. and {Harrington}, K. and {Harrison}, I. and {Hasegawa}, M. and {Hasselfield}, M. and {Haynes}, V. and {Hazumi}, M. and {He}, A. and {Healy}, E. and {Henderson}, S.~W. and {Hensley}, B.~S. and {Hertig}, E. and {Herv{\'\i}as-Caimapo}, C. and {Higuchi}, M. and {Hill}, C.~A. and {Hill}, J.~C. and {Hilton}, G. and {Hilton}, M. and {Hincks}, A.~D. and {Hinshaw}, G. and {Hlo{\v{z}}ek}, R. and {Ho}, A.~Y.~Q. and {Ho}, S. and {Ho}, S.~P. and {Hoang}, T.~D. and {Hoh}, J. and {Hornecker}, E. and {Hornsby}, A.~L. and {Hotinli}, S.~C. and {Huang}, Z. and {Huber}, Z.~B. and {Hubmayr}, J. and {Huffenberger}, K. and {Hughes}, J.~P. and {Idicherian Lonappan}, A. and {Ikape}, M. and {Irwin}, K. and {Iuliano}, J. and {Jaffe}, A.~H. and {Jain}, B. and {Jense}, H.~T. and {Jeong}, O. and {Johnson}, A. and {Johnson}, B.~R. and {Johnson}, M. and {Jones}, M. and {Jost}, B. and {Kaneko}, D. and {Karpel}, E.~D. and {Kasai}, Y. and {Katayama}, N. and {Keating}, B. and {Keller}, B. and {Keskitalo}, R. and {Kim}, J. and {Kisner}, T. and {Kiuchi}, K.},
        title = "{The Simons Observatory: science goals and forecasts for the enhanced Large Aperture Telescope}",
      journal = {\jcap},
         year = 2025,
        month = aug,
       volume = {2025},
       number = {8},
          eid = {034},
        pages = {034},
          doi = {10.1088/1475-7516/2025/08/034},
archivePrefix = {arXiv},
       eprint = {2503.00636},
 primaryClass = {astro-ph.IM},
       adsurl = {https://ui.adsabs.harvard.edu/abs/2025JCAP...08..034A}
}

@ARTICLE{2024arXiv240500809B,
       author = {{Bloch}, Richard and {Johnson}, Matthew C.},
        title = "{Kinetic Sunyaev Zel'dovich velocity reconstruction from Planck and unWISE}",
      journal = {arXiv e-prints},
         year = 2024,
        month = may,
          eid = {arXiv:2405.00809},
        pages = {arXiv:2405.00809},
          doi = {10.48550/arXiv.2405.00809},
archivePrefix = {arXiv},
       eprint = {2405.00809},
 primaryClass = {astro-ph.CO},
       adsurl = {https://ui.adsabs.harvard.edu/abs/2024arXiv240500809B}
}

@ARTICLE{2008PhRvD..77l3514D,
       author = {{Dalal}, Neal and {Dor{\'e}}, Olivier and {Huterer}, Dragan and {Shirokov}, Alexander},
        title = "{Imprints of primordial non-Gaussianities on large-scale structure: Scale-dependent bias and abundance of virialized objects}",
      journal = {\prd},
         year = 2008,
        month = jun,
       volume = {77},
       number = {12},
          eid = {123514},
        pages = {123514},
          doi = {10.1103/PhysRevD.77.123514},
archivePrefix = {arXiv},
       eprint = {0710.4560},
 primaryClass = {astro-ph},
       adsurl = {https://ui.adsabs.harvard.edu/abs/2008PhRvD..77l3514D}
}

@book{messiah61,
  author = {Messiah, Albert},
  publisher = {Elsevier Science B.V.},
  title = {Quantum Mechanics Volume II},
  year = 1961
}

@ARTICLE{2024PhRvD.110f3524K,
       author = {{Khek}, Brandon and {Gebhardt}, Henry Grasshorn and {Dor{\'e}}, Olivier},
        title = "{Fast theoretical predictions for spherical Fourier analysis of large-scale structures}",
      journal = {\prd},
         year = 2024,
        month = sep,
       volume = {110},
       number = {6},
          eid = {063524},
        pages = {063524},
          doi = {10.1103/PhysRevD.110.063524},
archivePrefix = {arXiv},
       eprint = {2212.05760},
 primaryClass = {astro-ph.CO},
       adsurl = {https://ui.adsabs.harvard.edu/abs/2024PhRvD.110f3524K}
}

@article{Bruton:2026ykj,
    author = "Bruton, Sean and Cheshire, James R. and Dor{\'e}, Olivier and Gebhardt, Henry S. Grasshorn and Wen, Robin Y.",
    title = "{Demonstrating the Use of the Spherical Fourier Bessel Basis for Large Scale Clustering Systematics Discovery and Mitigation with eBOSS}",
    eprint = "2605.10935",
    archivePrefix = "arXiv",
    primaryClass = "astro-ph.CO",
    month = "5",
    year = "2026"
}

@ARTICLE{2020A&A...641A...6P,
       author = {{Planck Collaboration} and {Aghanim}, N. and {Akrami}, Y. and {Ashdown}, M. and {Aumont}, J. and {Baccigalupi}, C. and {Ballardini}, M. and {Banday}, A.~J. and {Barreiro}, R.~B. and {Bartolo}, N. and {Basak}, S. and {Battye}, R. and {Benabed}, K. and {Bernard}, J.-P. and {Bersanelli}, M. and {Bielewicz}, P. and {Bock}, J.~J. and {Bond}, J.~R. and {Borrill}, J. and {Bouchet}, F.~R. and {Boulanger}, F. and {Bucher}, M. and {Burigana}, C. and {Butler}, R.~C. and {Calabrese}, E. and {Cardoso}, J.-F. and {Carron}, J. and {Challinor}, A. and {Chiang}, H.~C. and {Chluba}, J. and {Colombo}, L.~P.~L. and {Combet}, C. and {Contreras}, D. and {Crill}, B.~P. and {Cuttaia}, F. and {de Bernardis}, P. and {de Zotti}, G. and {Delabrouille}, J. and {Delouis}, J.-M. and {Di Valentino}, E. and {Diego}, J.~M. and {Dor{\'e}}, O. and {Douspis}, M. and {Ducout}, A. and {Dupac}, X. and {Dusini}, S. and {Efstathiou}, G. and {Elsner}, F. and {En{\ss}lin}, T.~A. and {Eriksen}, H.~K. and {Fantaye}, Y. and {Farhang}, M. and {Fergusson}, J. and {Fernandez-Cobos}, R. and {Finelli}, F. and {Forastieri}, F. and {Frailis}, M. and {Fraisse}, A.~A. and {Franceschi}, E. and {Frolov}, A. and {Galeotta}, S. and {Galli}, S. and {Ganga}, K. and {G{\'e}nova-Santos}, R.~T. and {Gerbino}, M. and {Ghosh}, T. and {Gonz{\'a}lez-Nuevo}, J. and {G{\'o}rski}, K.~M. and {Gratton}, S. and {Gruppuso}, A. and {Gudmundsson}, J.~E. and {Hamann}, J. and {Handley}, W. and {Hansen}, F.~K. and {Herranz}, D. and {Hildebrandt}, S.~R. and {Hivon}, E. and {Huang}, Z. and {Jaffe}, A.~H. and {Jones}, W.~C. and {Karakci}, A. and {Keih{\"a}nen}, E. and {Keskitalo}, R. and {Kiiveri}, K. and {Kim}, J. and {Kisner}, T.~S. and {Knox}, L. and {Krachmalnicoff}, N. and {Kunz}, M. and {Kurki-Suonio}, H. and {Lagache}, G. and {Lamarre}, J.-M. and {Lasenby}, A. and {Lattanzi}, M. and {Lawrence}, C.~R. and {Le Jeune}, M. and {Lemos}, P. and {Lesgourgues}, J. and {Levrier}, F. and {Lewis}, A. and {Liguori}, M. and {Lilje}, P.~B. and {Lilley}, M. and {Lindholm}, V. and {L{\'o}pez-Caniego}, M. and {Lubin}, P.~M. and {Ma}, Y.-Z. and {Mac{\'\i}as-P{\'e}rez}, J.~F. and {Maggio}, G. and {Maino}, D. and {Mandolesi}, N. and {Mangilli}, A. and {Marcos-Caballero}, A. and {Maris}, M. and {Martin}, P.~G. and {Martinelli}, M. and {Mart{\'\i}nez-Gonz{\'a}lez}, E. and {Matarrese}, S. and {Mauri}, N. and {McEwen}, J.~D. and {Meinhold}, P.~R. and {Melchiorri}, A. and {Mennella}, A. and {Migliaccio}, M. and {Millea}, M. and {Mitra}, S. and {Miville-Desch{\^e}nes}, M.-A. and {Molinari}, D. and {Montier}, L. and {Morgante}, G. and {Moss}, A. and {Natoli}, P. and {N{\o}rgaard-Nielsen}, H.~U. and {Pagano}, L. and {Paoletti}, D. and {Partridge}, B. and {Patanchon}, G. and {Peiris}, H.~V. and {Perrotta}, F. and {Pettorino}, V. and {Piacentini}, F. and {Polastri}, L. and {Polenta}, G. and {Puget}, J.-L. and {Rachen}, J.~P. and {Reinecke}, M. and {Remazeilles}, M. and {Renzi}, A. and {Rocha}, G. and {Rosset}, C. and {Roudier}, G. and {Rubi{\~n}o-Mart{\'\i}n}, J.~A. and {Ruiz-Granados}, B. and {Salvati}, L. and {Sandri}, M. and {Savelainen}, M. and {Scott}, D. and {Shellard}, E.~P.~S. and {Sirignano}, C. and {Sirri}, G. and {Spencer}, L.~D. and {Sunyaev}, R. and {Suur-Uski}, A.-S. and {Tauber}, J.~A. and {Tavagnacco}, D. and {Tenti}, M. and {Toffolatti}, L. and {Tomasi}, M. and {Trombetti}, T. and {Valenziano}, L. and {Valiviita}, J. and {Van Tent}, B. and {Vibert}, L. and {Vielva}, P. and {Villa}, F. and {Vittorio}, N. and {Wandelt}, B.~D. and {Wehus}, I.~K. and {White}, M. and {White}, S.~D.~M. and {Zacchei}, A. and {Zonca}, A.},
        title = "{Planck 2018 results. VI. Cosmological parameters}",
      journal = {\aap},
         year = 2020,
        month = sep,
       volume = {641},
          eid = {A6},
        pages = {A6},
          doi = {10.1051/0004-6361/201833910},
archivePrefix = {arXiv},
       eprint = {1807.06209},
 primaryClass = {astro-ph.CO},
       adsurl = {https://ui.adsabs.harvard.edu/abs/2020A&A...641A...6P}
}

@ARTICLE{2011JCAP...07..034B,
       author = {{Blas}, Diego and {Lesgourgues}, Julien and {Tram}, Thomas},
        title = "{The Cosmic Linear Anisotropy Solving System (CLASS). Part II: Approximation schemes}",
      journal = {\jcap},
         year = 2011,
        month = jul,
       volume = {2011},
       number = {7},
          eid = {034},
        pages = {034},
          doi = {10.1088/1475-7516/2011/07/034},
archivePrefix = {arXiv},
       eprint = {1104.2933},
 primaryClass = {astro-ph.CO},
       adsurl = {https://ui.adsabs.harvard.edu/abs/2011JCAP...07..034B}
}
